\documentclass[twoside,12pt]{article}
\usepackage{cite}
\usepackage{epsfig}
\usepackage{graphicx}
\usepackage{amsmath}
\usepackage{color}
\usepackage{comment}

\usepackage{lineno}

\renewcommand{\vec}[1]{\mbox{\boldmath $#1$}}

\newcommand{\be}{\begin{equation}}
\newcommand{\ee}{\end{equation}}
\newcommand{\bea}{\begin{eqnarray}}
\newcommand{\eea}{\end{eqnarray}}

\usepackage{xcolor}
\newcommand{\zbx}{Z_x(\vec{k}_b,\xi,\vecr_x)}
\newcommand{\psix}{\varphi_x(\vec{k}_b,\vecr_x)}

\newcommand{\psixaus}{\varphi^\mathrm{3b}_x} 

\newcommand{\PsiTB}{\Psi^\mathrm{3b(+)}}
\newcommand{\vecr}{{\vec r}}
\newcommand{\vecR}{{\vec R}}
\usepackage[normalem]{ulem}
\usepackage{xcolor}

\usepackage{amsfonts}
\usepackage{amssymb}

\begin{document}

\title{Coupled-channels calculations for nuclear reactions: from exotic nuclei 
to superheavy elements}

\author{K. Hagino$^a$, K. Ogata$^{b,c,d}$, A.M. Moro$^{e,f}$ \\
\\ 
$^a${\small Department of Physics, Kyoto University, Kyoto 606-8502, Japan}\\
$^b${\small Research Center for Nuclear Physics (RCNP), Osaka University, Ibaraki 567-0047, Japan}\\
$^c${\small Department of Physics, Osaka City University, Osaka 558-8585, Japan}\\
$^d${\small Nambu Yoichiro Institute of Theoretical and Experimental Physics (NITEP), Osaka City University,} \\{\small  Osaka 558-8585, Japan}\\
$^e${\small Department de FAMN, Universidad de Sevilla, Apartado 1065, E-41080 Sevilla, Spain} \\
$^f${\small Instituto Interuniversitario Carlos I de F\'{\i}sica Te\'{o}rica y Computacional (iC1), Apdo. 1065, E-41080 Sevilla, Spain}
}

\maketitle

\begin{abstract} 
Atomic nuclei are composite systems, and they may be dynamically excited 
during nuclear reactions. 
Such excitations are not only relevant to inelastic scattering but they also 
affect other reaction processes such as elastic scattering and fusion. 
The coupled-channels approach is a framework which can describe these 
reaction processes in a unified manner. It expands the total wave function of the 
system in terms of the ground and excited states of the colliding nuclei,  
and solves the coupled 
Shr\"odinger equations to obtain the $S$-matrix, from which several cross sections 
can be constructed. 
This approach has been a standard tool to analyze experimental 
data for nuclear reactions. 
In this paper, we review the present status and the recent developments of the coupled-channels 
approach. This includes the microscopic coupled-channels 
method and its application to cluster physics, the continuum discretized 
coupled-channels (CDCC) method for breakup reactions, the semi-microscopic 
approach to heavy-ion subbarrier fusion reactions, the channel coupling effects 
on nuclear astrophysics and 
syntheses of superheavy elements, and inclusive breakup and incomplete fusion 
reactions of weakly-bound nuclei. 
\end{abstract}
\tableofcontents

\section{Introduction}

The coupled-channels approach is a quantum mechanical reaction theory 
which takes into account internal excitations of the projectile and the target 
nuclei. 
For scattering of point particles, only elastic scattering takes place, which  
can be described by the Schr\"odinger equation for a two-body system with some 
potential between the colliding particles. 
In marked contrast, collisions of composite 
particles, such as atomic nuclei, exhibit a variety of phenomena, including 
elastic scattering, inelastic scattering, particle transfer, breakup, 
and fusion. Those process are not independent, but significantly affect each other. 
The two-body framework then has to be 
extended by explicitly taking into account  
the interplay of these processes. 
Such formalism is referred to as the coupled-channels (CC) approach, 
and has been a standard method in low-energy nuclear reactions. 
The framework is called the coupled-reaction-channels (CRC) approach 
when transfer reactions are involved. 

Inevitably, one must resort to several approximations when one would like to 
take into account the excitations during reaction processes, given that exact solutions of 
a many-body Hamiltonian for the ground state and excited states are unknown.  Nevertheless, 
one can either utilize to a large extent measured excitation energies and transition 
probabilities, or compute them theoretically with a good accuracy, providing 
a natural framework for low-energy nuclear reactions. 
Notice that 
an energy dependent complex optical potential for elastic scattering can also be 
formulated based on the coupled-channels approach. 
However, such formalism cannot describe inelastic channels, and 
the coupled-channels approach is indispensable when one
considers inelastic scattering, including breakup scattering. Also, 
as we shall see later, the so called fusion barrier distribution is intimately connected to the coupled-channels dynamics (see Sec. 5.2.2). 

Earlier developments of the coupled-channels approach are well summarized 
in a review article by Tamura \cite{tamura1965} as well as in several textbooks on nuclear 
reactions \cite{satchler1983,frobrich1996,broglia2004,bertulani2004,thompson2009,canto2013}.  
There are computer codes available for coupled-channels calculations, e.g., {\tt ECIS} \cite{ecis,lepine-szily2021}, 
{\tt FRESCO} \cite{thompson2009,fresco}, and {\tt CCFULL}\cite{ccfull}, 
which are still of current use in analysing experimental data of several nuclear reactions. 
We mention that the coupled-channels approach is not restricted to nuclear 
reaction studies but can also be applied to nuclear structure 
\cite{rost1967,hagino2020,kruppa2000,esbensen2000,davids2001,hagino2001,hagino2004,wang2017,wang2018}.  
In addition to those conventional approaches which the codes {\tt ECIS}, {\tt FRESCO}, 
and  {\tt CCFULL} follow, there have also been new developments in the physics of 
coupled-channels approach.
The first is a treatment of continuum states in connection to neutron-rich nuclei.
When a projectile (or a target) nucleus is weakly bound, it is likely that the excited states populated during the reaction are in the continuum 
spectrum. The coupled-channels formalism can be extended to this case as well by discretizing the continuum 
states. This method is referred to as the continuum discretized coupled-channels (CDCC) method 
\cite{thompson2009,kamimura1986,austern1987,yahiro2012}, and has been playing an important role in physics of 
neutron-rich nuclei \cite{yahiro2012}. In particular, nuclear structure information of 
exotic nuclei is often extracted using 
nuclear reactions, and the CDCC approach has provided a powerful tool for that purpose. 
Of course, the reaction dynamics of neutron-rich nuclei itself is intriguing, as several reaction processes are 
considerably affected by the breakup process \cite{canto2006,canto2015}. 

Another important recent direction of the coupled-channels approach 
is a development of a microscopic coupled-channels 
approach \cite{takashina2010,furumoto2013,minomo2016,kim2008}. See also 
Ref. \cite{descouvemont2013} for a related work. In the conventional coupled-channels approach, 
the coupling potentials are often constructed based on the phenomenological 
collective model \cite{broglia2004}. The imaginary part of an internuclear potential is also 
introduced phenomenologically. In contrast, in the microscopic coupled-channels approach, the 
coupling Hamiltonians are constructed using the double folding procedure \cite{satchler1979} with transition densities 
obtained with microscopic nuclear structure calculations. The internuclear potential is also 
constructed with the double folding procedure with a complex $G$-matrix interaction. 
An advantage of this method over the conventional coupled-channels method is that it can 
be applied with a larger reliability to nuclear reactions in an unknown region in which there is no experimental data. Moreover, the imaginary part of the optical potential does not have to be supplemented phenomenologically.  Recently, this approach has been successfully applied to elastic and inelastic 
scattering to discuss the cluster structure of light nuclei \cite{kanada-en'yo2019,kanada-en'yo2019b,kanada-en'yo2019c,kanada-en'yo2020,kanada-en'yo2020d,kanada-en'yo2021}. 

In this paper, we review the recent developments of the coupled-channels approach, including the CDCC 
and the microscopic coupled-channels approach are 
important recent developments. 
We shall also cover heavy-ion fusion 
reactions at energies around the Coulomb barrier, at which the channel coupling effects play a crucial 
role. In particular, we shall discuss the role of channel couplings in synthesis of superheavy elements.  

The paper is organized as follows. In the next section, we shall detail the basic formalism of the coupled-channels method. To this end, we will start with a simple one-dimensional system and then discuss 
three-dimensional systems. In the latter case, the angular momentum coupling has to be properly taken 
into account. We will also discuss the semi-classical approximation to the coupled-channels approach, 
in which the internal excitations are described as a time evolution of intrinsic wave functions whereas the 
relative motion is treated classically. 
In Sec. 3, we will apply the coupled-channels approach to direct reactions. 
We will focus on the microscopic CC description of direct reactions based on the bare nucleon-nucleon interaction. The framework will be briefly reviewed in Sec.~\ref{sec31} and its applications to elastic and inelastic scattering of nuclei will be discussed in Sec.~\ref{sec32}.
As an accurate and efficient reaction model applicable to reactions of loosely bound nuclei, we will briefly introduce the CDCC 
in Sec.~\ref{sec33}. 
Its recent applications to studies for revealing structures and dynamical properties of unstable nuclei will be reviewed in Sec.~\ref{sec34}. Finally, an extension of CDCC to include the excitation of the core and target nuclei will be discussed in Sec.~\ref{sec:coretargex}.
In Sec. 4, we will discuss recent developments on the theory 
of inclusive breakup reactions, putting some emphasis on the Ichimura-Austern-Vincent (IAV) method.  
In Sec. 5, we will discuss heavy-ion fusion reactions. After a brief discussion on the coupled-channels 
approach to nuclear astrophysical reactions, we will discuss the fusion reactions in medium-heavy systems. 
Here, it is known that the channel coupling effects considerably enhance fusion cross sections as compared 
to a prediction of a simple potential model. 
We will then discuss the role of channel couplings, especially the role of deformation of a target nucleus, 
in fusion reaction for superheavy elements.  
Fusion of weakly bound nuclei will also be discussed in this section.
We will then summarize the paper in Sec. 6. 

\section{Coupled-channels approach}
\label{sec:cca}

\subsection{One-dimensional systems}

Before we consider the coupled-channels formalism in three-dimensional systems, it is instructive 
to discuss its application to one-dimensional systems. For this purpose, let us denote the one-dimensional coordinate as $x$ and consider the problem in which a particle 
with mass $m$ approaches a potential barrier $V(x)$ from the right hand side. 
If one denotes the intrinsic degree of freedom of the particle as $\xi$, the total Hamiltonian 
for this system reads,
\begin{equation}
H=-\frac{\hbar^2}{2m}\frac{d^2}{dx^2}+V(x)+H_0(\xi)+V_{\rm coup}(x,\xi),
\label{eq:totalH1}
\end{equation}
where $H_0(\xi)$ is the Hamiltonian for the intrinsic motion and $V_{\rm coup}$ represents 
the coupling term between $x$ and $\xi$. 
Let us assume that the eigenenergies and the eigenfunctions of $H_0$ are all known and are given 
by 
\begin{equation}
H_0(\xi)\phi_n(\xi)=\epsilon_n\phi_n(\xi), 
\label{eq:channel}
\end{equation}
where $n=0$ denotes the ground state. We shall scale the energy such that the ground state 
energy, $\epsilon_0$, is zero.  

We expand the total wave function of the system, $\Psi(x,\xi)$, using the eigenfunctions of $H_0$ as basis 
functions. This leads to 
\begin{equation}
\Psi(x,\xi)=\sum_nu_n(x)\phi_n(\xi).
\end{equation}
Substituting this expression into the Schr\"odinger equation for the whole system and 
projecting it onto a particular state $\phi_n$, one obtains coupled equations given by 
\begin{equation}
0=\langle\phi_n|H-E|\Psi\rangle=\left[-\frac{\hbar^2}{2m}\frac{d^2}{dx^2}+V(x)+\epsilon_n-E\right]u_n(x)
+\sum_{n'}F_{nn'}(x)u_{n'}(x), 
\label{cc-eq1d}
\end{equation}
where $E$ is the total energy, and the coupling potential $F_{nn'}(x)$ is given by 
\begin{equation}
F_{nn'}(x)=\langle\phi_n|V_{\rm coup}(x,\xi)|\phi_{n'}\rangle=\int d\xi\, \phi_n^*(\xi)V_{\rm coup}(x,\xi)\phi_{n'}(\xi).
\end{equation}
Each $n$ is referred to as a channel, and thus 
the equations (\ref{cc-eq1d}) are called the coupled-channels equations. 

The coupled-channels equations (\ref{cc-eq1d}) are solved under the boundary conditions given by 
\begin{eqnarray}
u_n(x)&\to& e^{-ik_nx}\delta_{n,0}-\sqrt{\frac{k_0}{k_n}}\,R_{n}e^{ik_nx}~~~~(x\to\infty), \\
&\to& \sqrt{\frac{k_0}{k_n}}
\,T_{n}e^{-ik_nx}~~~~~~~~~~~~~~~~~~~(x\to -\infty),
\end{eqnarray}
where $k_n=\sqrt{2m(E-\epsilon_n)/\hbar^2}$ is the wave number for the channel $n$. 
We have assumed that the system is in the ground state before the scattering takes place.
The penetration and the reflection probabilities are given by 
\begin{equation}
P(E)=\sum_n|T_{n}|^2, 
\label{eq:p1d}
\end{equation}
and
\begin{equation}
R(E)=\sum_n|R_{n}|^2, 
\end{equation}
respectively. Because of the flux conservation, these are related to each other 
with the relation $P(E)+R(E)=1$. 
The multi-channel penetrability, Eq.~(\ref{eq:p1d}), can be evaluated in the WKB approximation as well \cite{hagino2004b}. 

\begin{figure}[tb]
\begin{center}
\includegraphics*[width=0.5\textwidth]{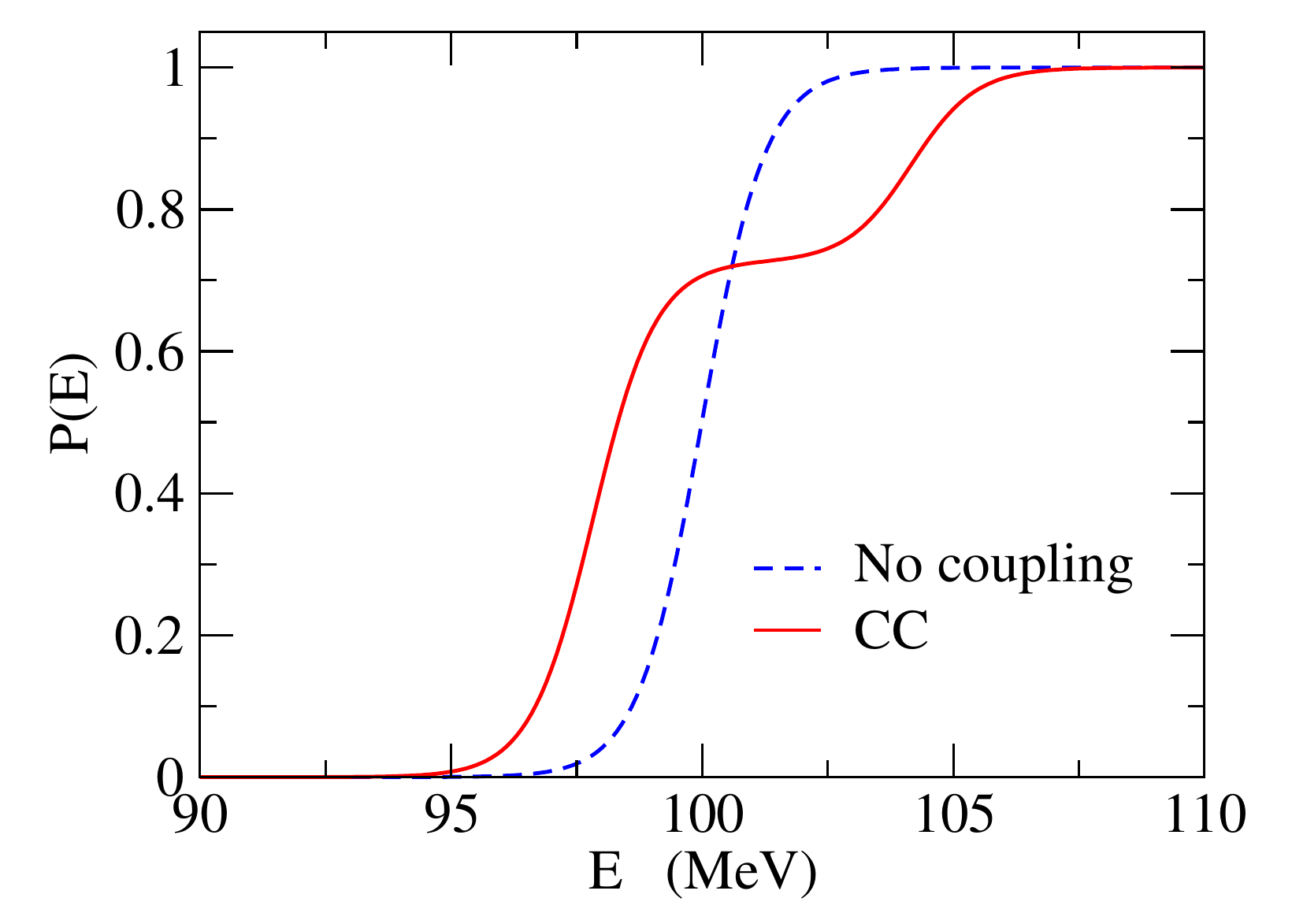}
\caption{
The penetrability for a two-channel system in one-dimension. 
The potential barrier is assumed to have a Gaussian form, 
$V(x)=V_0\,e^{-x^2/2\sigma^2}$ with 
$V_0=100$ MeV and $\sigma=3$ fm. 
The coupling potential is also assumed to have a Gaussian form, 
$F_{01}(x)=F_{10}(x)=F_0\,e^{-x^2/2\sigma^2}$ with $F_0=3$ MeV. 
The mass of the particle is assumed to be $m=29m_N$ with 
$m_Nc^2=938$ MeV, and the energy of the excited channel is set to be $\epsilon_1=2$ MeV. 
The dashed line denotes the result without the channel coupling, while the solid line shows the result of the 
coupled-channels calculations.}
\label{fig:cc2ch}
\end{center}
\end{figure}

Figure \ref{fig:cc2ch} shows the penetrability $P(E)$ for a two-channel problem with $n=0$ and 1 in one dimension. 
To draw this figure, we assume a Gaussian 
potential barrier, $V(x)=V_0\,e^{-x^2/2\sigma^2}$, as well as 
a Gaussian form of the coupling potential, 
$F_{01}(x)=F_{10}(x)=F_0\,e^{-x^2/2\sigma^2}$. 
We have set $F_{00}(x)=F_{11}(x)=0$. 
The dashed line shows the result without the coupling (i.e., the case with $F_0=0$), while the solid line is obtained by solving 
the coupled-channels equations. As has been pointed out by Dasso 
{\it et al.}~\cite{dasso1983a,dasso1983b}, the penetrability is enhanced 
by the coupling at energies below the barrier while it is hindered at energies above the barrier. The former effect is intimately related to the subbarrier enhancement 
of fusion cross sections, which we shall discuss in Sec. 5. 

\subsection{Three-dimensional systems}

Let us now discuss more realistic three-dimensional systems, in which 
the relative motion $R$ between two nuclei couples to an internal degree of 
freedom, $\xi$.
To this end, we consider the following total Hamiltonian: 
\begin{equation}
H=-\frac{\hbar^2}{2\mu}\vec{\nabla}^2+V(R)-iW(R)+H_0(\xi)+V_{\rm coup}(\vec{R},\xi),
\label{eq:H3d}
\end{equation}
where $\mu$ is the reduced mass. 
Here, we have introduced the 
imaginary part $-iW(R)$ to the potential term in order to simulate reaction 
processes going outside the model space given by $\xi$.
In the three-dimensional case, the eigenstates of the internal Hamiltonian, $H_0$, in general have a finite spin.  
The eigenstates of $H_0$ are thus labeled as $\varphi_{n I m_I}(\xi)$,
where $I$ is the intrinsic angular momentum, $m_I$ is its $z$-component, and $n$ denotes 
quantum numbers besides the angular momentum. 
Those states are coupled to the orbital angular momentum $\vec{L}$ of the relative motion between two nuclei 
and form the channel wave functions given by 
\begin{equation}
\langle \hat{\vec{R}}\xi\vert(n LI)J_TM_T\rangle
=\sum_{m_L,m_I}i^L\langle Lm_LIm_I\vert J_TM_T\rangle 
Y_{Lm_L}(\hat{\vec{R}})\varphi_{n Im_I}(\xi),
\label{eq:channelwf}
\end{equation}
where $J_T$ is the total angular momentum, $\vec{J}_T=\vec{L}+\vec{I}$, and $M_T$ is its $z$-component. 
The total wave function is then expanded as
\begin{equation}
\Psi_{J_TM_T}(\vec{R},\xi)=
\sum_{n,L,I}\frac{\chi^{J_T}_{cc_0}(R)}{KR}
\langle \hat{\vec{R}}\xi\vert(n LI)J_TM_T\rangle, 
\end{equation}
where $K=\sqrt{2\mu E/\hbar^2}$ is the wave number of the entrance channel, and $c$ is a simplified notation for $\{n,L,I\}$. Here, we explicitly denote the entrance channel $c_0=\{n_0,L_0,I_0\}$ in the radial wave function.  
We expand the coupling potential $V_{\rm coup}(\vec{R},\xi)$ into multipoles, 
\begin{equation}
V_{\rm coup}(\vec{R},\xi)
=\sum_{\lambda,\mu}f_{\lambda}(R)Y^*_{\lambda\mu}
(\hat{\vec{R}})
T_{\lambda\mu}(\xi),
\label{eq:Vcoup}
\end{equation}
where $Y_{\lambda\mu}(\hat{\vec{R}})$ 
and $T_{\lambda\mu}(\xi)$ are the spherical harmonics and 
the spherical tensors for the intrinsic degree of freedom, respectively. 
The matrix elements of the coupling potential between the channel wave functions read (see Eq.~(7.1.6) in Ref.~\cite{edmonds})
\begin{eqnarray}
V^{J_T}_{cc'}(R)
&=&\langle (n LI)J_TM_T\vert V_{\rm coup}(\vec{R},\xi)\vert 
(n'L'I')J_TM_T\rangle, \\
&=&\sum_{\lambda}(-)^{L'+I+J_T}
f_{\lambda}(R)\langle L||Y_{\lambda}||L'\rangle\langle 
n I||T_{\lambda}||n'I'\rangle \left\{
\begin{array}{ccc}
J_T&I&L \\
\lambda&L' &I'
\end{array}\right\},  
\end{eqnarray}
where $\langle L||Y_{\lambda}||L'\rangle$ is the reduced matrix element and similar for 
$\langle n I||T_{\lambda}||n'I'\rangle$. 
Notice that these matrix elements do not depend on $M_T$. 

With these coupling matrix elements, the coupled-channels equations for $\chi^{J_T}_{cc_0}(R)$ read,
\begin{equation}
\left[-\frac{\hbar^2}{2\mu}\frac{d^2}{dR^2}
+\frac{L_c(L_c+1)\hbar^2}{2\mu R^2}+V(R)-iW(R)
-E+\epsilon_{n I}\right]\chi^{J_T}_{cc_0}(R)
+\sum_{c'}V^{J_T}_{cc'}(R)\chi^{J_T}_{c'c_0}(R)=0, 
\label{eq:cc3d}
\end{equation}
where $\epsilon_{n I}$ is the energy of the 
intrinsic state,  $\varphi_{n I m_I}(\xi)$. 
These equations are solved with the boundary condition of 
\begin{equation}
\chi^{J_T}_{cc_0}(R)\to 
H_{L_c}^{(-)}(K_cR)\delta_{c,c_0}
-\sqrt{\frac{K}{K_c}}\,S^{J_T}_{cc_0}H_{L_c}^{(+)}(K_cR)
~~~~~~~~~~~~~~(R\to\infty)
\label{eq:ccboundary3}
\end{equation}
where $S^{J_T}_{cc_0}$ is the nuclear $S$-matrix, 
and $H_l^{(+)}(KR)$ and  $H_l^{(-)}(KR)$ are the 
outgoing and the incoming Coulomb wave functions with $K_c=\sqrt{2\mu(E-\epsilon_c)/\hbar^2}$, 
respectively. This boundary condition is supplemented 
by the regular boundary 
condition at the origin, which is sometimes 
replaced by the incoming wave boundary condition (IWBC) under an assumption of strong absorption 
inside the Coulomb barrier \cite{dasso1983a,dasso1983b,landowne1984,hagino2012,wen2020}. 
From the nuclear $S$-matrix, the scattering amplitude to populate the intrinsic state with $\{n_c,I_c,m_c\}$  is constructed as \cite{esbensen1987}   
\begin{eqnarray}
f_{n_c I_c m_c}(\theta,\phi)
&=&\frac{i\sqrt{\pi}}{K}\,\sum_{J_T}\sum_{L_c}\,i^{J_T-L_c}\hat{J}_T\langle L_c -m_c I_c m_c|J_T0\rangle
Y_{L_c,-m_c}(\theta,\phi) 
\left(\delta_{c,c_0}-e^{i(\sigma_{J_T}(E)+\sigma_{l_c}(E-\epsilon_c)}S^{J_T}_{cc_0}\right) \nonumber \\
&&+f_C(\theta)\delta_{c,c_0},
\end{eqnarray}
with $\hat{J}\equiv\sqrt{2J+1}$. Here, $\sigma_l(E)$ and $f_C(\theta)$ are the Coulomb phase shift and the 
Coulomb scattering amplitude, respectively. 
The differential cross section is then computed as,
\begin{equation}
\frac{d\sigma_{n_c I_c m_c}}{d\Omega}=
|f_{n_c I_c m_c}(\theta,\phi)|^2.
\end{equation}
The total absorption cross section is expressed as, 
\begin{equation}
\sigma_{\rm abs}=\frac{\pi}{K^2}\sum_{J_T} (2J_T+1)\left(1-\sum_c |S^{J_T}_{cc_0}|^2\right). 
\label{eq:sigma_abs}
\end{equation}
Notice that the absorption cross section can also be expressed as \cite{canto2013}
\begin{equation}
\sigma_{\rm abs}=\frac{\pi}{K^2}\,\frac{2\mu}{K\hbar^2}\,\sum_{J_T}\sum_c (2J_T+1)
\int_0^\infty dR\,W(R)\left|\chi^{J_T}_{cc_0}(R)\right|^2,
\end{equation}
with the boundary condition given by Eq.~(\ref{eq:ccboundary3}). 
This expression is numerically more convenient than Eq.~(\ref{eq:sigma_abs}) when the absolute value of $S$ matrix is close 
to 1, e.g., at energies far below the Coulomb barrier, even though the wave functions $\chi^{J_T}_{cc_0}(R)$ have to be stored. 
Incidentally, the formulation with the incoming wave boundary condition directly evaluates 
the quantity $1-|S^{J_T}_{nn_i}|^2$ \cite{hagino2012}, 
and the problem of a round-off error can be avoided also with this method. 

If the imaginary part of the potential, $W(R)$, is well localized inside the Coulomb barrier, 
the absorption due to $W(R)$ corresponds to  
a compound nucleus formation, and  
absorption cross sections $\sigma_{\rm abs}$ is  regarded as fusion cross sections, $\sigma_{\rm fus}$.
The formulation of fusion process with the incoming wave boundary condition is based on 
this idea. 

\subsection{Coupling potential with the collective model}
\label{sec:collect}

In conventional coupled-channels calculations, the coupling potential $V_{\rm coup}(\vec{R},\xi)$ is 
constructed based on the macroscopic collective model \cite{broglia2004,hagino2012}. For simplicity, let us consider 
excitations of a target nucleus while a projectile remains in the ground state. The radius of the target nucleus 
is expanded into multipoles as \cite{RS80}, 
\begin{equation}
R(\theta,\phi) = R_T \left(1+\sum_{\lambda\mu}\alpha_{\lambda\mu}
Y_{\lambda\mu}^*(\theta,\phi)
\right),
\end{equation}
where $R_T$ is the equivalent sharp surface radius. 
The nuclear part of the potential $V$ is then given by, 
\begin{equation}
V^{(N)}(\vec{R},\alpha_{\lambda\mu})
=V_N\left(R-R_T
\sum_{\lambda\mu}\alpha_{\lambda\mu}
Y_{\lambda\mu}^*(\hat{\vec{R}}) \right), 
\label{nucl-full}
\end{equation}
and similar for $W(R)$. To the first order of $\alpha_{\lambda\mu}$, the potential is expanded as, 
\begin{equation}
V^{(N)}(\vec{R},\alpha_{\lambda\mu})
=V_N(R)-R_T\frac{dV_N(R)}{dR}\sum_{\mu}\alpha_{\lambda\mu}
Y_{\lambda\mu}^*(\hat{\vec{R}}), 
\label{coupN}
\end{equation}
even though the higher order of $\alpha_{\lambda\mu}$ may play an important role, especially in 
heavy-ion subbarrier fusion reactions \cite{ccfull,hagino2012,hagino1997,EL87}. 
On the other hand, the Coulomb part of the potential 
is constructed as 
\begin{equation}
V_C(\vec{R})=\int d\vec{R}'
\frac{Z_PZ_Te^2}{|\vec{R}-\vec{R}'|}
\rho_T(\vec{R}')
=\frac{Z_PZ_Te^2}{R}+\sum_{\lambda\ne 0}\sum_{\mu}
\frac{4\pi Z_Pe}{2\lambda+1}
Q_{\lambda\mu}Y_{\lambda\mu}^*(\hat{\vec{R}})\frac{1}{R^{\lambda+1}},
\label{coupC}
\end{equation}
where $\rho_T$ is the charge density of the target nucleus, and $Z_P$ and $Z_T$ are 
the atomic numbers of the projectile and the target, respectively. 
$Q_{\lambda\mu}$ is the electric multipole operator defined by
\begin{equation}
Q_{\lambda\mu}=\int d\vec{R} Z_Te\rho_T(\vec{R})
R^{\lambda}Y_{\lambda\mu}(\hat{\vec{R}}). 
\end{equation}
To the first order of $\alpha_{\lambda\mu}$, $Q_{\lambda\mu}$ is approximated as 
\begin{equation}
Q_{\lambda\mu}=\frac{3e}{4\pi}Z_T R_T^{\lambda}
\alpha_{\lambda\mu},
\label{multipole}
\end{equation}
using the sharp-cut density for $\rho_T$. Combining the nuclear and the Coulomb parts of the 
potential, the coupling potential for an excitation mode with multipolarity $\lambda$ reads, 
\begin{equation}
V_{\rm coup}(\vec{R},\alpha_{\lambda}) = f_{\lambda}(R)
\sum_{\mu}\alpha_{\lambda\mu}
Y_{\lambda\mu}^*(\hat{\vec{R}}),
\label{vcoup}
\end{equation}
with 
\begin{equation}
f_{\lambda}(R)=-R_T\frac{dV_N}{dR}
+\frac{3}{2\lambda+1}Z_PZ_Te^2\frac{R_T^{\lambda}}
{R^{\lambda+1}}. 
\end{equation}
Notice that this is in the same form as Eq.~(\ref{eq:Vcoup}). 

For a vibrational motion in spherical nuclei, $\alpha_{\lambda\mu}$ is expressed in terms of phonon creation 
and annihilation  operators as,
\begin{equation}
\alpha_{\lambda\mu}=
\alpha_0 \left(a_{\lambda\mu}^{\dagger}+(-)^{\mu}
a_{\lambda\mu} \right)
=\frac{\beta_\lambda}{\sqrt{2\lambda+1}}\,
\left(a_{\lambda\mu}^{\dagger}+(-)^{\mu}
a_{\lambda\mu} \right),
\end{equation}
where
$\alpha_0=\beta_{\lambda}/\sqrt{2\lambda+1}$
is the amplitude of the zero-point motion\cite{bohr-mottelson2}.
The parameter $\beta_\lambda$ is often referred to as a deformation 
parameter, and can be estimated
from an experimental transition probability as 
\begin{equation}
\beta_\lambda=\frac{4\pi}{3Z_TR_T^{\lambda}}
\sqrt{\frac{B(E\lambda)\uparrow}{e^{2}}}.
\end{equation}
Assuming a harmonic vibration, and after subtracting the zero point energy, 
the intrinsic Hamiltonian, $H_0(\xi)$, is given by 
\begin{equation}
H_0=\hbar\omega_\lambda \sum_{\mu} a_{\lambda\mu}^{\dagger}a_{\lambda\mu}, 
\end{equation}
where $\hbar\omega_\lambda$ is the excitation energy of a one phonon state. 

For a rotational motion of axially deformed nuclei, $\alpha_{\lambda\mu}$ is expressed as 
\begin{equation}
\alpha_{\lambda\mu}=\beta_\lambda \sqrt{\frac{4\pi}{2\lambda+1}}\,Y_{\lambda\mu}(\hat{\vec{r}}_d), 
\label{eq:rot}
\end{equation}
where $\hat{\vec{r}}_d$ is the angle to specify the orientation of the symmetry axis 
of a deformed nucleus in the laboratory frame. The intrinsic Hamiltonian is given by 
\begin{equation}
H_0(\xi)=\frac{\vec{I}^2\hbar^2}{2{\cal J}}, 
\end{equation}
where $\vec{I}$ is the angular momentum associated with the angle $\hat{\vec{r}}_d$ and ${\cal J}$ 
is the moment of inertia. For the ground state rotational band of even-even nuclei, this provides the 
rotational energy of 
\begin{equation}
\epsilon_I=\frac{I(I+1)\hbar^2}{2{\cal J}}.  
\end{equation}

\subsection{Isocentrifugal approximation}

The dimension of the coupled-channels equations, Eq.~(\ref{eq:cc3d}), can be largely reduced if the 
isocentrifugal approximation is introduced \cite{hagino2012,esbensen1987,esbensen1987b}. 
In this approximation, the orbital angular momentum $L$ in the centrifugal potential 
in Eq.~(\ref{eq:cc3d}) is firstly replaced by the 
total angular momentum $J_T$, that is, 
\begin{equation}
\frac{L(L+1)\hbar^2}{2\mu R^2}\to \frac{J_T(J_T+1)\hbar^2}{2\mu R^2}, 
\end{equation}
and then the whole system is rotated such that the vector $\vec{R}$ is always along the $z$-axis. 
The coupling potential (\ref{eq:Vcoup}) is then transformed to  
\begin{equation}
V_{\rm coup}(\hat{\vec{R}}=0,\xi)
=\sum_{\lambda}\sqrt{\frac{2\lambda+1}{4\pi}}\,f_{\lambda}(R)T_{\lambda 0}(\xi).
\label{eq:vcoup-isocentrifugal}
\end{equation}
The coupled-channels equations (\ref{eq:cc3d}) are also transformed to \cite{hagino2012}
\begin{eqnarray}
&&\left(-\frac{\hbar^2}{2\mu}\frac{d^2}{dR^2}+
\frac{J_T(J_T+1)\hbar^2}{2\mu R^2}+V(R)-E+\epsilon_{n I}\right)
\bar{\chi}_{nI}(R)  \nonumber \\
&&~~~~~~~~~~+\sum_{n',I'}\sum_\lambda
\sqrt{\frac{2\lambda+1}{4\pi}}f_\lambda(R)
\langle\varphi_{\alpha I0}|T_{\lambda 0}|\varphi_{\alpha' I'0}\rangle
\bar{\chi}_{n' I'}(R)=0,
\label{eq:ccisocentrifugal}
\end{eqnarray}
with the wave functions $\bar{\chi}_{n I}(R)$ defined by 
\begin{equation}
\bar{\chi}_{n I}(r)=(-)^{I}\sum_{L}\langle J_T 0 I 0|L 0\rangle \chi_{n LI,c_0}(R). 
\label{eq:barwf}
\end{equation}
Notice that Eq.~(\ref{eq:barwf}) can be inverted as
\begin{equation}
\chi_{\alpha LI,c_0}(R)=
(-)^{I}\sum_{L}\langle J_T 0 I 0|L 0\rangle \bar{\chi}_{n I}(R). 
\label{eq:barwf2}
\end{equation}
The reduced coupled-channels equation, (\ref{eq:ccisocentrifugal}), 
are solved with the boundary condition of 
\begin{equation}
\bar{\chi}_{n I}(R)\to
H_{J_T}^{(-)}(K_IR)\delta_{n I,n_0I_0}
-\sqrt{\frac{K_{I_0}}{K_I}}\,\bar{S}^{J_T}_{n I,n_0I_0}H_{J_T}^{(+)}(K_IR),
~~~~~~~~~~~~~~R\to\infty. 
\label{eq:ccboundary3-isocentrifugal}
\end{equation}
Using the asymptotic form of the Coulomb wave functions in Eq.~(\ref{eq:barwf2}), 
one finds that the total (the nuclear + the Coulomb) $S$-matrix 
is expressed as,
\begin{equation}
e^{i(\sigma_{J_T}(E)+\sigma_{l}(E-\epsilon_{n_c I_c})}S^{J_T}_{cc_0}
=i^{L_c+L_0-2J}\,(-)^{I_c}\langle J_T 0 I_c 0|L_c 0\rangle\,
e^{i(\sigma_{J_T}(E)+\sigma_{J_T}(E-\epsilon_{n_c I_c}))}
\bar{S}^{J_T}_{n_c I_c,n_0I_0}.
\end{equation}

In the isocentrifugal approximation, the differential cross sections are given by 
\begin{equation}
\frac{d\sigma_{n I}}{d\Omega}
=|f_{n I}(\theta)|^2
\end{equation}
with the scattering amplitude given by 
\begin{equation}
f_{n I}(\theta)=\sum_{J_T}
e^{i[\sigma_{J_T}(E)+\sigma_{J_T}(E-\epsilon_{n I})]}
\sqrt{\frac{2J+1}{4\pi}}\,Y_{J_T0}(\theta)\,
\frac{-2i\pi}{K}(\bar{S}^{J_T}_{n I,n_0I_0}
-\delta_{n I,n_0I_0})
+f_C(\theta)\delta_{n I,n_0 I_0}. 
\end{equation}
On the other hand, the absorption cross sections are given by 
\begin{equation}
\sigma_{\rm abs}=\frac{\pi}{K^2}\sum_{J_T}(2J_T+1)\left(1-\sum_{n,I}\left|\bar{S}^{J_T}_{nI,n_0I_0}\right|^2\right).
\end{equation}

The isocentrifugal approximation works well when large values of the angular momentum $L$ do not 
contribute to a reaction process \cite{esbensen1987,HR04,T87}. 
Therefore, 
the approximation is suitable for heavy-ion fusion reactions, to  
which only small values of the angular momentum contribute \cite{ccfull,hagino2012}. 

The coupled-channels problem becomes particularly simple in the isocentrifugal approximation when 
the excitation energies of the intrinsic degree of freedom can be ignored. 
This is often the case for medium-heavy and heavy deformed nuclei, for which the energy of the first 2$^+$ state 
in the rotational band 
is considerably small. 
In this case, reaction processes can be formulated as an average of contributions 
of different orientation angles, 
with the angle dependent potential given by 
\begin{equation}
V(R,\theta_d)=V(R)-iW(R)+\sum_\lambda \beta_\lambda f_\lambda(R)Y_{\lambda 0}(\theta_d), 
\end{equation}
where $\theta_d$ denotes the angle between $\hat{\vec{R}}$ and $\hat{\vec{r}}_d$ 
(see Eqs. (\ref{eq:rot}) and (\ref{eq:vcoup-isocentrifugal})). 
For reaction systems with 
even-even nuclei, where the ground spin is zero, 
the cross sections for the elastic scattering, the quasi-elastic scattering, and the absorption (fusion) reactions 
are given by \cite{hagino2012,HR04,ARN88,NBT1986,TMBR1991}
\begin{eqnarray}
\frac{d\sigma_{\rm el}}{d\Omega}&=&\left|\int^1_0d(\cos\theta_d)\,f_{\rm el}(\theta,\theta_d)\right|^2, \\
\frac{d\sigma_{\rm qel}}{d\Omega}&=&\sum_{I}\frac{d\sigma_I}{d\Omega}
=\int^1_0d(\cos\theta_d)\,\frac{d\sigma_{\rm el}(\theta_d)}{d\Omega},
\end{eqnarray}
and
\begin{equation}
\sigma_{\rm fus}(E)=\int^1_0d(\cos\theta_d)\,\sigma_{\rm fus}(E;\theta_d),
\label{eq:fus-orientation}
\end{equation}
respectively. 
Here, $f_{\rm el}(\theta,\theta_d)$ is a single-channel scattering amplitude with the potential $V(R,\theta_d)$, while 
$d\sigma_{\rm el}(\theta_d)/d\Omega$ and $\sigma_{\rm fus}(E;\theta_d)$ are corresponding 
single-channel elastic and fusion 
cross sections, respectively. An interpretation of these formulas is that 
the moment of inertia for the rotational motion is so large that 
the orientation angle of the target 
nucleus does not change during the reactions, and thus cross sections can be evaluated by averaging the 
contributions of each orientation angle $\theta_d$. 

These formulas can be extended to the case with odd-mass nuclei as well  \cite{CJT1994,Hinde2003,HS2019}. 
The cross sections for a magnetic substate $M$, which is a projection of the 
nuclear spin of a odd-mass nucleus on the $z$-axis, read,  
\begin{eqnarray}
\frac{d\sigma^{(M)}_{\rm el}}{d\Omega}&=&\left|\int^1_0d(\cos\theta_d)w_M(\theta)
\,f_{\rm el}(\theta,\theta_d)\right|^2, \\
\frac{d\sigma^{(M)}_{\rm qel}}{d\Omega}&=&\sum_{I}\frac{d\sigma_I}{d\Omega}
=\int^1_0d(\cos\theta_d)w_M(\theta)
\,\frac{d\sigma_{\rm el}(\theta_d)}{d\Omega},
\end{eqnarray}
and
\begin{equation}
\sigma^{(M)}_{\rm fus}(E)=\int^1_0d(\cos\theta_d)w_M(\theta)
\,\sigma_{\rm fus}(E;\theta_d),
\end{equation}
with 
\begin{equation}
w_M(\theta)=\frac{2I_0+1}{2}
\left[\left|d^{I_0}_{MK_0}(\theta)\right|^2+\left|d^{I_0}_{M-K_0}(\theta)\right|^2\right],
\end{equation}
where $I_0$ and $K_0$ are the ground state spin and the $K$ quantum number 
for the ground state rotational band. 
For a magnetic substate $M'$ with a quantization axis which is along the direction of 
$\theta_{\rm ax}$ from the $z$-axis, the weight factor is given by 
\begin{equation}
\tilde{w}_{M'}(\theta)=\sum_M \left|d^{I_0}_{M'M}(-\theta_{\rm ax})\right|^2
w_M(\theta). 
\end{equation}
It has been pointed out that one can 
effectively change the sign of the hexadecapole deformation parameter $\beta_4$ 
by aligning deformed target nuclei\cite{HS2019}. 
Notice that for an unpolarized target, using the relation $\sum_M|d^I_{MK}(\theta)|^2=1$, 
the weight factor reads,
\begin{equation}
w^{(\rm unpol)}(\theta)=\frac{1}{2I_0+1}\sum_Mw_M(\theta)=1, 
\end{equation}
that is equivalent to the weight factor for even-even nuclei with $I_0=K_0=0$.

\subsection{Numerical methods}

\subsubsection{Direct integration}

The coupled-channels equations, (\ref{eq:cc3d}) or (\ref{eq:ccisocentrifugal}), can be computed in several ways. 
The most 
straightforward way to solve the coupled-channels equations 
is to integrate the equations directly with e.g., the Numerov \cite{koonin1990} or the modified Numerov 
\cite{melkanoff1966} methods. In this approach, one constructs $N$-linearly independent solutions of the coupled-channels 
equations, where $N$ is the dimension of the equations, and obtains the physical solution by superposing these $N$ solutions. 
The coefficients of the superposition are determined so that the physical solution satisfies the boundary condition, (\ref{eq:ccboundary3}). 
The computer code {\tt CCFULL} adopts this method. 

When the coupling strengths are large and/or when the classically forbidden regions are involved, 
it may be numerically difficult to maintain the linear independence of the $N$ solutions. This is particularly the case 
when wave functions in some channels are different from those in the other channels by a large order of 
magnitude.  In such cases, the liner independence can be easily violated  within a numerical accuracy, but it can be recovered using e.g., methods proposed in 
Refs. \cite{levine1984,RO1982,baylis1982}. 
Alternatively, one may also employ the method of the local transmission and reflection matrices \cite{BBGR1993,BR1994} 
or the multi-channel logderivative method \cite{johnson1973,johnson1985}.  
In these methods, the channel wave functions are first 
transformed  in such a way that they are similar in magnitude for all the channels and 
then the equations for such functions 
are solved numerically. 
These methods have been shown to provide numerically stable solutions of the coupled-channels 
equations. 
 
\subsubsection{Iterative method}

In the iterative method, 
one treats the last term on the left hand side of Eq.~(\ref{eq:cc3d}) as a source term, 
and solves the uncoupled equations for each channel wave function \cite{Rhoades-Brown1980}. 
That is, those last terms are evaluated with 
the wave functions at the $i$-th step $\{\chi^{(i)}_{cc_0}(R)\}$ and solves the uncoupled equations for the $(i+1)$-th step given by 
\begin{equation}
\left[-\frac{\hbar^2}{2\mu}\frac{d^2}{dR^2}
+\frac{L_c(L_c+1)\hbar^2}{2\mu R^2}+V(R)-iW(R)
-E+\epsilon_{c}\right]\chi^{(i+1)}_{cc_0}(R)+S^{(i)}_{c}(R)=0,
\end{equation}
with 
\begin{equation}
S^{(i)}_c(R)=
\sum_{c'}V_{cc'}(R)\chi^{(i)}_{c'c_0}(R).
\end{equation}
The iteration scheme can also be formulated 
by expressing the channel wave functions as a linear superposition of the uncoupled equations with variable coefficients 
\cite{ichimura1977,ARM1977} (see also Ref. \cite{EVZ11}). 

The iterative method is well suitable when the coupling potential is non-local, e.g., for transfer channels, 
and the computer code {\tt FRESCO} adopts this method \cite{thompson2009,fresco}. 
Another advantage of the iterative method is that it provides a clear connection to the distorted wave Born approximation (DWBA). 

\subsubsection{Variational method}

The variational method \cite{kamimura1977,matsuse1978} is based on the 
Kohn-Hulth\'en-Kato variational principle \cite{kohn1948,hulthen1948,kato1951}. 
In this method, the channel wave function $\chi_{cc_0}(R)$ is expanded as 
\begin{equation}
\chi_{cc_0}(R)=\sum_{i=1}^{N_{\rm basis}}C_{i}^{(cc_0)}\phi_{i}(R)
+
\left[H_{L_c}^{(-)}(K_c R)\delta_{c,c_0}
-\sqrt{\frac{K}{K_c}}\, \tilde{S}_{cc_0} H^{(+)}_{L_c}(K_c R)\right]
\theta(E-\epsilon_c), 
\label{eq:variational}
\end{equation}
where $\{\phi_{i}(R)\}$ are basis functions which vanish as $R\to\infty$ and $N_{\rm basis}$ is 
the number of basis. The expansion coefficients $C_{i}^{(cc_0)}$ as well 
as the factor  $\tilde{S}_{cc_0}$ are determined so that 
the quantity 
\begin{equation}
J_{c_0b_0}=
\tilde{S}_{c_0b_0}
+i\frac{\mu}{\sqrt{K_{c_0}K_{b_0}}\hbar^2}
\sum_{\gamma,\gamma'}
\int_0^\infty dR\,
\chi_{\gamma c_0}(R) 
(h_{\gamma\gamma'}(R)-E\delta_{\gamma,\gamma'})\chi_{\gamma'b_0}(R) 
\label{eq:J}
\end{equation}
with 
\begin{equation}
h_{\gamma\gamma'}(R)
\equiv
\left[-\frac{\hbar^2}{2\mu}\frac{d^2}{dR^2}
+\frac{L_\gamma(L_\gamma+1)\hbar^2}{2\mu R^2}+V(R)-iW(R)
+\epsilon_{\gamma}\right]\delta_{\gamma,\gamma'}
+V_{\gamma\gamma'}(R)
\label{eq:h_mn}
\end{equation}
is stationary with respect to the variation of 
$C_{i}^{(cc_0)}$ and $\tilde{S}_{cc_0}$.  
Notice that the wave function in the second term on the right hand side of 
Eq.~(\ref{eq:J}) is not $\chi_{\gamma c_0}^*(R)$ but $\chi_{\gamma c_0}(R)$ itself.
After the coefficients are so obtained, the $S$-matrix components are obtained as 
\begin{equation}
S_{cc_0}
=
\tilde{S}_{cc_0}
+i\frac{\mu}{\sqrt{K_{c}K_{c_0}}\hbar^2}
\sum_{\gamma,\gamma'}
\int_0^\infty dR\,
\chi_{\gamma c_0}(R) 
(h_{\gamma\gamma'}(R)-E\delta_{\gamma,\gamma'})\chi_{\gamma'\alpha}(R). 
\end{equation}

An advantage of this method is that open channels ($E>\epsilon_c$) 
and closed channels ($E<\epsilon_c$) can be treated on equal footing, as 
is evidenced in the expansion (\ref{eq:variational}). 
The method has been applied e.g., to the $^{12}$C+$^{12}$C fusion reactions at 
astrophysical energies \cite{kondo1978} (see Sec. 5.1).  

\subsubsection{R-matrix method}

The R-matrix method \cite{HSVB1998,D2016} 
follows a similar philosophy to the variational method. 
In this method, the region of $R$ is divided into the internal and the 
external regions at the channel radius $R=a$. 
The channel wave functions in the internal region are expanded with basis 
functions $\{\phi_i(R)\}$ as 
\begin{equation}
\chi^{({\rm int})}_{cc_0}(R)
=\sum_{i=1}^{N_{\rm basis}}C_{i}^{(cc_0)}\phi_{i}(R)~~~(R\leq a),
\label{eq:R-exp1}
\end{equation}
while the wave functions in the external region are assumed to take the 
asymptotic form, 
\begin{equation}
\chi^{({\rm ext})}_{cc_0}(R)
=
C_{c_0}\left[ H_{L_c}^{(-)}(K_c R)\delta_{c,c_0}
-\sqrt{\frac{K_{c_0}}{K_{c}}}\, S_{cc_0}H^{(+)}_{L_c}(K_c R)
\right]~~~(R> a).
\label{eq:R-exp2}
\end{equation}
The wave functions and their derivatives are smoothly matched at $R=a$, that is, 
$\chi^{({\rm int})}_{cc_0}(a)=\chi^{({\rm ext})}_{cc_0}(a)$  and 
$\chi'^{({\rm int})}_{cc_0}(a)=\chi'^{({\rm ext})}_{cc_0}(a)$, where 
the prime denotes the derivative with respect to $R$. 
Using this property,  
the coupled-channels equations (\ref{eq:cc3d}) can be transformed 
to 
\begin{equation}
\sum_{c'}\left(h_{cc'}+{\cal L}_c\delta_{c,c'}
-E\delta_{c,c'}\right)\chi^{({\rm int})}_{cc_0}(R)
={\cal L}_\alpha \chi^{({\rm ext})}_{cc_0}(R),
\end{equation}
where $h_{cc'}$ is given by Eq.~(\ref{eq:h_mn}) and the Bloch operator 
${\cal L}_\alpha$ is defined as\footnote{
Here, the Bloch operator is introduced because the 
kinetic operator is not Hermitian in a finite interval.}
\begin{equation}
{\cal L}_c=\frac{\hbar^2}{2\mu}\delta(R-a)\left(\frac{d}{dR}-\frac{B_c}{R}\right).
\end{equation}
The constant $B_c$ may be taken to be zero for open channels \cite{D2016}. 
 Substituting Eq.~(\ref{eq:R-exp1}) into this equation, one finds 
\begin{equation}
\sum_{c'}\sum_{i'} A_{c i,c'i'}
C_{i'}^{(c'c_0)}
=\phi_i(a)
\cdot
\frac{\hbar^2}{2\mu a}\left(a
\chi'^{({\rm ext})}_{cc_0}(a)-B_c \chi^{({\rm ext})}_{cc_0}(a)
\right),
\end{equation}
with 
\begin{equation}
A_{c i,c'i'}\equiv
\langle\phi_i|h_{cc'}+{\cal L}_c\delta_{c,c'}
-E\delta_{c,c'}|\phi_{i'}\rangle.
\end{equation}
Inverting this equation, the expansion coefficients $C_{i}^{(cc_0)}$ are 
found to be 
\begin{equation}
C_{i}^{(cc_0)}
=\frac{\hbar^2}{2\mu a}
\sum_{c'}\sum_{i'} \left(A^{-1}\right)_{c i,c'i'}
\phi_{i'}(a)
\left(a
\chi'^{({\rm ext})}_{c'c_0}(a)-B_{c'} u^{({\rm ext})}_{c'c_0}(a)
\right).
\end{equation}
One therefore finds that the internal wave functions at $R=a$ are given by 
\begin{equation}
u^{({\rm int})}_{cc_0}(a)
=\sum_{c'}
R_{cc'}(E)
\left(a
\chi'^{({\rm ext})}_{c'c_0}(a)-B_{c'} u^{({\rm ext})}_{c'c_0}(a)
\right),
\label{eq:R-wf}
\end{equation}
with the R-matrix defined as 
\begin{equation}
R_{cc'}(E)
\equiv
\frac{\hbar^2}{2\mu a}
\sum_{c'}\sum_{i,i'} \left(A^{-1}\right)_{c i,c'i'}
\phi_{i}(a)\phi_{i'}(a). 
\end{equation}
Using Eq.~(\ref{eq:R-exp1}) and the condition 
$\chi^{({\rm int})}_{cc_0}(a)=\chi^{({\rm ext})}_{cc_0}(a)$ in 
Eq.~(\ref{eq:R-wf}), the $S$-matrix components then read,
\begin{equation}
\sqrt{\frac{K_{c_0}}{K_{c}}}\, S_{cc_0}
=\sum_{c'}\left(Z_+^{-1}\right)_{cc'}\left(Z_-\right)_{c'c_0},
\end{equation}
with 
\begin{equation}
\left(Z_\pm\right)_{cc'}
=
H^{(\pm)}_{L_c}(K_c a)\delta_{c,c'}
-aR_{cc'}H'^{(\pm)}_{L_{c'}}(K_{c'} a)
-R_{cc'}B_{c'}H^{(\pm)}_{L_{c'}}(K_{c'} a).
\end{equation}
By combining the Lagrange mesh technique \cite{BH1986} 
to construct the basis functions, the R-matrix method provides 
a stable and efficient method 
both for single-channel 
\cite{BHSV1998,HRB2002,BGS2002} 
and coupled-channels \cite{HSVB1998,D2016} 
problems.  

\subsection{Semi-classical coupled-channels method}

In heavy-ion reactions, for which the reduced mass $\mu$ is large, 
the semi-classical approximation works well, providing a convenient way to 
understand the reaction dynamics \cite{Brink1985}. 
The coupled-channels equations have been solved in the semi-classical 
approximation in the following way \cite{broglia2004}. In this method, the relative 
motion is assumed to follow the classical trajectory, $R_{\rm cl}(t)$. On the other hand, 
the time evolution of the intrinsic motion is solved quantum mechanically using 
the time-dependent Hamiltonian given by 
\begin{equation} 
H_{\rm int}(t)= 
H_0(\xi)+V_{\rm coup}(R_{\rm cl}(t),\hat{\vec{R}},\xi).
\end{equation}
That is, by expanding the intrinsic wave function with the channel wave functions 
$|\varphi_c\rangle$ as 
\begin{equation}
|\Psi_{\rm int}\rangle=
\sum_{c}C_c(t)|\varphi_c\rangle,
\end{equation}
the time-dependent expansion coefficients are solved as, 
\begin{equation}
i\hbar\frac{dC_c}{dt}=\sum_{c'}\left[
\epsilon_c\delta_{c,c'}+V_{cc'}(R_{\rm cl}(t))\right]
C_{c'}(t),
\end{equation}
with the initial condition $C_{c}(t=-\infty)=\delta_{c,c_0}$. 
The separation of the treatment between the relative and the intrinsic 
motions can be justified using the influence functional technique 
in the path integral approach \cite{Brink1985,pechukas1969,BT1985,TAB1992,THA1995}. 
The cross section to populate the channel $c$ is then given by 
\begin{equation}
\frac{d\sigma_c}{d\Omega}
=\frac{d\sigma_{\rm el}}{d\Omega}\,|C_c(t=\infty)|^2,
\end{equation}
where $d\sigma_{\rm el}/d\Omega$ is the elastic cross section, and the 
coefficient $C_c(t=\infty)$ is evaluated using the classical 
trajectory for the corresponding scattering angle. 

The semi-classical coupled-channels method was first developed 
by Alder and Winther for 
Coulomb excitations \cite{alder1967,alder1956}.  Subsequently, it has been 
applied to various reaction processes, such as inelastic scattering \cite{broglia2004,bertulani1996}, one-particle and two-particle transfers  \cite{broglia2004,maglione1985,Oertzen2001,LFV2014,SH2015}, multi-nucleon 
transfer \cite{winther1994,DPW1994,winther1995}, subbarrier fusion reactions \cite{PA2000}, breakup reactions\cite{MCDL2002,MCD2008}, and 
fusion of weakly bound nuclei\cite{CDM2006,MCD2014,KCDS2018}. 
See also Refs. \cite{diaz-torres2020,diaz-torres2021,kino2001} 
for a related approach with a time-dependent Gaussian wave packet. 

\section{Direct reactions}
\label{sec3}

\subsection{Briew overview of the microscopic coupled-channels method}
\label{sec31}

In this subsection, we briefly review the framework of the microscopic coupled-channel (MCC) calculation for direct reactions, elastic and inelastic scattering in particular, based on the multiple scattering theory (MST). All details of the MCC framework are given in Appendix~\ref{secappa}.

According to the MST for nucleon-nucleus (NA) scattering developed by Foldy~\cite{foldy1945}, Watson~\cite{watson1953}, and Kerman, Mc-Manus, and Thaler~\cite{kerman1959}, one can describe the scattering process in terms of an {\it effective} nucleon-nucleon (NN) interaction $\tau$, rather than a bare NN interaction $v$. Because $\tau$ does not contain a short-range repulsive core, it is easy to handle compared with $v$, which makes the microscopic description of nuclear reactions feasible. However, it should be kept in mind that the MST assumes the coupling to rearrangement channels to be less significant. Therefore, the MST is expected to work at higher incident energies.

The key ingredient of the MST is $\tau$, which is an NN effective interaction in a many-body system, hence obtaining $\tau$ is as difficult as solving the NA many-body problem. In many applications, therefore, an approximated $\tau$ is adopted. The simplest way is to use an NN transition matrix $t$ in free space, which can be obtained by solving the Lippmann-Schwinger (LS) equation for the NN system. If the on-shell (on-the-energy-shell) approximation is further made, a phase-shift equivalent $t$~\cite{arndt2007,workman2016} can be employed. One may also use parameterized $t$-matrix interactions in a functional form~\cite{franey1985,horowitz1985}. The replacement of $\tau$ with $t$ will, however, become less accurate at lower incident energies. In that case, the G-matrix interaction $g$, a solution to the Br\"{u}ckner-Bethe-Goldstone equation for the NN scattering in an infinite nuclear matter, is often adopted. In most cases discussed in Sec.~\ref{sec32}, we take this G-matrix approach~\cite{amos2000} to $\tau$. The MST has also been applied to formulate inelastic and knockout processes within the distorted-wave approach, which is called the distorted-wave impulse approximation (DWIA).

The extension of the MST to nucleus-nucleus (AA) scattering is rather straightforward \cite{yahiro2008}. Here, we put a remark on the series of works by Crespo and collaborators~\cite{crespo1992,crespo1999,crespo2002}. They evaluated the second-order term of the optical potential derived from the MST, and applied it to proton scattering of light nuclei displaying few-body structures, e.g., $^{11}$Li. The MST-based full folding models in momentum space can be found in Refs.~\cite{arellano1995,arellano2007,gennari2018,burrows2020}.

Then, as described in Sec.~\ref{sec:cca}, the total scattering wave function is expanded in terms of eigenstates of the nucleus A. The coupling potentials can be obtained in a straightforward manner, once transition densities are given by a nuclear structure calculation. When nucleus-nucleus (AA) scattering is considered, a double-folding (DF) model is used. An important aspect here is what density should be used as an argument of the G-matrix. Phenomenologically, it is sometimes argued that the frozen density approximation (FDA), in which the sum of the projectile and target densities is used, is favored.
An interesting aspect of the recent MCC studies on AA scattering is the effect of three-nucleon force (3NF) on reaction observables. In Refs.~\cite{takashina2010,furumoto2013,minomo2016,furumoto2009,furumoto2009b,toyokawa2015}, the 3NF is included in the calculation of the G-matrix interaction in infinite nuclear matter; the matrix element of the 3NF is averaged over the third nucleon in the Fermi sea, and its role is included as a modification of the two-body G-matrix interaction. Some examples of such studies are shown in Sec.~\ref{sec324}.

When one of the two nuclei is $\alpha$ ($^4$He), the extended version of the nucleon-nucleus folding (NAF) model, referred to as the extended NAF model, has been employed in a series of recent MCC studies of $\alpha$ elastic and inelastic scattering~\cite{kanada-en'yo2019,kanada-en'yo2019b,kanada-en'yo2020,kanada-en'yo2020b,kanada-en'yo2020c,kanada-en'yo2020d,kanada-en'yo2021,kanada-en'yo2021b} as a practical MCC framework. In Ref.~\cite{egashira2014}, the NAF model was shown to be a good alternative to the double-folding (DF) model using the density of the target nucleus as an argument of the G matrix, which is called the target density approximation (TDA). The DF-TDA model was shown to reasonably reproduce the $\alpha$-$^{58}$Ni elastic and total reaction cross sections in the wide range of incident energies, in contrast to the DF model with the FDA. Justification of the DF-TDA model in view of the tightly-bound property of an $\alpha$ particle  is given in Ref.~\cite{egashira2014}.

\subsection{Application of the MCC framework to elastic and inelastic scattering}
\label{sec32}

There have been many applications of the MCC framework to elastic and inelastic scattering. Among them, in this article, we review some works in which not only the real part but also the imaginary part of the coupling potential are evaluated microscopically. It is worth pointing out that, until quite recently, a phenomenological treatment of the imaginary potential has widely been adopted in {\lq\lq}microscopic'' calculations. Although such a strategy accelerated the (semi-)microscopic approach to nuclear reactions in the pioneering era, nowadays, it is recommended to treat both the real and imaginary parts microscopically. This will be particularly important when one discusses reactions of unstable nuclei because phenomenological optical potentials are not available in such cases; one does not have a clear guideline even for a functional form of the imaginary potential. Furthermore, constructing an imaginary part of the coupling (nondiagonal) potential will be completely nontrivial in the phenomenological approach unless a macroscopic structure model is assumed.
Note, however, that the MCC framework discussed in this article is still an approximate approach for describing direct reactions of many-body systems. When {\it ab initio} calculations for various reaction processes will become feasible in future, one would get a deeper understanding of the reliability, accuracy, and limitation of the MCC framework.

\subsubsection{Applicability of the MST}
\label{sec321}

\begin{figure}[!htbp]
\begin{center}
\includegraphics*[width=0.45\textwidth]{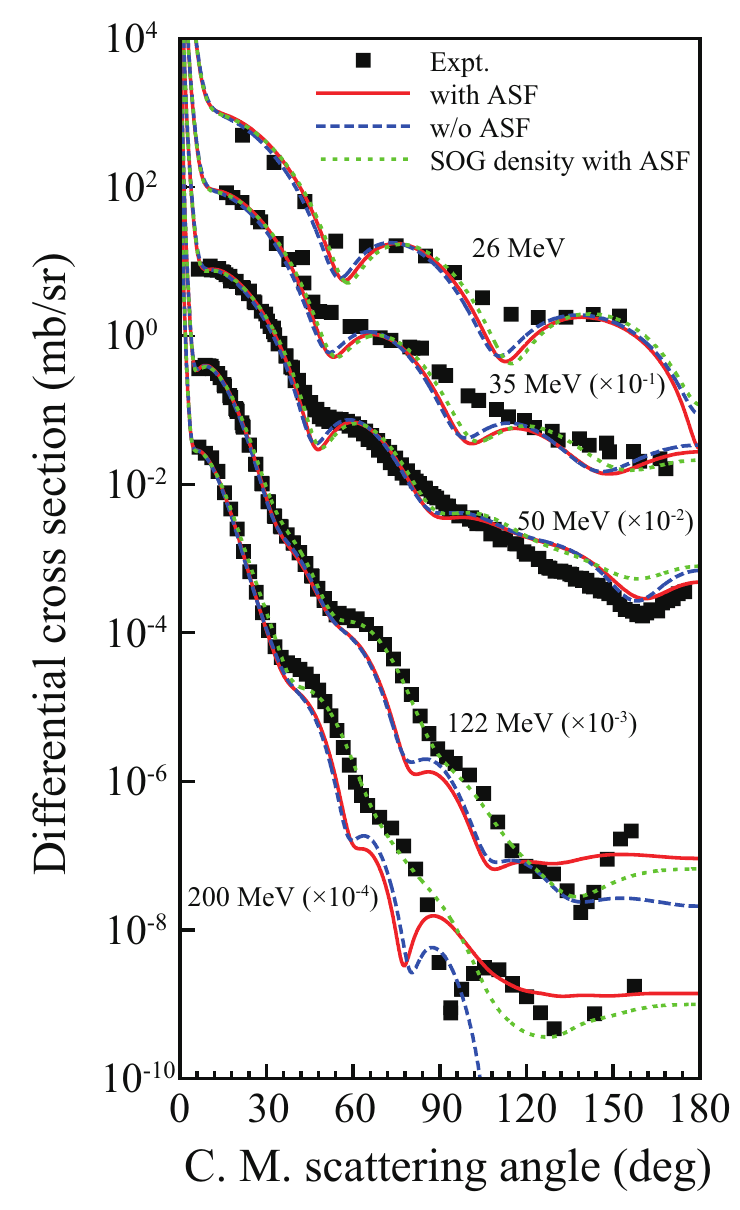}
\includegraphics*[width=0.45\textwidth]{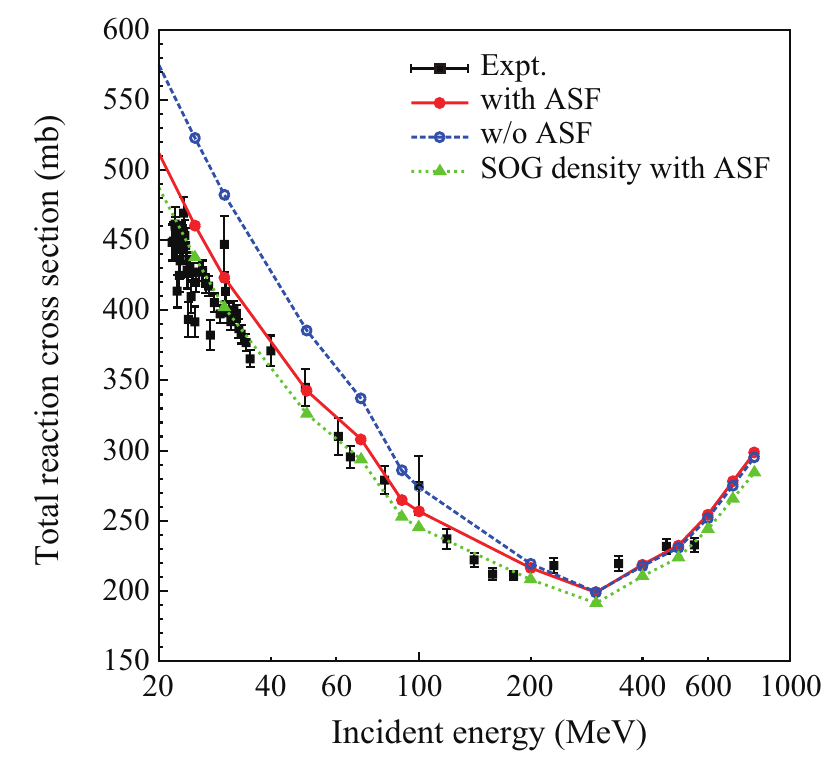}
\caption{
(a) Differential cross sections for $p$-$^{12}$C elastic scattering as a function of the c.m.\ scattering angle. The solid and dashed lines represent the results obtained with the microscopic density with and without the ASFs, respectively, whereas the dotted lines correspond to the result with the phenomenological density and the ASF. (b) Energy dependence of the total reaction cross sections for $p$-$^{12}$C scattering. The closed (open) circles correspond to the result with (without) the ASF and the triangles to the results with the phenomenological density including the ASF. Taken from Ref.~\cite{minomo2017}.
}
\label{fig3-1}
\end{center}
\end{figure}

Before going to CC calculations, we briefly discuss the applicability of the MST to NA elastic scattering regarding the incident energy. Although the MST becomes less reliable as the incident energy decreases, it will be difficult to specify its lower limit from theory. Thus, a phenomenological investigation on the applicability of the MST has been performed in Ref.~\cite{minomo2017}. The Melbourne G matrix~\cite{amos2000} was adopted as $\tau$ and the central and spin-orbit interactions were both taken into account. Figure \ref{fig3-1}(a) represents the $p$-$^{12}$C elastic scattering cross sections from 26~MeV to 200~MeV as a function of the scattering angle in the center-of-mass (c.m.) frame. The solid (dashed) lines show the results with the SLy4 Skyrme-type Hartree-Fock-Bogoliubov (SLy4-HFB) densities with (without) the antisymmetrization factor (ASF) in the MST, $A/\left( A-1\right)$ in Eq.~(\ref{3tbar}) and $(A-1)/A$ in Eq.~(\ref{3taubar}), whereas the dotted lines show the results with phenomenological nuclear densities with the ASF. The calculation has no free adjustable parameter. The dotted lines reproduce well the experimental data. At 122~MeV and 200~MeV, the results at middle and backward angles depend on the choice of the nuclear density. This indicates that the accuracy of the HFB calculation is not enough for this light nucleus. Note, however, that for the results of proton scattering off $^{40}$Ca and $^{208}$Pb, the two densities are found to provide very similar results \cite{minomo2017}. As one sees from Fig.~\ref{fig3-1}(a), the single-channel calculation based on the MST works well even at 26~MeV, which is also the case with $^{40}$Ca and $^{208}$Pb targets. Phenomenologically, therefore, one may conclude that the MST will be applicable to NA scattering down to about 25~MeV.

In Fig.~\ref{fig3-1}(b), we show the $p$-$^{12}$C total reaction cross section as a function of incident energy. The difference between the solid and dashed lines shows the importance of the ASF, at low energies in particular. The dotted line reproduces the data in a wide region of energies. This is also the case with the $p$-$^{40}$Ca and $p$-$^{208}$Pb systems, though the agreement with data 
is slightly worse than for the $p$-$^{12}$C system; see Figs.~6 and 7 in Ref.~\cite{minomo2017}.

The effect of the BR localization and the dependence of the choice of the positions at which the nuclear density is evaluated in the G-matrix approach to $\tau$ are also discussed in Ref.~\cite{minomo2017}.

\subsubsection{Proton scattering}
\label{sec322}

Proton inelastic, $(p,p')$, scattering has been used to study properties of excited states of nuclei and/or the responses of nuclei, form factors or transition densities, for the transition induced by the proton interaction. In contrast to $\alpha$ inelastic, $(\alpha,\alpha')$, scattering, both isoscalar and isovector transitions are expected to be probed. When the CC effect is important, however, the observables are not directly proportional to the square of the form factors and their relationship becomes nontrivial. Even in such cases, 
from a comparison of results of CC calculations with experimental 
data, using different nuclear models and/or different model parameters, 
properties of nuclear excited states can be deduced.

Nowadays, $(p,p')$ scattering for unstable nuclei has widely been measured. One of the most important characteristics of unstable nuclei is the {\lq\lq}decoupling'' of proton and neutron properties. In this view, the isovector nature of $(p,p')$ scattering plays an important role. One should, however, be careful that some empirical methods established for stable nuclei may not work for unstable nuclei.

\begin{figure}[!ht]
\begin{center}
\includegraphics*[width=0.48\textwidth]{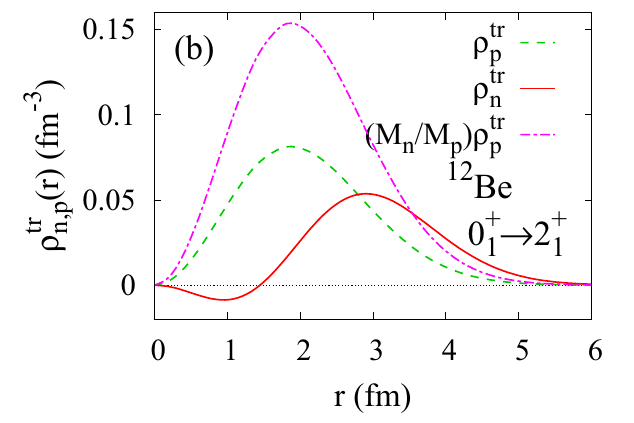}
\includegraphics*[width=0.40\textwidth]{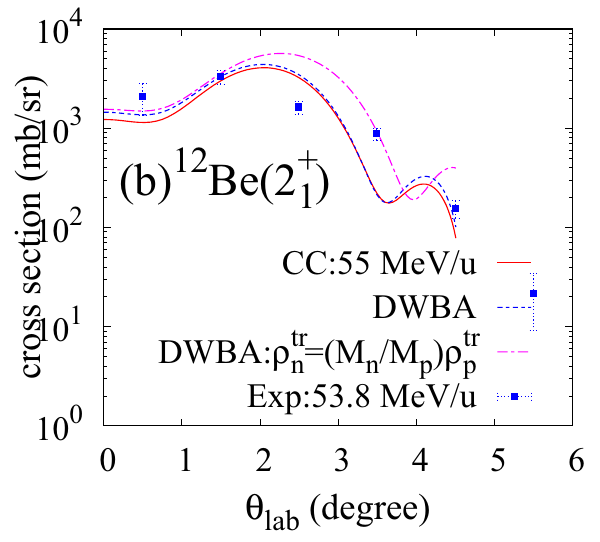}
\caption{
(a) Neutron and proton transition densities for the $0_1^+ \to 2_1^+$ transition of $^{12}$Be calculated with AMD. (b) Cross sections of inelastic proton scattering to the $2_1^+$ state of $^{12}$Be at 55~MeV/nucleon obtained with MCC calculations (solid line). The result of one-step DWBA calculation is shown by the dotted line. The DWBA result using the neutron transition density $\rho_n^{\rm tr}(r)=(M_n/M_p)\rho_p^{\rm tr}(r)$ is also shown for comparison (dash-dotted line). Taken from Ref.~\cite{kanada-en'yo2019c}.
}
\label{fig3-2}
\end{center}
\end{figure}
Recently, the MCC framework using the Melbourne G matrix~\cite{amos2000} and antisymmetrized molecular dynamics (AMD)~\cite{kanada-en'yo1995,kanada-en'yo1995a,kanada-en'yo2012} has successfully been applied to $(p,p')$ scattering off $^{18}$O, $^{10}$Be, $^{12}$Be, $^{16}$C~\cite{kanada-en'yo2019c}. In Fig.~\ref{fig3-2}(a), we show the proton (dashed line) and neutron (solid line) transition densities of $^{12}$Be from the ground state to the $2_1^+$ state. A clear difference in the shape of the two densities can be seen. In a {\lq\lq}standard'' manner, the neutron transition density is approximated by that of proton multiplied by $M_n/M_p$, where $M_p$ ($M_n$) stands for the proton (neutron) transition matrix element. This prescription results in the dash-dotted line in Fig.~\ref{fig3-2}(a), severely deviating from the correct result (solid line). This significantly affects the corresponding $(p,p')$ cross section at 55~MeV/nucleon shown in Fig.~\ref{fig3-2}(b). This was found to be the case also with $^{16}$C.

\begin{figure}[tb]
\begin{center}
\includegraphics*[width=0.4\textwidth]{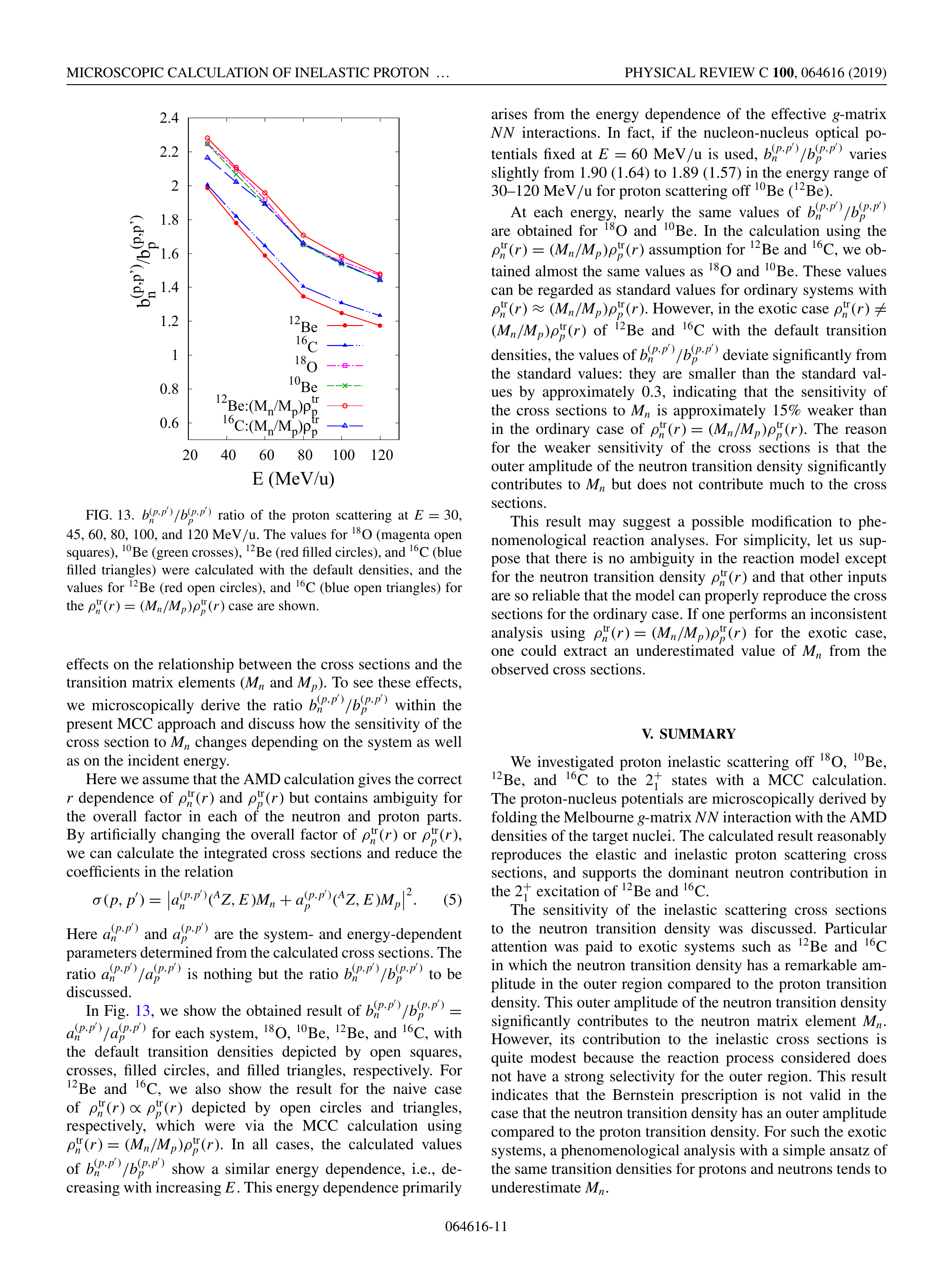}
\caption{
$b_n/b_p$ ratio for proton inelastic scattering. Open squares, crosses, filled circles, and filled triangles represent the results for $^{18}$O, $^{10}$Be, $^{12}$Be, and $^{16}$C calculated with the AMD densities, respectively. Open circles (open triangles) show the results for $^{12}$Be ($^{16}$C) with $\rho_n^{\rm tr}(r)=(M_n/M_p)\rho_p^{\rm tr}(r)$. Taken from Ref.~\cite{kanada-en'yo2019c}.
}
\label{fig3-3}
\end{center}
\end{figure}
In analyses of $(p,p')$ cross sections, the Bernstein prescription has widely been employed, which usually assumes
\begin{equation}
\sigma \propto |b_n M_n + b_p M_p|^2.
\label{3bern}
\end{equation}
Here, $\sigma$ is the ($p,p'$) cross section and $b_p$ ($b_n$) is the strength for the transition interaction for proton (neutron). $b_p$ and $b_n$ depend on the colliding partners and on the incident energy, but a standard value of around 3 is often used as $b_n/b_p$ for $(p,p')$ scattering at 10--50~MeV/nucleon. We show in Fig.~\ref{fig3-3} $b_n/b_p$ theoretically obtained using Eq.~(\ref{3bern}). One sees that for $^{12}$Be and $^{16}$C, for which the proportionality between the proton and neutron transition densities do not hold, $b_n/b_p$ deviates from the results for $^{18}$O and $^{10}$Be having the proportionality. In fact, if we artificially assume the proportionality for $^{12}$Be and $^{16}$C, the results well agree with those for $^{18}$O and $^{10}$Be. An important message from this figure is that $M_n$ can be underestimated if the proportionality between the proton and neutron transition density is assumed by hand for nuclei having no proportionality. A similar discussion was made in Refs.~\cite{marechal1999,scheit2000,takashina2008} with a semi-microscopic CC calculation; the Jeukenne-Lejeune-Mahaux (JLM)~\cite{jeukenne1977} G-matrix interaction was adopted and the strengths and range parameters of both real and imaginary parts were determined phenomenologically.

\begin{figure}[!ht]
\begin{center}
\includegraphics*[width=0.45\textwidth]{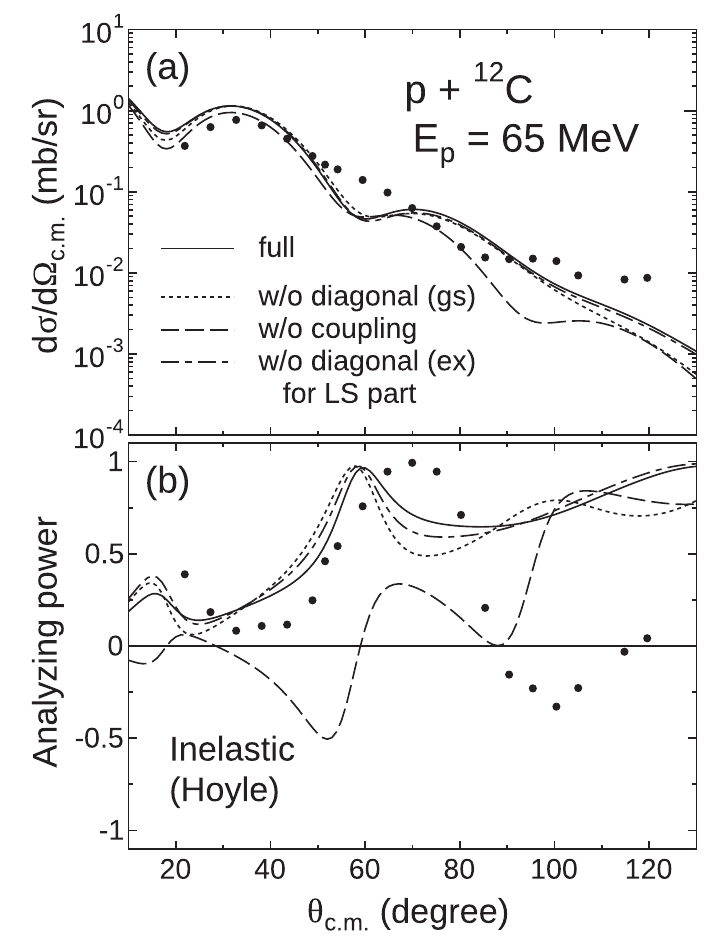}
\includegraphics*[width=0.45\textwidth]{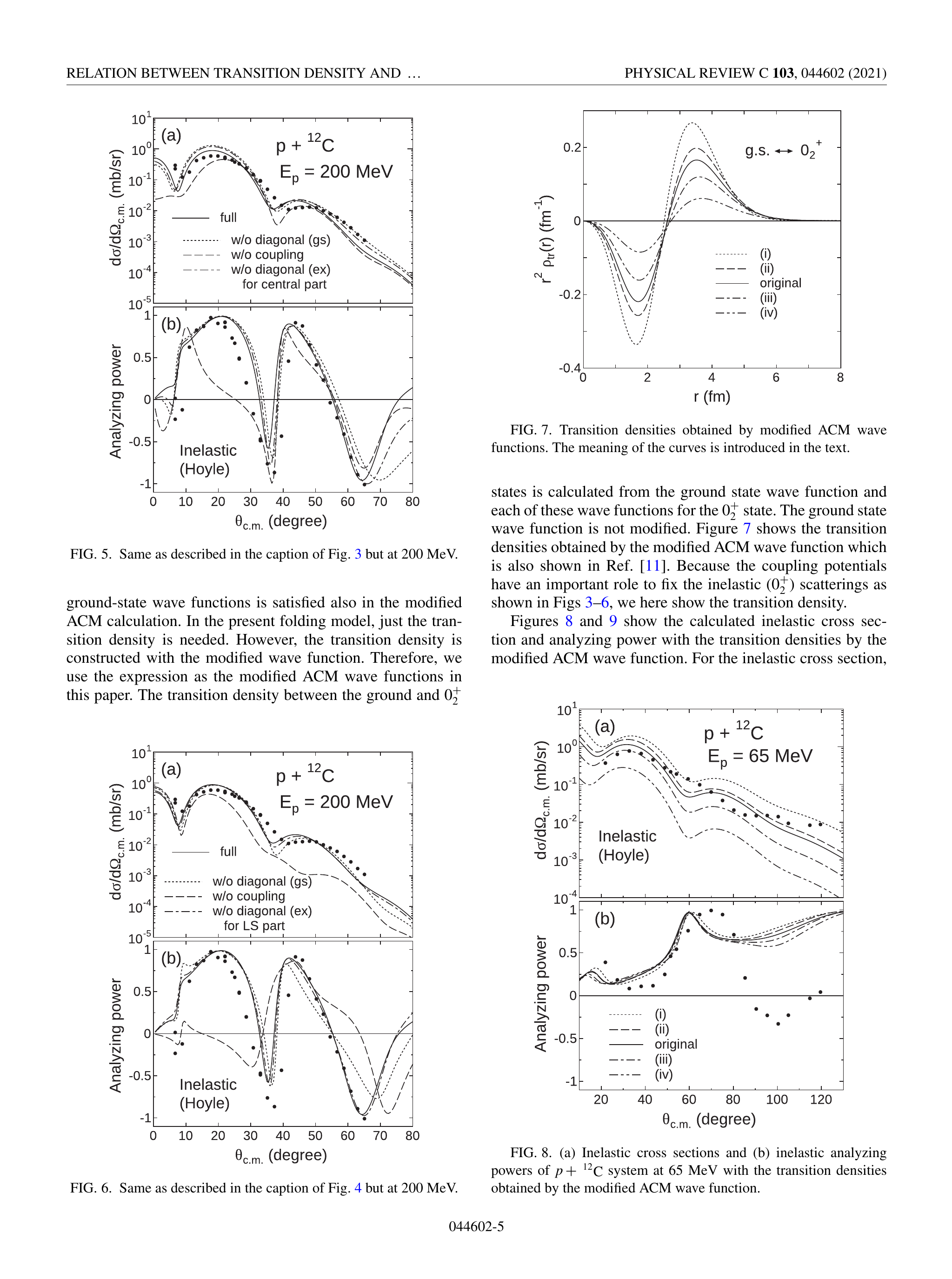}
\caption{
(a) Cross sections and (b) analyzing power of inelastic $p$-$^{12}$C scattering to the $0_2^+$ (Hoyle) state at 65 MeV. The solid curves show the results of the MCC calculation. The dotted (dash-dotted) curves represent the results without the spin-orbit part of the diagonal potential of the elastic (excited) channel. The dashed curves are obtained without the spin-orbit part of the coupling potential between the ground and excited states.
(c) and (d) are the same as (a) and (b), respectively, but at 200~MeV. Teken from Ref.~\cite{furumoto2021}.
}
\label{fig3-4}
\end{center}
\end{figure}
An ingredient of the MCC framework that has been disregarded until recently is the spin-orbit coupling potential. In Ref.~\cite{furumoto2021}, Furumoto and Takashina investigated the importance of the diagonal and nondiagonal spin-orbit potentials in the description of the $^{12}$C$(p,p')$ scattering cross section and analyzing power $A_y$ to the $0_2^+$ state. As shown by the dashed lines in Fig.~\ref{fig3-4}(a), the role of the nondiagonal spin-orbit potential is crucial for $A_y$ but rather modest for the cross section at 65~MeV. On the other hand, at 200~MeV, both the cross section and $A_y$ are significantly affected by it except at forward angles as shown in Fig.~\ref{fig3-4}(b). A more systematic investigation on the role of the spin-orbit coupling potential in inelastic scattering will be 
interesting and important, as well as the dependence on the choice of the G-matrix interactions and structure models. Note that in Ref.~\cite{furumoto2021}, the so-called MPa G-matrix interaction~\cite{yamamoto2003,yamamoto2014} derived from the ESC08 nucleon-nucleon interaction~\cite{rijken2010} was adopted with a renormalization factor $N_{\rm w}$ for central and spin-orbit imaginary potentials. They adopted in Ref.~\cite{furumoto2021} the energy-dependent $N_{\rm w}$ determined from a systematic analysis of NA scattering data using the MPa G matrix.

Recently, a first step to the MCC calculation for nuclei in heavy-mass region was reported to investigate a large-amplitude shape mixing in nuclei~\cite{sato2019}. Although a phenomenological model based on the 5-dimensional quadrupole collective Hamiltonian was adopted in Ref.~\cite{sato2019}, a possibility for observing the strong $\beta$-$\gamma$ coupling and large-amplitude quadrupole shape mixing in spherical-to-prolate transitional nuclei, for example, $^{154}$Sm, via the ($p,p'$) scattering to the $2_2^+$ state was pointed out. A microscopic description of the transition densities with a constrained Hartree-Fock-Bogoliubov plus local quasiparticle random phase approximation (CHFB+LQRPA) method is ongoing and an MCC calculation result will be reported elsewhere.

\subsubsection{Alpha scattering}
\label{sec323}

Inelastic scattering of nuclei with a $^4$He
target, the $\alpha$ inelastic scattering, has been utilized as a probe for isoscalar transitions~\cite{youngblood1997,youngblood1999,youngblood1999b,itoh2002,itoh2003,uchida2004,li07,itoh2013,bagchi2015,vandebrouck2015,gupta2016,peach2016,adachi2018}. In particular, the isoscalar monopole transition of nuclei, which is expected to be deduced from $\alpha$ inelastic scattering data, has been considered as an indicator of the $\alpha$ clustering of nuclei~\cite{suzuki1976,yamada2008}.

A well-known big puzzle regarding this subject is the so-called missing monopole strength of the $0_2^+$ Hoyle state in the $^{12}$C$(\alpha,\alpha')$ scattering. Khoa and Cuong~\cite{khoa2008} demonstrated that the $^{12}$C$(\alpha,\alpha')$ cross sections to the $0_2^+$ state evaluated with their microscopic calculation severely overshoot the experimental data by about a factor of three. They suggested that a large part of the isoscalar transition strength to the $0_2^+$ state predicted by structure model calculations seemed to be {\it missing}.

\begin{figure}[!ht]
\begin{center}
\includegraphics*[width=0.9\textwidth]{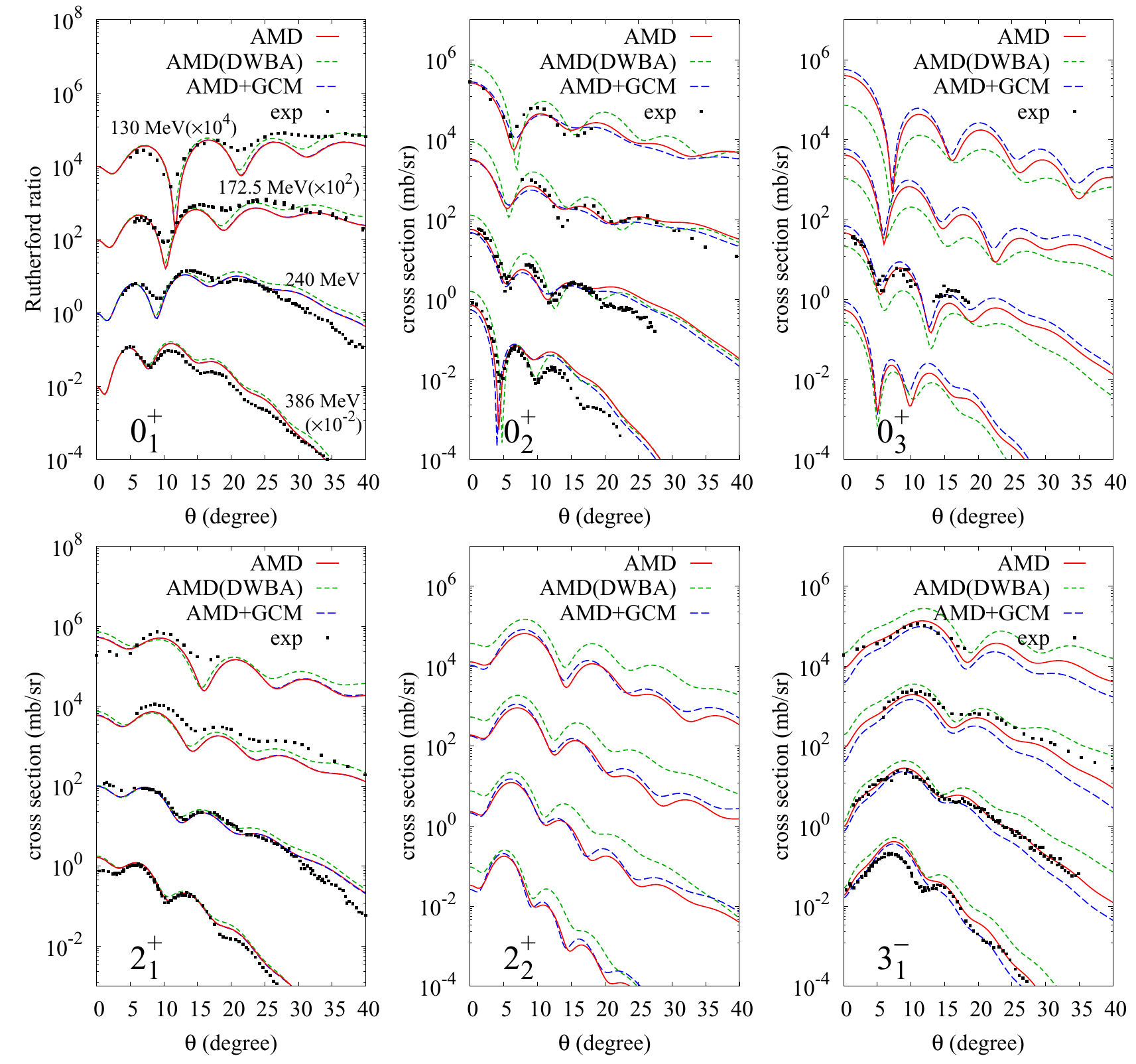}
\caption{
$\alpha$ scattering cross sections on $^{12}$C at 130, 172.5, 240, and 386~MeV. The solid and the dotted lines represent the results of the MCC calculations with the AMD and ${\rm AMD} + {\rm GCM}$ transition densities, respectively. The cross sections obtained with the DWBA calculation with the AMD transition densities are also shown for comparison by the dashed lines. Taken from Ref.~\cite{kanada-en'yo2019}.
}
\label{fig3-5}
\end{center}
\end{figure}
Recently, the MCC framework was applied to the $^{12}$C$(\alpha,\alpha')$ scattering and the result is shown in Fig.~\ref{fig3-5}. The MCC calculation (solid lines) reasonably reproduces the data for elastic and inelastic cross sections with no adjustable parameters for the reaction part. As one can see, the $0_2^+$ cross sections are well reproduced except at backward angles, suggesting that there is no missing monopole strength issue. This finding is essentially the same as that in Ref.~\cite{minomo2016b}, in which coupling potentials were calculated with the DF-TDA approach adopting the resonating group method (RGM) transition densities~\cite{kamimura1981}. It is worth mentioning that in Ref.~\cite{kanada-en'yo2019}, transition densities are renormalized to reproduce experimental results of $B(E\lambda)$ or charge form factors, whereas in Ref.~\cite{minomo2016b} such renormalization 
was not introduced. 
In any case, however, the overshooting problem for the $0^+_2$ cross sections by about a factor of three disappears when an MCC framework is adopted. A key ingredient of the success of the MCC calculations is the imaginary part of the coupling potential, which was treated phenomenologically in Ref.~\cite{khoa2008}. Another important point is the strong channel coupling between the $0_2^+$ and $2_2^+$ states of $^{12}$C.

Quite recently, a systematic measurement for $(\alpha,\alpha')$ scattering off $A=4n$ nuclei was carried out~\cite{adachi2018}. The authors adopted as their standard method of analysis a semi-microscopic DWBA calculation with transition densities evaluated with a macroscopic model; a phenomenological $\alpha$-N potential having a density dependence was folded by the transition densities to obtain the coupling potentials. An interesting finding of their analysis is that the use of density-independent $\alpha$-N potential gives a better description of the data than with the density-dependent one. It should be noted, however, that this does not necessarily mean the absence of the density dependence of effective interactions because their finding may strongly be model-dependent. In fact, it was found in Ref.~\cite{minomo2016b} that the transition potential between the $0_1^+$ and $0_2^+$ states of $^{12}$C obtained with the Melbourne G matrix has a similar behaviour to that obtained with the density-independent $\alpha$-N potential in Ref.~\cite{adachi2018}. At this stage, there is no clear explanation for this similarity and hence it might be merely accidental; the frameworks adopted in Refs.~\cite{minomo2016b} and \cite{adachi2018} are too different to be related in a simple way.

The MCC framework has been applied systematically to $(\alpha,\alpha')$ scattering to study the cluster and band structures of low-lying states of many nuclei as well as their transition properties~\cite{kanada-en'yo2019,kanada-en'yo2019b,kanada-en'yo2020,kanada-en'yo2020b,kanada-en'yo2020c,kanada-en'yo2020d,kanada-en'yo2021,kanada-en'yo2021b}. In some of the works, a combined analysis of $(\alpha,\alpha')$ and ($p,p'$) scattering was carried out.

\begin{figure}[!ht]
\begin{center}
\includegraphics*[width=0.84\textwidth]{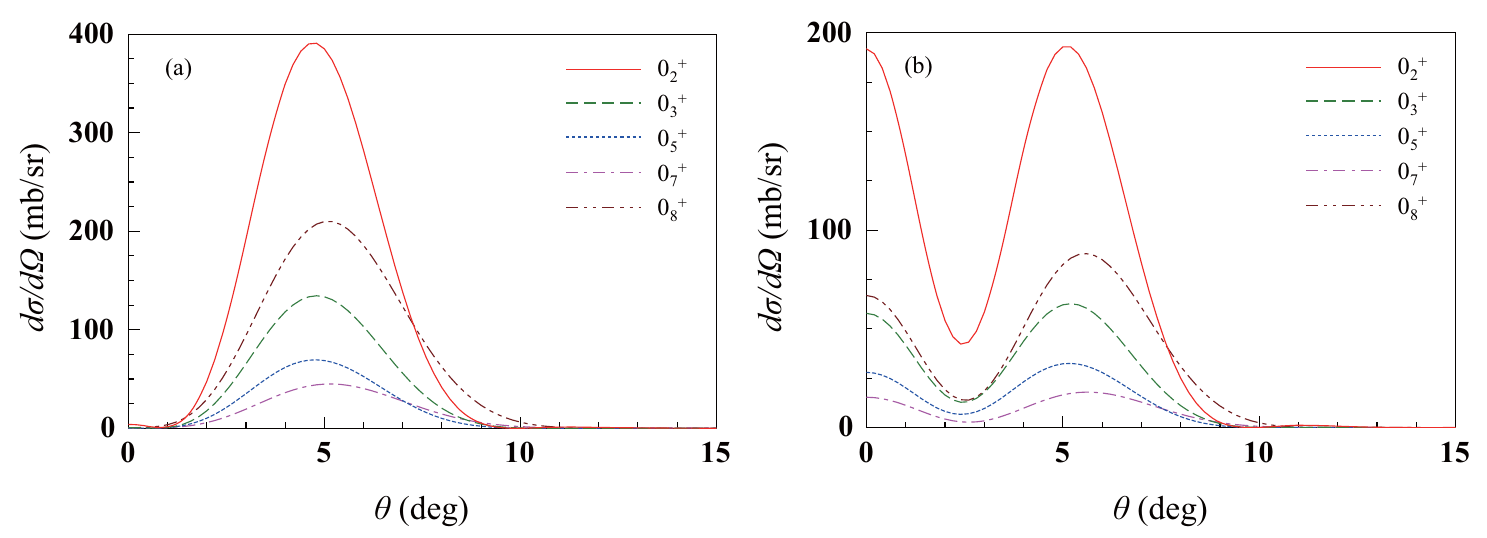}
\includegraphics*[width=0.42\textwidth]{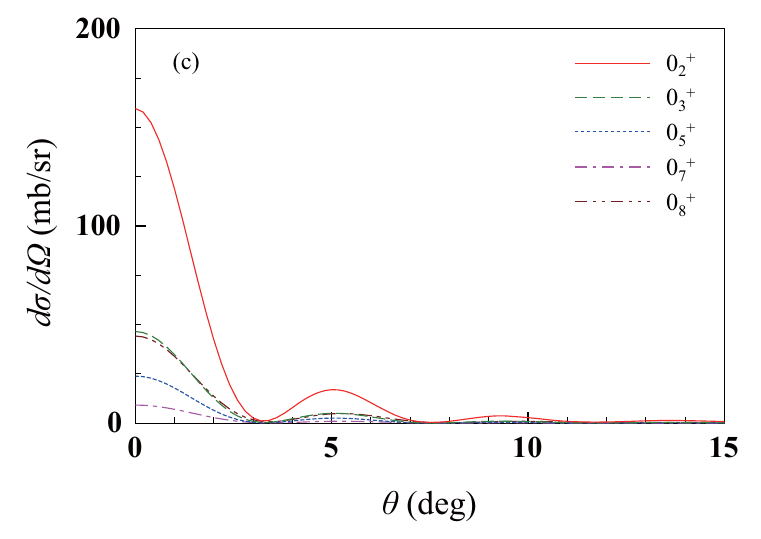}
\includegraphics*[width=0.42\textwidth]{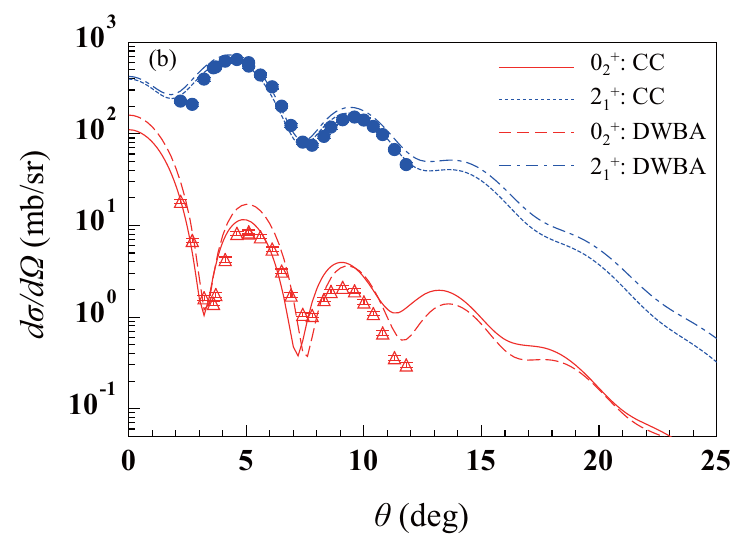}
\caption{
$\alpha$ inelastic cross sections on $^{24}$Mg to $0^+$ excited states at 386~MeV obtained with (a) PWBA based on a free NN effective interaction (PWBA-$t$), (b) PWBA employing a G-matrix interaction (PWBA-$g$), and (c) DWBA with a G-matrix interaction (DWBA-$g$). In each panel, the solid, dashed, dotted, dot-dashed, dot-dot-dashed lines correspond to the cross sections to the $0_2^+$, $0_3^+$, $0_5^+$, $0_7^+$, and $0_8^+$ states, respectively. (d) Comparison with experimental data for the inelastic cross sections; the solid (dashed) and dotted (dot-dashed) lines show the results of the MCC (DWBA) calculation to the $0^+_2$ and $2^+_1$ states, respectively. Taken from Ref.~\cite{ogata2021}.
}
\label{fig3-6}
\end{center}
\end{figure}
As was mentioned above, $(\alpha,\alpha')$ scattering is believed to be a good probe of the isoscalar monopole (IS0) strength $B({\rm IS0})$ in analogy to the electron inelastic ($e,e'$) scattering. Recently, the $B({\rm IS0})$-$(\alpha,\alpha')$ correspondence has been tested in the MCC framework and it turned out that the 
correspondence is significantly affected by the in-medium modifications to the NN effective interaction, nuclear distortion, and CC effect~\cite{ogata2021}. As is shown in Fig.~\ref{fig3-6}, the textbook explanation about the $B({\rm IS0})$-$(\alpha,\alpha')$ correspondence holds only when a plane-wave calculation with a density-independent NN effective interaction $\tau$ is performed. Inclusion of the density dependence of $\tau$ and nuclear distortion completely change the angular distribution of the $(\alpha,\alpha')$ cross section. Therefore, the $B({\rm IS0})$-$(\alpha,\alpha')$ correspondence must be discussed with caution, even though a strong constraint on the monopole transition density helps the correspondence to hold~\cite{ogata2021}. In Table~\ref{tab3-1}, we show the relative strengths of $B({\rm IS0})$ of $^{24}$Mg and the $(\alpha,\alpha')$ cross sections to those for the $0_2^+$ state. The $B({\rm IS0})$-$(\alpha,\alpha')$ correspondence holds, with about 20--30~\% error, at 386~MeV, whereas it breaks down at 130~MeV. A similar discussion was made for $(p,p')$ scattering in Ref.~\cite{miskimen1988}.
%
\begin{table}[hbtp]
 \caption{IS0 transition strengths and $\alpha$ inelastic cross sections relative to those for the $0^+_2$ state.}
 \label{tab3-1}
 \centering
 \begin{tabular}{cc|cccc}
 \hline
  Quantity & reaction model & $0_3^+$ & $0_5^+$ & $0_7^+$ & $0_8^+$ \\
  \hline
  $B({\rm IS0})_i/B({\rm IS0})_2$                              & ------   & 0.33 & 0.18 & 0.09 & 0.42 \\
  \hline
  $({d\sigma_i}/{d\Omega})/({d\sigma_2}/{d\Omega})$ at 130~MeV & ${\rm PWBA}$-$t$ & 0.34 & 0.18 & 0.12 & 0.54 \\
                                                               & ${\rm PWBA}$-$g$ & 0.32 & 0.17 & 0.09 & 0.46 \\
                                                               & ${\rm DWBA}$-$g$ & 0.26 & 0.13 & 0.03 & 0.15 \\
                                                               & ${\rm CC}$       & 0.20 & 0.10 & 0.04 & 0.19 \\

  \hline
  $({d\sigma_i}/{d\Omega})/({d\sigma_2}/{d\Omega})$ at 386~MeV & ${\rm PWBA}$-$t$ & 0.34 & 0.18 & 0.12 & 0.53 \\
                                                               & ${\rm PWBA}$-$g$ & 0.33 & 0.17 & 0.09 & 0.46 \\
                                                               & ${\rm DWBA}$-$g$ & 0.30 & 0.15 & 0.06 & 0.29 \\
                                                               & ${\rm CC}$       & 0.26 & 0.14 & 0.08 & 0.33 \\
 \hline
 \end{tabular}
\end{table}
%

\subsubsection{Nucleus-nucleus scattering}
\label{sec324}

\begin{figure}[!ht]
\begin{center}
\includegraphics*[width=0.45\textwidth]{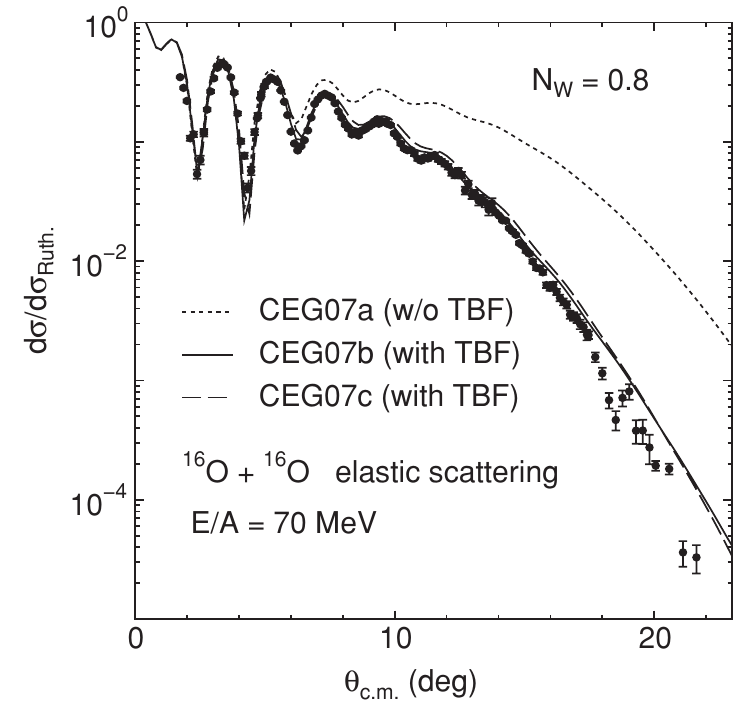}
\includegraphics*[width=0.48\textwidth]{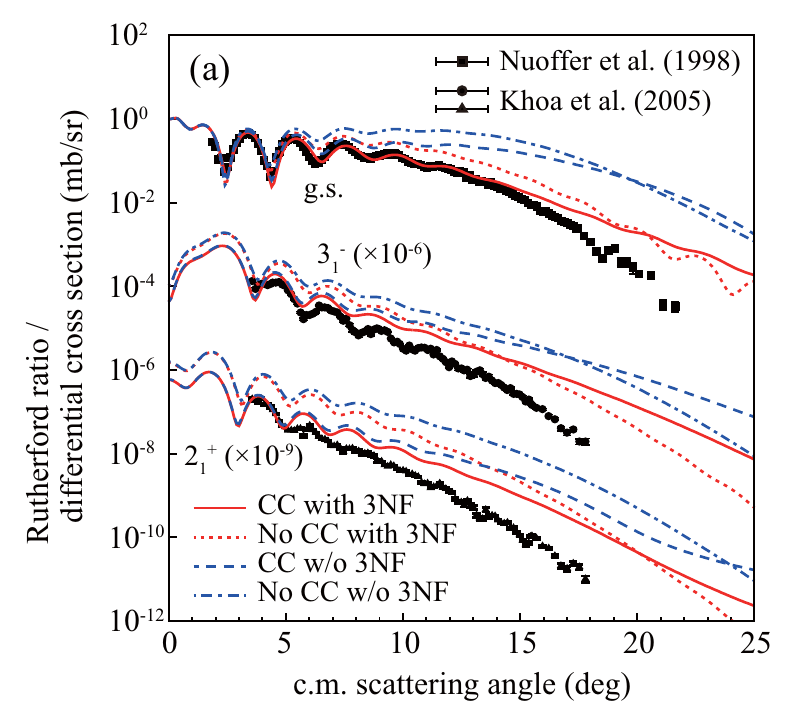}
\caption{
(a) $^{16}{\rm O}+{}^{16}{\rm O}$ elastic cross section at 70~MeV/nucleon obtained with a single-channel calculation employing the so-called CEG07 G-matrix interactions~\cite{furumoto2008}. The solid and dashed lines show the results with the TBF effect, whereas the dotted line without the TBF effect. A phenomenological renormalization factor $N_{\rm w}$ of 0.8 is included in all calculations. Taken from Ref.~\cite{furumoto2009b}. (b) Differential cross sections of elastic and inelastic $^{16}{\rm O}+{}^{16}{\rm O}$ scattering at (a) 70MeV/nucleon obtained with the MCC calculation adopting the Melbourne G-matrix interaction modified with the chiral 3NF effects (solid lines). The dashed lines represent the results without the 3NF effects. The dotted (dot-dashed) lines correspond to the result of the one-step calculation with (without) the 3NF effects. No adjustable parameter is included. Taken from Ref.~\cite{minomo2016}.
}
\label{fig3-7}
\end{center}
\end{figure}
An interesting feature of AA scattering within the MCC framework is that NN effective interactions at high-density are expected to become relevant. As a result, one may expect that the 3NF effect can be probed via AA scattering cross sections. This was first pointed out by Furumoto and collaborators~\cite{furumoto2009b} for $^{16}$O-$^{16}$O elastic scattering at 70~MeV/nucleon calculated with a single-channel calculation with the CEG07 G-matrix interaction including the effect of 3NF, or three-body force (TBF) in their terminology~\cite{furumoto2008}; the phenomenological renormalization factor of 0.8 was adopted for the imaginary part of the microscopic potential. As seen from Fig.~\ref{fig3-7}(a), a clear signature of the role of the TBF is indicated at backward angles. Later, an MCC calculation with the Melbourne G matrix has been applied to the same reaction system~\cite{minomo2016} not only for elastic scattering but also for inelastic scattering to the $3_1^-$ and $2_1^+$ states, including mutual excitation. As shown in Fig.~\ref{fig3-7}(b), the inclusion of the 3NF effect gives a remarkable improvement in reproducing the elastic scattering data at backward angles. For the inelastic cross sections, something seems to be needed to explain the experimental data. It should be noted that the MCC calculation of Ref.~\cite{minomo2016} does not contain free adjustable parameters. As mentioned, in Ref.~\cite{furumoto2009b}, a renormalization factor $N_{\rm w}$ for the imaginary potential was introduced to reproduce the data with the TBF effect in a single-channel calculation. It can be understood that the CC effect shown in Fig.~\ref{fig3-7}(b) for the elastic cross section, 
that is, the difference between the solid and dotted lines, was effectively included by (a part) of $N_{\rm w}$. To draw a more definite conclusion, however, a more detailed analysis of the difference between the two G-matrix interactions will be necessary.

Quite recently, Furumoto, Suhara, and Itagaki~\cite{furumoto2018} performed an MCC calculation, with $N_{\rm w}$, for scattering of $^{6\mbox{--}9}$Li off $^{12}$C and $^{28}$Si at around 50~MeV/nucleon adopting a stochastic multi-configuration mixing method. A decrease in the CC effect on the elastic cross sections as increasing the number of valence neutrons is found, thereby a glue-like effect of valence neutrons in lithium isotopes was discussed. Another important finding is a somewhat large CC effect via the excitation of the target nucleus. This feature may make an MCC analysis of AA elastic, inelastic, breakup reactions rather complicated in general. A more systematic investigation of the importance of the target excitation regarding the incident energy will be interesting and important.

\subsection{CDCC: treatment of continuum}
\label{sec33}

When one is interested in breakup processes to nonresonant continuum states of the projectile and/or the target nucleus, one needs to prepare a model space that is large enough to describe the physics phenomena, observables in particular, of interest. Sometimes, a description of resonant states of a particle embedded in the continuum region becomes important. In such cases, one needs to treat the resonant and nonresonant states on the same footing in the CC framework. CDCC~\cite{kamimura1986,austern1987,yahiro2012} enables one to achieve this efficiently and with high accuracy. In this subsection, we briefly recapitulate the CDCC formalism. In Sec.~\ref{sec331}, the theoretical foundation of CDCC is reviewed and in Sec.~\ref{sec332}, we briefly discuss how one can interpret/modify the MCC framework mentioned in Sec.~\ref{sec31} for describing breakup reactions.

\subsubsection{Theoretical foundation of CDCC ---three-body reaction theory in a model space}
\label{sec331}

Let us consider for a while a reaction process between a projectile consisting of two particles, particles 1 and 2, a target nucleus A, which is regarded as a three-body scattering problem. It is known that in this case, using the LS equation causes nonuniqueness of the scattering solution. Instead, we must use a set of three LS equations, which leads to the Faddeev equations after making a Faddeev decomposition of the scattering wave function~\cite{faddeev1960}. Faddeev equations give the exact solution of the three-body scattering problem. CDCC, on the other hand, solves a single LS equation in a model space characterized by the maximum $l_{\max}$ of the orbital angular momentum $l$ between particles 1 and 2. As shown in Refs.~\cite{austern1989,austern1996}, the solution of the LS equation in the model space is an approximate solution to the distorted-wave (DW) Faddeev equations~\cite{birse1982} that are rigorously equivalent to the original Faddeev equations, with the correction to the approximate solution becoming negligibly small when we take sufficiently large $l_{\max}$. The crucial point is that when a model space corresponding to large but finite $l_{\max}$ is taken, the distorting potential, that is, the auxiliary potential in the DW Faddeev formalism, between A and particle 1 or 2 becomes a three-body interaction, not a pair interaction. In this situation, there is no disconnected diagram and the solution of the single LS equation is physically meaningful. Therefore, the most important approximation made in the CDCC formalism is the $l$ truncation, as concluded in Refs.~\cite{yahiro2012,austern1989,austern1996}. An important aspect is that the solution to the LS equation with the $l$-truncation makes sense only if the result of the CDCC calculation converges with a certain value of $l_{\max}$. If this is not the case, it indicates the necessity of \ including a wave-function component corresponding to a rearrangement channel. This is certainly the case with transfer reactions, which will not be discussed in this review article.

\subsubsection{Description of breakup reactions with CDCC}
\label{sec332}

Let us consider NA scattering with the intrinsic spin of N disregarded. Henceforth, for a transparent correspondence with standard breakup experiments, we regard A as a projectile to be broken up and N as a target being inert. After the $l$-truncation, what is needed is to prepare a set $\left\{\Phi_{nIm_{I}}\right\}$ of A, not only bound and resonant states, but also for  nonresonant continuum states, which requires much larger model space than for the former two. If a set $\left\{\Phi_{nIm_{I}}\right\}$ covering sufficiently large space is prepared by diagonalizing the Hamiltonian of A, one can follow the MCC framework mentioned above. The discretization of the continuum in this manner is called the pseudostate (PS) discretization. 

Except for few-nucleon systems, however, it is extremely difficult to prepare an appropriate set $\left\{\Phi_{nIm_{I}}\right\}$ of a nucleus for breakup reactions with $A$-body structure model calculations. In many cases, therefore, a few-body (cluster) model is adopted to describe A. For instance, a two-neutron halo nucleus $^{6}$He is usually described by an
$\alpha+n+n$ three-body model assuming $\alpha$ to be inert. A set $\left\{\Phi_{nIm_{I}}\right\}$ is obtained by diagonalizing a three-body Hamiltonian of $^{6}$He in this case; the Gaussian basis functions~\cite{hiyama2003} and transformed harmonic oscillator basis functions~\cite{perez-bernal2001,mor2006} are widely used.

When a few-body model is applied to the description of A, the transition densities are evaluated accordingly. Nucleon transition densities can be obtained once a one-body density of the inert particle is specified. Then, the coupling-potentials are calculated in the same way as in the MCC framework. On the other hand, one may respect the clustering (few-body) nature of A in evaluating the coupling potentials. In this case, the coupling potentials are obtained by folding the interaction $U_{\mathrm{c}_{i}\mathrm{A}}$\ between the target and each constituent c$_{i}$ of A with the transition densities regarding the inter-cluster coordinate(s). In the latter approach, if available, one can also use a phenomenological $U_{\mathrm{c}_{i}\mathrm{A}}$.

\subsection{Breakup reaction studies with CDCC}
\label{sec34}

In this subsection, we will introduce some selected studies employing CDCC. Because there are numerous applications of CDCC to breakup reactions, we will focus on results that appeared after the recent review of CDCC~\cite{yahiro2012}. Furthermore, we will emphasize studies of new subjects which have not been investigated with CDCC before, rather than those for 
more standard applications of CDCC to analyses of breakup observables.

\subsubsection{Applicability of CDCC to low-energy breakup reactions}
\label{sec341}

\begin{figure}[!ht]
\begin{center}
\includegraphics*[width=0.6\textwidth]{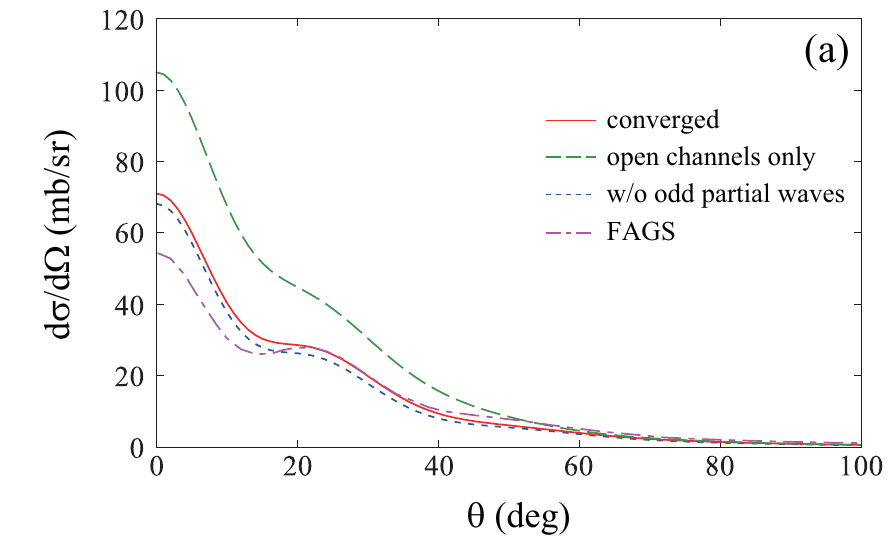}
\includegraphics*[width=0.6\textwidth]{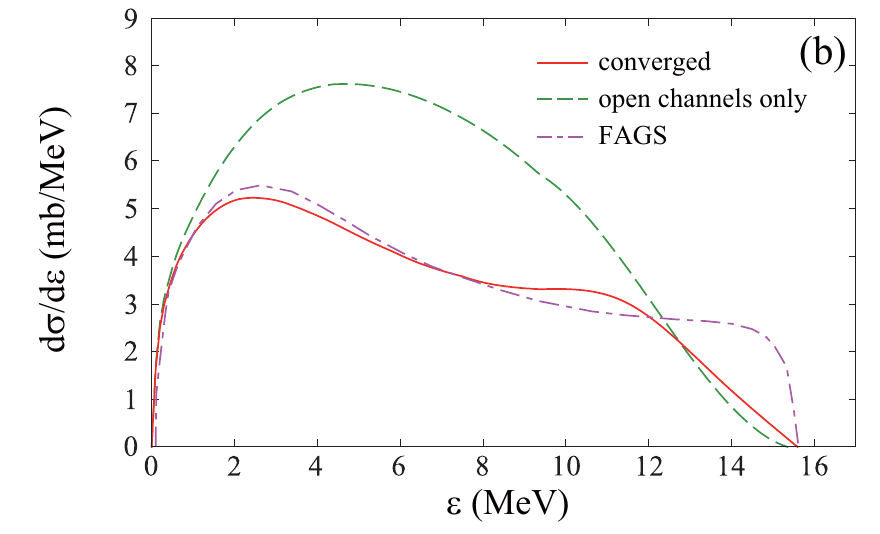}
\caption{
(a) Angular distribution and (b) breakup energy distribution of the elastic breakup cross section for $^{12}$C($d,pn$)$^{12}$C at 12~MeV. The solid, dashed, and dash-dotted lines in each panel show the converged CDCC result, the result of the CDCC method calculated with including only the open channels, and the result of FAGS theory taken from Ref.~\cite{upadhyay2012}, respectively. The dotted line in (a) is the same as the solid line but omitting the odd partial waves between $p$ and $n$. Taken from Ref.~\cite{ogata2016}.
}
\label{fig3-8}
\end{center}
\end{figure}
About 20 years after the theoretical validation of CDCC~\cite{austern1989,austern1996}, direct comparison between CDCC and the Faddeev-Alt-Grassberger-Sandhas (FAGS) theory~\cite{faddeev1960,alt1967} became feasible~\cite{deltuva2007,upadhyay2012} because of a novel approach to the Coulomb interaction in the FAGS calculation~\cite{deltuva2005}. In most cases, results obtained with the two frameworks reasonably agreed with each other, showing the reliability of CDCC at the numerical level. Although a striking deviation of CDCC results from FAGS ones for low-energy deuteron breakup reactions was reported as an exception in Ref.~\cite{upadhyay2012}, later, one of the authors (K.O.) and Yoshida~\cite{ogata2016} showed that it is due to the lack of closed channels in the CDCC calculation of Ref.~\cite{upadhyay2012}. Closed channels,which are characterized by the negative scattering energy between the projectile and the target, are regarded as virtual breakup states\footnote{In Ref.~\cite{ogata2016}, deuteron breakup with a target nucleus was described by CDCC adopting a nucleon-nucleus phenomenological optical potential. It is a three-body model calculation and does not rely on the MST. Thus, the role of closed channels can be discussed within the three-body reaction model.}. At low energies, couplings with the closed channels are crucially important~\cite{austern1987}, which is shown by the change from the dashed lines to the solid lines in Fig.~\ref{fig3-8}. Although a small difference between the FAGS (dash-dotted line) and converged CDCC (solid line) results remains, the applicability of CDCC to low-energy breakup reactions has been essentially confirmed~\cite{ogata2016}.

One may infer that the neglect of closed channels results in significant overshooting of the wave function near the three-body threshold at low incident energies. This will be closely related to the tremendous increase in triple-$\alpha$ reaction rates at the low temperature suggested by one of the authors (K.O.) and collaborators~\cite{ogata2009}. In Ref.~\cite{ogata2009}, a CDCC calculation including only the open channels was performed for the triple-$\alpha$ reaction, expecting that the effect of the closed channels could effectively be included by setting $\alpha$-$\alpha$ interaction to reproduce the position and width of the Hoyle ($0_2^+$) state. Later, Akahori and collaborators~\cite{akahori2015} have clarified with the imaginary-time formalism that such an increase in triple-$\alpha$ reaction rates was not realized, showing the importance of the closed-channel components of their three-$\alpha$ wave function. Although it will be extremely difficult to confirm this finding of Ref.~\cite{akahori2015} by performing CDCC calculations with closed channels, one may expect that inclusion of closed-channels can significantly change the result of Ref.~\cite{ogata2009}.

\subsubsection{Selection of decay mode and breakup channel}
\label{sec342}

In recent studies of the breakup of unstable nuclei, the specification of the final channel becomes more important to understand their many-body structures. For example, breakup of $^{16}$C, which will be well described by a ${}^{14}{\rm C}+n+n$ model, to the ${}^{15}{\rm C}+n$ channel will reveal to what extent $^{16}$C contains a halo nucleus $^{15}$C inside. Similarly, the decay mode study of a resonant state of an unstable nucleus is crucially important to understand its structure.

In principle, to impose a specific boundary condition for scattering states of a system consisting of more than two particles, we need an exact solution to the many-body scattering problem; for three-body systems, solving Faddeev equations will be the most rigorous way. As an alternative approach for describing observables of three-body scattering, Kikuchi and collaborators~\cite{kikuchi2009} proposed to use the complex-scaled solutions of the LS equation, which is referred to as the CSLS method. The CSLS method describes the three-body scattering states with correct boundary conditions in a finite space needed to describe breakup observables. In Ref.~\cite{kikuchi2013}, the CSLS method was implemented in CDCC and the decay mode of the $2_1^+$ state of $^6$He was studied. 

\begin{figure}[!ht]
\begin{center}
\includegraphics*[width=0.45\textwidth]{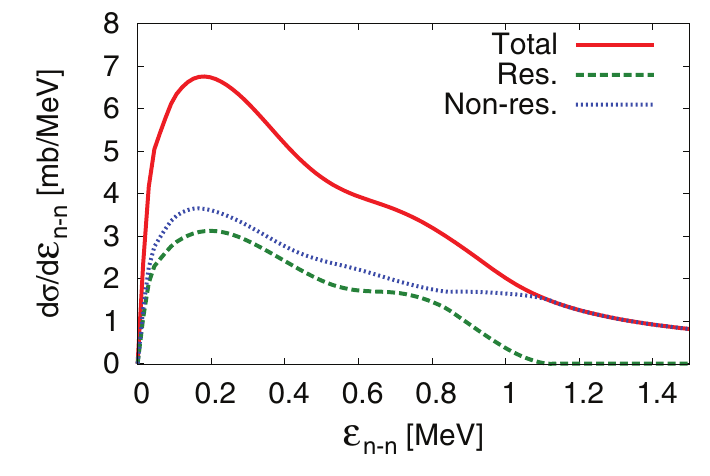}
\caption{
Invariant mass spectra with respect to $\epsilon_{nn}$ (solid line). The dashed and dotted lines show the resonant and nonresonant parts, respectively. Taken from Ref.~\cite{kikuchi2013}.
}
\label{fig3-9}
\end{center}
\end{figure}
The solid line in Fig.~\ref{fig3-9} represents the $nn$ invariant mass spectrum $d\sigma/d\epsilon_{nn}$ for breakup of $^6$He with a $^{12}$C target at 240~MeV/nucleon. A peak at low $nn$ energy $\epsilon_{nn}$ at 0.2~MeV and a shoulder structure around 0.8~MeV are seen. The former is due to the virtual state of the $nn$ system that has been reported in breakup by an electric dipole field~\cite{kikuchi2010}, whereas the latter not. If one selects the three-body energy of the $\alpha+n+n$ system around the $2_1^+$ resonance energy (0.98~MeV with the structure model adopted) of $^6$He, the dashed line is obtained, in which the shoulder structure remains. The rest of the solid line, the dotted line, has no clear structure around $\epsilon_{nn}=0.8$~MeV. Because the three-body energy is restricted at around 0.98~MeV in obtaining the dashed line, at $\epsilon_{nn}=0.8$~MeV, almost all energy is exhausted by the $nn$ relative motion. Therefore, it suggests that two neutrons are emitted in the opposite directions, i.e., the back-to-back decay is realized. Because the back-to-back decay mode is hardly affected by the $nn$ final-state interaction (FSI), it will carry structural information about the two neutrons in the $2_1^+$ state of $^6$He. A relatively high $nn$ momentum indicates a spatially correlated $nn$ pair. Therefore, the shoulder in $d\sigma/d\epsilon_{nn}$ for the decay from the $2_1^+$ resonant state is a possible signature of a dineutron in the $2_1^+$ state of $^6$He. Very recently, an indication of the shoulder structure in $d\sigma/d\epsilon_{nn}$ has been reported for $^6$He breakup with a $^{12}$C target at 184~MeV/nucleon~\cite{saito2021}.

The CSLS method has also been applied to $\alpha+d$ scattering with the $\alpha+p+n$ three-body model. In this case, the incident wave in the CSLS method is set to the $\alpha+d$ two-body Coulomb wave function. The three-body scattering wave function generated from this incident wave in a finite space is expressed in terms of eigenstates of the complex-scaled $\alpha+p+n$ Hamiltonian. The CSLS method was shown to  reproduce 
successfully the $d(\alpha,\gamma){}^{6}$Li capture cross section~\cite{kikuchi2011}.

Specification of the incident wave in solving three-body scattering problems with an outgoing boundary condition corresponds to the selection of the observed channel in breakup reaction of a three-body system, to which three-body scattering wave functions with an incoming boundary condition are relevant. Although the implementation of the CSLS method in CDCC has been done~\cite{kikuchi2013}, it is rather demanding computationally. In this situation, recently, Watanabe and collaborators~\cite{watanabe2021} proposed an approximated method, the P-separation method, for decomposing discretized breakup cross sections to individual PSs of a three-body system into the components corresponding to specific breakup channels. The P-separation method uses a probability $P$ of each PS having the component of interest, the discretized breakup cross section to the PS is multiplied by $P$. This method was applied to the breakup of $^6$Li with $^{208}$Pb at 39 and 210~MeV. The total breakup cross section at 31 (210)~MeV is 68.7 (137.0)~mb and the breakup cross section to the $\alpha+d$ and $\alpha+p+n$ channels were found to be 45.3 (89.9)~mb and 23.4 (47.1)~mb; about one-third of the $^6$Li breakup cross section goes to the $\alpha+p+n$ three-body channel. On the other hand, it was found in Ref.~\cite{watanabe2015} that the $\alpha+p+n$ channel had little effect on the $^6$Li elastic cross sections with $^{208}$Pb at both energies. A possible explanation of these two findings might be the strong coupling between the $\alpha+p+n$ and $\alpha+d$ channels, namely, $^6$Li first breaks up into $\alpha$ and $d$, then $d$ is broken up into $p$ and $n$. Further investigation including comparison with results of the CSLS method will be highly important.

\subsubsection{Interplay between resonant and nonresonant states}
\label{sec343}

Although there are a number of studies on resonant states of nuclei, we introduce here a few characteristic resonant phenomena (to be) found in breakup reactions of unstable nuclei. Emphasis is put on the interplay between resonant and nonresonant states of unstable nuclei.

In Ref.~\cite{ogata2013}, CDCC was applied to the breakup reaction of $^{22}$C with $^{12}$C at 250~MeV/nucleon; the cluster-orbital shell model (COSM)~\cite{suzuki1988} was adopted assuming a $^{20}{\rm C}+n+n$ three-body structure. COSM predicted a $0_2^+$ state at 1.02~MeV above the three-body threshold. To obtain a smooth breakup cross section $d\sigma_{\rm BU}/d\epsilon$ regarding the breakup energy $\epsilon$ from discrete cross sections obtained with CDCC, we employed the complex-scaling smoothing method proposed by Matsumoto and collaborators~\cite{matsumoto2010}. Because the $d\sigma_{\rm BU}/d\epsilon$ is expressed as an incoherent sum of the contributions from individual eigenstates of a complex-scaled Hamiltonian, one can isolate the breakup cross section corresponding to a resonant state, which is identified by a pole on the complex-energy plane. One must be careful, however, that a breakup cross section to a resonant state thus defined can be  significantly affected by the interference with nonresonant components and sometimes become negative.

\begin{figure}[!ht]
\begin{center}
\includegraphics*[width=0.45\textwidth]{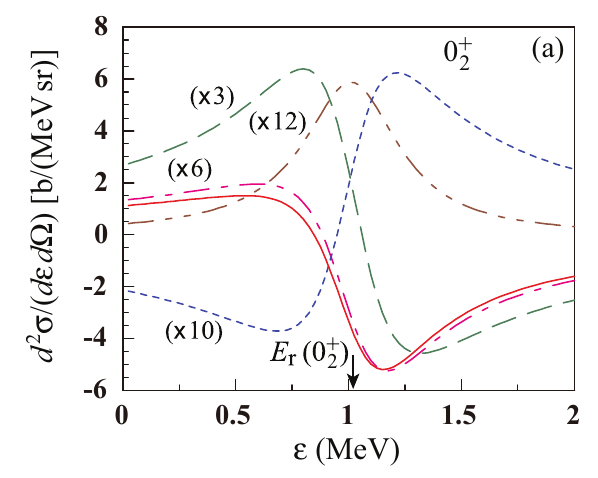}
\includegraphics*[width=0.45\textwidth]{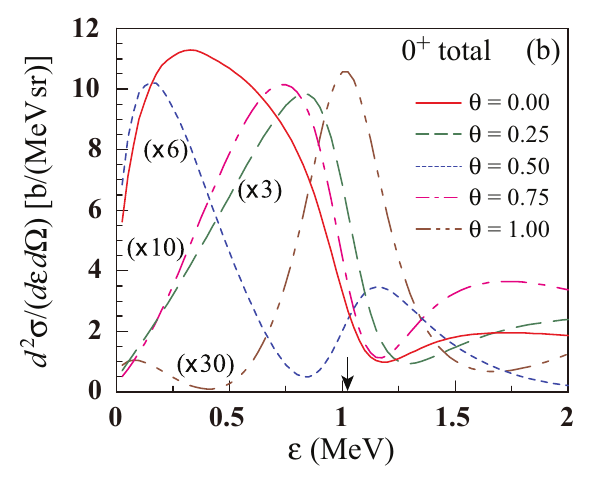}
\caption{
(a) $0_2^+$ contribution to the DDBUX of $^{22}$C by $^{12}$C at 250~MeV/nucleon and (b) total $0^+$ DDBUX. All results except for the solid lines are multiplied by the numbers beside the lines. Taken from Ref.~\cite{ogata2013}.
}
\label{fig3-10}
\end{center}
\end{figure}
Figure~\ref{fig3-10}(a) represents the double-differential breakup cross sections (DDBUXs) to the $0_2^+$ state of $^{22}$C. The shape of the DDBUX strongly depends on $\theta$; 
for $\theta=1$ deg. (the dash-dotted line) the distribution has the 
standard Breit-Wigner shape, whereas at 0.25 and 0.50 deg. (the 
dashed and dotted lines) the distribution almost vanishes at the resonant energy $E_{\rm r}$. 
This is well known to be a consequence of the background phase effect (BPE) on a resonance, or the Fano resonance~\cite{fano1961}. Although the $0_2^+$ cross section itself is not an observable, the total (resonant and nonresonant) $0^+$ BUDDX also has a strong dependence on $\theta$ as shown in Fig.~\ref{fig3-10}(b). The appearance of the Fano resonance in the $^{22}$C breakup can be explained as follows. First, the ground state of $^{22}$C has a large amount of the $(1s_{1/2})^2$ configuration which is responsible for its halo structure and large breakup cross section. This configuration is the main component of the low-lying $0^+$ continuum because of the absence of the centrifugal barrier. Second, the ground state of $^{22}$C also contains a $(0d_{3/2})^2$ configuration, which remains when excited by a monopole operator to the $0^+$ continuum. In fact, COSM suggests that the $0_2^+$ resonant state has the $(0d_{3/2})^2$ configuration. Thus, a resonant $(0d_{3/2})^2$ state and nonresonant $(1s_{1/2})^2$ states generated from each of the two components of the $^{22}$C ground state coexist and strongly affect each other in the low-lying $0^+$ continuum. This may also be the case with other two-neutron halo nuclei having a $(1s_{1/2})^2$ configuration, as the main component, a low-lying $0^+$ resonance. Note, however, that the calculation in Ref.~\cite{ogata2013} assumes that $^{20}$C is a core nucleus. Some indications of the core excitation in $^{22}$C have been discussed in Ref.~\cite{suzuki2016}.

\begin{figure}[!ht]
\begin{center}
\includegraphics*[width=0.7\textwidth]{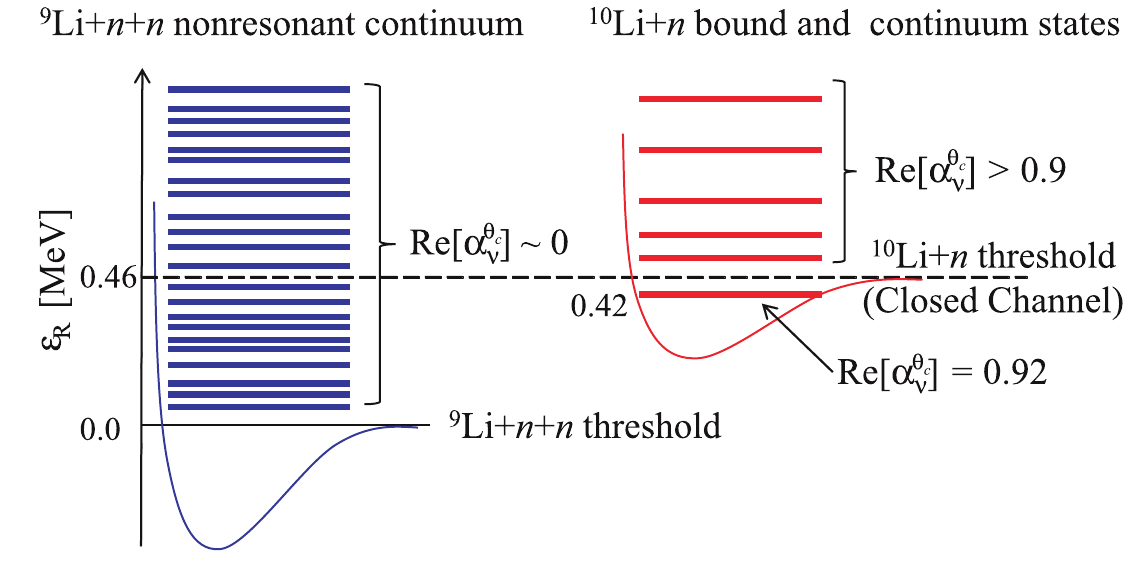}
\caption{
Schematic representation of complex-scaled states of $^{11}$Li for the $1^-$ state. Taken from Ref.~\cite{matsumoto2019}.
}
\label{fig3-11}
\end{center}
\end{figure}
Another interesting phenomenon related to nuclear resonances is the Borromean Feshbach resonance (BFR) in $^{11}$Li suggested by Matsumoto and collaborators~\cite{matsumoto2019}. It has been a long-standing issue whether $^{11}$Li has a soft dipole resonance or not. Recently, proton and deuteron inelastic scattering experiments of $^{11}$Li with high resolution have been conducted at TRIUMF~\cite{kanungo2015,tanaka2017}, 
and a clear peak was observed at around 1~MeV with a narrow width. In Ref.~\cite{matsumoto2019}, with a four-body CDCC calculation combined with the complex-scaling method (CSM)~\cite{aguilar1971,balslev1971}, a resonance in the $1^-$ three-body continuum state was identified and found to be responsible for the peak observed in the $^{11}$Li($p,p'$) experiment; note that the intrinsic spin of $^9$Li was disregarded as in the analyses of the experiments~\cite{kanungo2015,tanaka2017}. The resonance is suggested to have a structure consisting of the $^{10}$Li resonance and a neutron with negative energy with respect to the $^{10}{\rm Li}+n$ threshold, which can be regarded as a Feshbach resonance~\cite{feshbach1958,feshbach1962}. The $1^-$ resonant and nonresonant states classified with the CSM are summarized in Fig.~\ref{fig3-11}. As a distinctive character of Borromean nuclei, the three-body threshold is lower than the two-body one. The $1^-$ resonant state is a Feshbach resonance reflecting the Borromean nature of the three-body system, which is thus referred to be a BFR.

There have been several studies on $^{11}$Li with three-body models similar to that adopted in Ref.~\cite{matsumoto2019}. In Ref.~\cite{pinilla2012}, a resonance state in the $1^-$ state has been suggested and its role in Coulomb breakup observables was discussed. An indication of the existence of a $1^-$ resonance was obtained also in Ref.~\cite{cubero2012} by investigating $^{11}{\rm Li}+{}^{208}{\rm Pb}$ elastic scattering data. In view of this, the key finding of Ref.~\cite{matsumoto2019} will be the identification of the distinguishable feature of the $1^-$ resonant state of $^{11}$Li. This has been achieved by the implementation of the CSM.

\begin{figure}[!ht]
\begin{center}
\includegraphics*[width=0.45\textwidth]{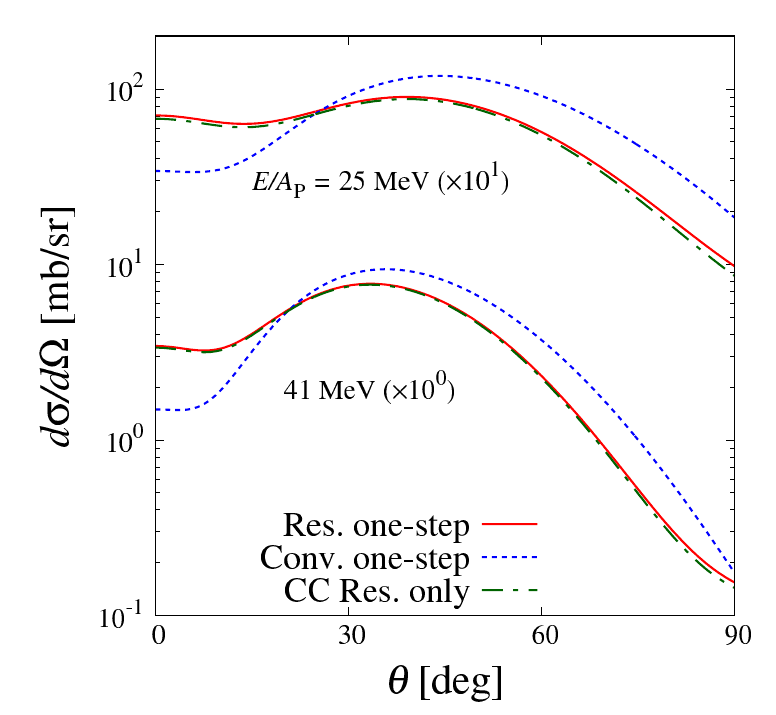}
\includegraphics*[width=0.45\textwidth]{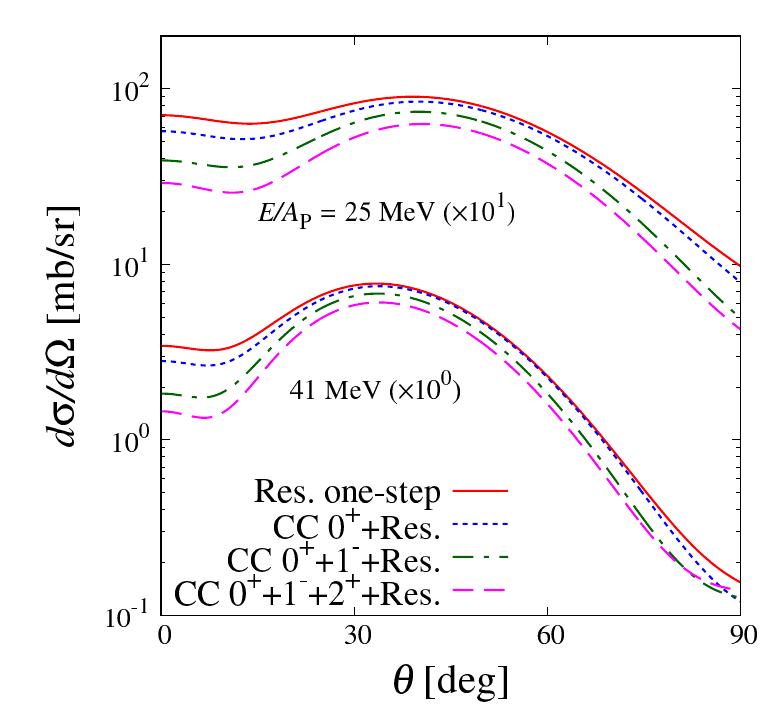}
\caption{
(a) Inelastic cross sections for the $^{6}{\rm He} + p$ reaction at 41 and 25~MeV/nucleon obtained with CDCC including only the fragmented resonant states (solid lines). The dotted lines represent the results of the one-step calculation in a conventional manner. The dashed lines are obtained by assuming a one-step transition from the ground state to each fragmented resonant state but the couplings among the fragmented resonant states are taken into account to all orders. (b) The solid line is the same as in (a). The dotted (dot-dashed) line represents the CDCC result including the ground state, fragmented resonant states, and the $0^+$ ($0^+$ and $1^-$) nonresonant states. The dashed line corresponds to the result of the CDCC with all states in the model space. Taken from Ref.~\cite{ogawa2021}.
}
\label{fig3-12}
\end{center}
\end{figure}
As mentioned, CDCC can treat both resonant and nonresonant states on the same footing by adopting a large model space that can cover the latter. As a result, in some cases, a resonant state is fragmented into several PSs obtained in the diagonalization of a Hamiltonian. Note that CDCC does not adopt eigenstates of a complex-scaled Hamiltonian and classification of PSs into resonant and nonresonant states is difficult unless a special effort~\cite{casal2019} is made. However, it is expected that only a limited number of states contain the resonant component as their main components. Let us consider inelastic scattering of $^6$He to the $2_1^+$ state and assume for simplicity that there are a few fragmented resonant states (FRSs) and many discretized nonresonant states. In this situation, a one-step (DWBA-type) calculation can give a result that significantly deviates from the result of CDCC. In fact, a significant CC effect on the $^6$He($p,p'$) cross section to the $2_1^+$ state, which changes even the shape of the angular distribution, was reported in Ref.~\cite{ogawa2020}. It was found, however, that the strong couplings among the FRSs are the primary source of the large CC effect~\cite{ogawa2021}. In Fig.~\ref{fig3-12}(a), the dash-dotted lines represent the results of CC calculation taking the ground state and the FRSs, whereas the dotted lines the one-step calculation results. If only the couplings among the FRSs are treated nonperturbatively and others are taken up to the first-order, i.e., a resonant one-step calculation is performed, the solid lines are obtained showing an excellent agreement with the CC results. This suggests that the coupling between the ground state and the $2_1^+$ resonant state is weak enough to be treated perturbatively. This is consistent with the finding of Ref.~\cite{kanada-en'yo2020e} in which only resonant continuum states obtained with AMD were included. It should be noted, however, that the coupling between the $2_1^+$ state and nonresonant continuum states are not negligible as shown in Fig.~\ref{fig3-12}(b), which affects the magnitude of the $2_1^+$ cross section by 20--30~\% and is extremely difficult to take into account with usual many-body calculations like AMD.

\subsubsection{Related subjects}
\label{sec344}

In recent years, four-body CDCC calculation of the breakup of a projectile having a three-body structure \cite{matsumoto2004,Rod09} becomes more popular than before \cite{descouvemont2013,watanabe2015,Cas13,Cas14,Cas15,ogawa2020,singh2021}. In a different direction, recently, a new four ($=2+2$) body CDCC framework was proposed by Descouvemont~\cite{descouvemont2017,descouvemont2018} and applied to the $d+{}^{11}$Be elastic scattering. It was shown that breakup states of both $d$ and $^{11}$Be affect the elastic cross section. This work can be regarded as an extension of the studies on mutual excitation discussed in Sec.~\ref{sec324}. Sometimes, a deuteron target is used to induce isoscalar transition as in the $^{11}$Li inelastic experiment mentioned above~\cite{kanungo2015}. The finding in Refs.~\cite{descouvemont2017,descouvemont2018} evidences the importance of deuteron breakup in such inelastic scattering (or breakup) of unstable nuclei.

An important ingredient that had not been explored until quite recently is the dynamical relativistic effect (DRE) on breakup reactions. At intermediate energies, imposing a Lorentz covariance of nuclear and Coulomb coupling potentials was found to affect breakup observables of unstable nuclei with a heavy target like $^{208}$Pb~\cite{bertulani2005,ogata2009b}; an eikonal CC framework was needed to implement the Lorentz covariance of the potentials. In Ref.~\cite{ogata2010}, the DRE was shown to affect the breakup amplitude only for large impact parameters $b$, and thus large projectile-target orbital angular momenta $L$. Combining the breakup amplitude with DRE calculated with the eikonal CDCC (E-CDCC)~\cite{ogata2003} for large $L$ and a fully quantum-mechanical amplitude obtained with CDCC without DRE for small $L$ allows one to perform a relativistic CDCC calculation. Very recently, Moschini and Capel~\cite{moschini2019} investigated the DRE in the framework of dynamical eikonal approximation (DEA)~\cite{goldstein2006}, finding a sizable DRE on the $^{11}$Be breakup cross section with $^{208}$Pb at 520~MeV/nucleon. The DRE in DEA seems considerably larger than that evaluated with E-CDCC~\cite{ogata2010}, though the latter was evaluated at 250~MeV/nucleon. Because DEA was shown to be formally equivalent to E-CDCC if a semi-adiabatic assumption is made~\cite{fukui2014}, further investigation on the difference in the DREs suggested by the two models will be encouraged.

Recently, proton-induced nucleon knockout, $(p,pN)$, reactions have been measured for many unstable nuclei to reveal their single-particle nature, the magicity in particular. An interesting {\lq\lq}application'' of CDCC is made for such studies, namely, the transfer to the continuum (TC) method~\cite{moro2015}. The main idea of the TC method is to use a $p+N+{\rm B}$ three-body wave function, where B is the residual nucleus, calculated with CDCC, as a final-state wave function in the ($p,pN$) transition matrix. In contrast to the standard DWIA description~\cite{jacob1966,chant1983,wakasa2017}, the TC method does not rely on the impulse approximation. Moreover, the TC method can describe $(p,pn)$ reactions and ($p,d$) transfer reactions simultaneously and consistently. The TC method has successfully been benchmarked with FAGS~\cite{yoshida2018,gomez-ramos2020} and applied to investigate proton-induced knockout reactions~\cite{gomez-ramos2017,gomez-ramos2018}; in Ref.~\cite{yoshida2018}, DWIA has also been benchmarked.
Note that the TC method discussed here is based on a purely quantum-mechanical treatment of the reaction. A semiclassical TC method (STC) was also proposed by Bonaccorso and Brink and successfully applied to a number of reactions~\cite{bonaccorso1988,bonaccorso1991,bonaccorso2001,flavigny2012}. Although both methods treat the breakup process as a transfer of one part of the projectile to the target continuum, the STC treats the projectile-target relative motion classically (see Sec. 2.6).

CDCC has been applied also to nuclear data science as introduced in the recent review article~\cite{yahiro2012}. Subsequently, a systematic calculation of deuteron total reaction cross sections $\sigma_{\rm R}^{(d)}$ has been conducted with CDCC~\cite{minomo2017b} and implemented in the particle and heavy ion transport code system (PHITS)~\cite{sato2013}. The $\sigma_{\rm R}^{(d)}$ formula used before was extrapolated from AA reaction data and found to severely undershoot the $\sigma_{\rm R}^{(d)}$ data. Implementing the $\sigma_{\rm R}^{(d)}$ calculated with CDCC significantly increased the reaction probability of all processes induced by the deuteron, which was crucial to design a nuclear transmutation scenario by deuteron. A simple functional form of $\sigma_{\rm R}^{(d)}$ is given in Ref.~\cite{minomo2017b}. CDCC has also been used for evaluating the elastic breakup component in the inclusive neutron production process by deuteron, which is of crucial importance for the engineering design of the international fusion materials irradiation facility (IFMIF)~\cite{garin2011}. A new deuteron-induced reaction analysis code system named DEURACS was developed by Nakayama and collaborators~\cite{nakayama2016} and a new deuteron nuclear data library JENDLE-DEU20 has been released~\cite{nakayama2021}. DEURACS takes into account not only the elastic breakup but also nucleon transfer, pre-equilibrium and compound processes, and nonelastic breakup induced by deuteron. The nonelastic breakup of deuteron is described with the formula by Hencken, Bertsch, and Esbensen~\cite{hencken1996} based on the Glauber model~\cite{glauber1959}. Details of the nonelastic breakup process and its theoretical description will be given in Sec.~\ref{sec:inclbu}.

Very recently, CDCC has been applied also to hadron physics. In Ref.~\cite{ogata2021b}, the $d$-$\Xi^-$ correlation function was studied with CDCC adopting a nucleon-$\Xi$ potential determined with lattice QCD. Although the effect of deuteron breakup on the correlation function was not found very significant, the framework developed will be helpful to investigate correlation functions for three-body (two protons and a baryon) systems to be measured at LHC.

\subsection{Core and target excitations\label{sec:coretargex}}
\subsubsection{Core excitations\label{sec:corex}}
As discussed in Sec.~3.3, for weakly-bound projectiles it is
convenient to describe the structure using a few-body model. Let us consider for simplicity the case of a two-body projectile $a$ composed of fragments $b$+$x$ impinging on a target $A$. In the standard CDCC formulation, possible excitations of the target nucleus $A$ are not considered explicitly, although they are taken into account effectively through the $x+A$ and $b+A$ optical potentials. Likewise, if  either $b$ or $x$ are composite systems themselves, possible excitations of these fragments are also possible and will be therefore also accounted for by the fragment-target optical potentials. 

In some cases, a proper description of the reaction may require however the explicit inclusion of such fragment excitations. For example, for the scattering of halo nuclei,
core excitations may  affect the structure of the projectile since projectile states will contain in general admixtures of core-excited components, which are not included in the standard single-particle description of these nuclei. Additionally, the interaction of the core with the target will produce excitations and deexcitations of the former during the collision, and this will  modify the reaction observables to some extent. These two effects (structural and dynamical) have been recently investigated within extended versions of the DWBA and CDCC methods \cite{Cre11,Mor12,Sum06,Die14}. 

\begin{figure}[ht]
{\par\centering \resizebox*{0.4\textwidth}{!}{\includegraphics{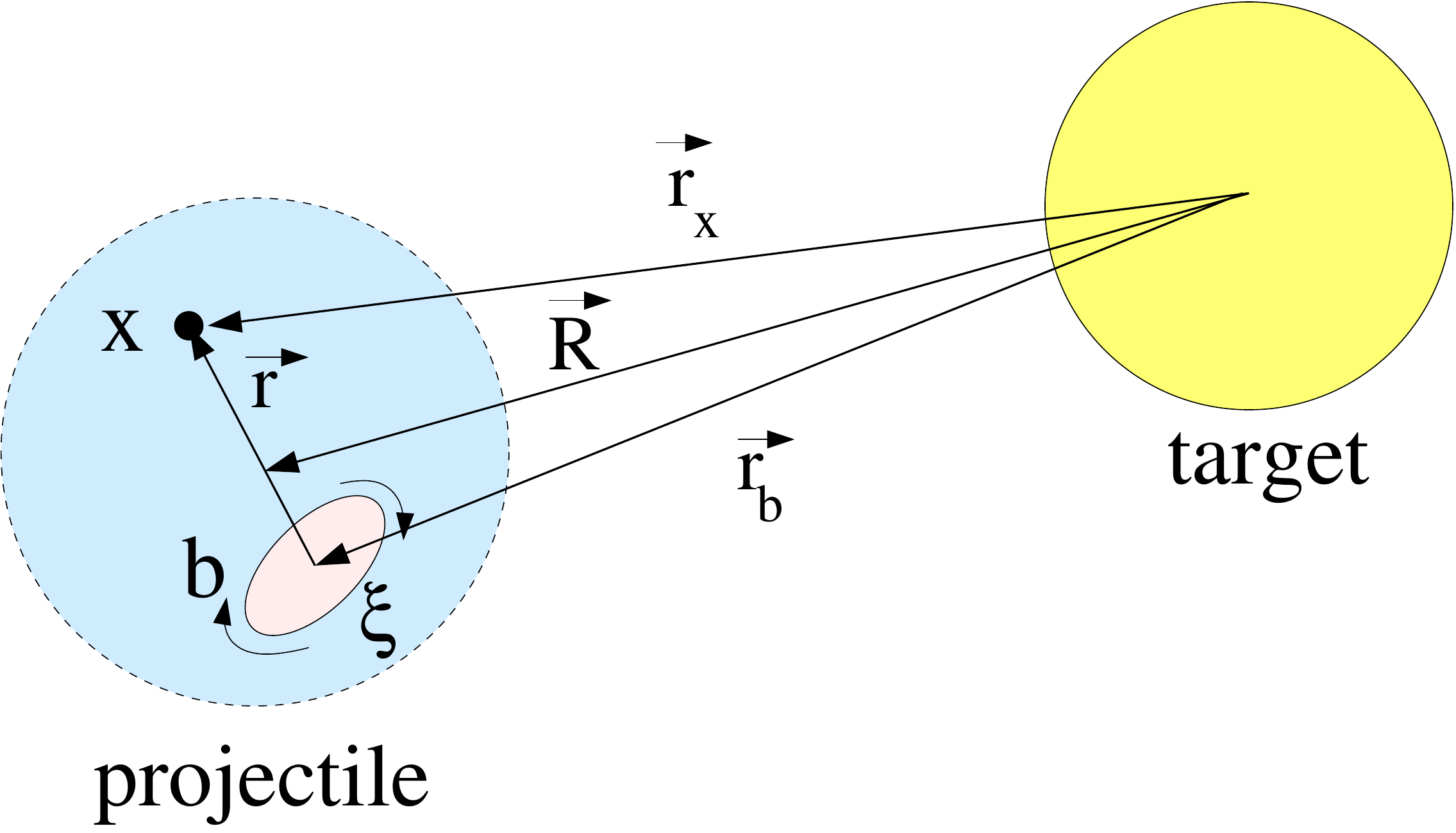}}\par}
 \caption{\label{fig:coord_corex} (Color online) Schematic sketch of the
   weakly-bound projectile composed by a core ($b$) and a valence
   particle ($x$). To study the scattering of the composite projectile with a inert target, within 
a three-body model, the relevant coordinates are the relative coordinate of the valence particle 
with respect to the core ($\vec r$) and that between the center of mass of the projectile and the target ($\vec R$). 
}   
\end{figure}

Considering for definiteness the case of core excitations in two-body halo nuclei, the CDCC Hamiltonian is conveniently generalized as follows (see Fig.~\ref{fig:coord_corex} for a sketch of  the relevant coordinates):
\begin{equation}
\label{eq:Heff_corex}
H= H_{\rm proj}(\vecr,\xi_b)   + \hat{T}_{\vecR} + U_{bA}(\vecr_{bA},\xi_b) + U_{xA}(\vecr_{xA}) 
\end{equation}
where $\hat{T}_{\vecR}$ is the kinetic energy operator for the projectile-target relative motion, 
$U_{bA}$ and $U_{xA}$ are the optical potentials for the $b+A$ and $x+A$ systems, with $\xi_b$ denoting the internal degrees of freedom of the core, and $H_{\rm proj}(\vecr,\xi_b)$ is the internal projectile Hamiltonian. The potential $U_{bA}(\vecr_{bA},\xi_b)$ is meant to describe both elastic and inelastic scattering  of the $b+A$ system (for example, it could be represented by a deformed potential such as those used in the context of inelastic scattering with collective models, as described in Sec.~2.3).  Note that the core degrees of freedom ($\xi_b$) appear in the projectile Hamiltonian (structure effect) as well as in the core-target interaction (dynamical effect). 

In the weak coupling limit, the projectile Hamiltonian can be written more explicitly as
\be
H_{\rm proj}= \hat{T}_{\vecr} + V_{bx}(\vecr,\xi_b) + h_{\rm core}(\xi_b) , 
\label{Hproj}
\ee
where $h_{\rm core}(\xi_b)$ is the internal Hamiltonian of the core. The eigenstates of this Hamiltonian are of the form 
\be
\Phi_{n j_p m_p }(\xi_b,\vecr) 
= \sum_{\alpha}  
\left[   \varphi_{n,\alpha}(\vecr) \otimes \phi_{I}(\xi_b) \right]_{j_p m_p} ,
\label{wfrot}
\ee
where $n$ is an index labeling the states with angular momentum $j_p$,    $\alpha \equiv \{\ell,s, j,I \}$, with $I$ and $s$ the core and valence intrinsic spins, $\vec{j}=\vec{\ell} + \vec{s}$ and $\vec{j}_p=\vec{j}+\vec{I}$. 
The functions $\phi_{I}(\xi_b)$  and  $\varphi_{n,\alpha}(\vecr)$ describe, respectively, the core states and the valence--core relative motion.  For continuum states, a procedure of continuum discretization is used, similar to that employed in standard CDCC. Calculations published so far have made use of either multi-channel bins  \cite{Sum06,Pes17,Dip19} or pseudo-states (PS) \cite{Die14,Lay16,Che16a,Dua20,Mor20}.  For further details on the calculation of the functions $\Phi_{n j_p m_p }(\xi_b,\vecr)$ we refer to Refs.~\cite{Sum06,Die14}.

Once the projectile states (\ref{wfrot}) have been calculated, the three-body wave function for a total angular momentum $J_T$ is expanded in a truncated basis of such  states, as in the standard CDCC method [c.f.~Eq.~(\ref{3psijm})],
\begin{equation}
\Psi^\mathrm{CDCC}_{c_0,J_T, M_T}(\vec{R},\vec{r},\xi)=\sum_{c}
\frac{\chi_{c,c_0}^{J_T}(R)}{K R}  \left [i^L Y_L({\hat{R}})\otimes \Phi_{n j_p }(\xi_b,\vecr) \right]_{J_T,M_T},  
\label{f3b}
\end{equation}
with $c=\{n,L,j_p \}$ and likewise for $c_0$, the incident channel.

Early calculations using this extended CDCC method (referred to as XCDCC) were first performed by Summers {\it et al.}~\cite{Sum06,Sum07} for  $^{11}$Be and $^{17}$C on $^{9}$Be and  $^{11}$Be+$p$, finding a very little core excitation effect in all these cases. However, later calculations for the $^{11}$Be+$p$ reaction based on a alternative implementation of the XCDCC method using a PS representation of the projectile states \cite{Die14} suggested much larger effects.  The discrepancy was found to be due to an inconsistency in the numerical implementation of the XCDCC formalism presented in Ref.~\cite{Sum06}, as clarified in Ref. \cite{Sum14}. For heavier targets, such as $^{64}$Zn or $^{208}$Pb, the calculations of Refs. \cite{Die14,Pes17} suggest that the core excitation mechanism plays a minor role, although its effect on the structure of the projectile is still important.  

\begin{figure}[!ht]
\begin{center}\includegraphics[width=0.6\columnwidth]{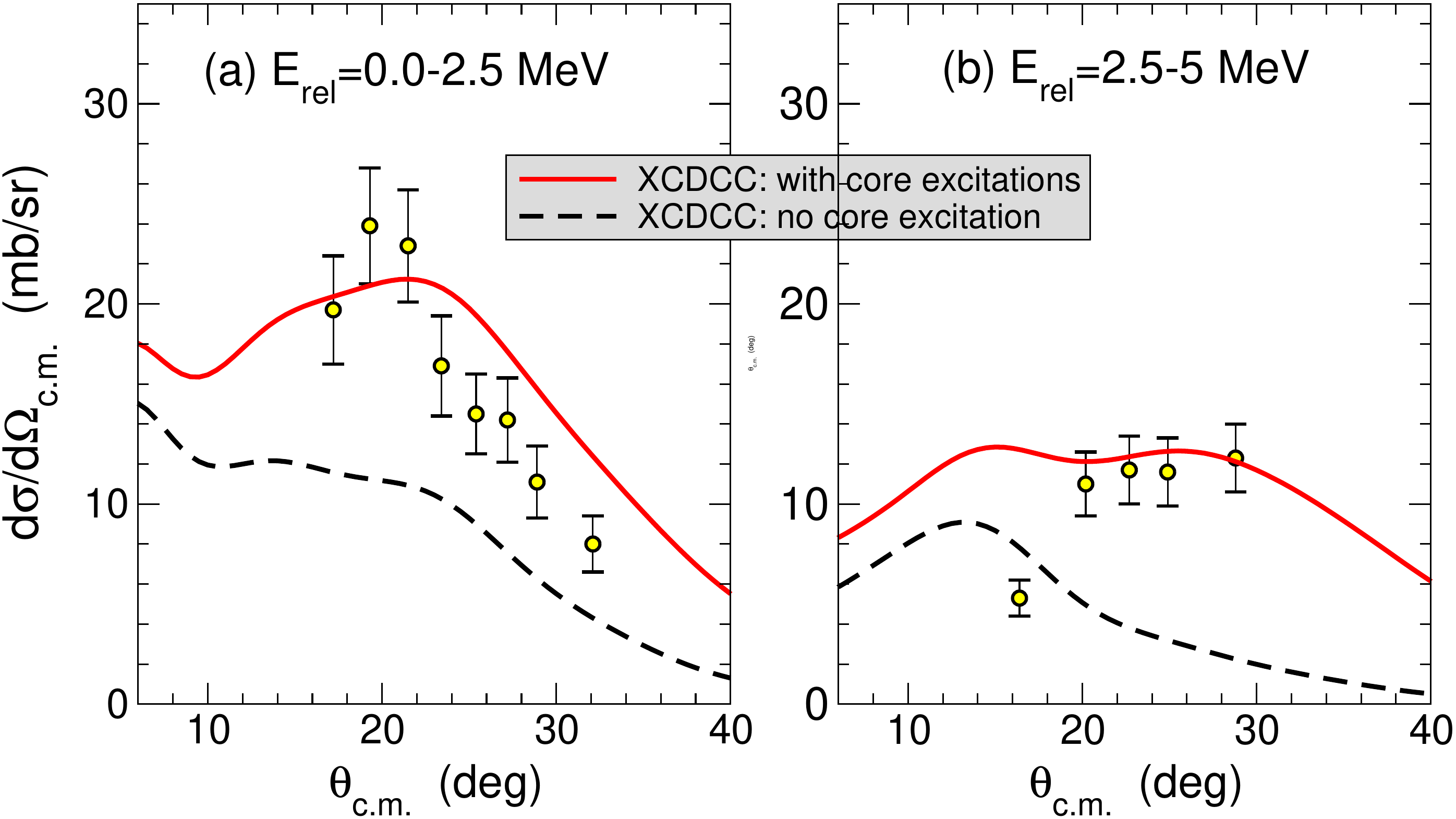}\end{center}
\caption{\label{fig:be11p_xcdcc}
Differential breakup cross  sections, with respect to  $^{11}$Be c.m.\ scattering angle,  for the breakup of $^{11}$Be on protons at 63.7~MeV/nucleon.  XCDCC calculations with (solid lines) and without (dashed lines) are shown. The circles  are the data from Ref.~\cite{Shr04}. Adapted from Ref.~\cite{Die14}.}
\end{figure}

As an example of these XCDCC calculations we show in  Fig.~\ref{fig:be11p_xcdcc} the differential breakup cross section, as a function of the $^{11}$Be c.m.\ scattering angle,  for the reaction $^{11}$Be+$p$ at 63.7~MeV/nucleon. Continuum states with angular momentum/parity $j_p=1/2^\pm$, $3/2^\pm$ and $5/2^+$ were included using a  PS basis of transformed harmonic oscillator (THO) functions \cite{Lay12}. Further details of the structure model and potentials are given  in Ref.~\cite{Die14}. The two panels correspond to different relative-energy intervals of the $^{10}$Be+$n$ continuum, as specified by the labels. The solid and dashed lines are the XCDCC calculations with and without core excitations, respectively. The circles are the data from Shrivastava {\it et al.}~\cite{Shr04}. It is seen that the inclusion of core excitations is crucial for a correct description of these data. 

\begin{figure}[!ht]
\begin{center}
\begin{minipage}[c]{.32\textwidth}
{\par \resizebox*{0.88\textwidth}{!}
{\includegraphics{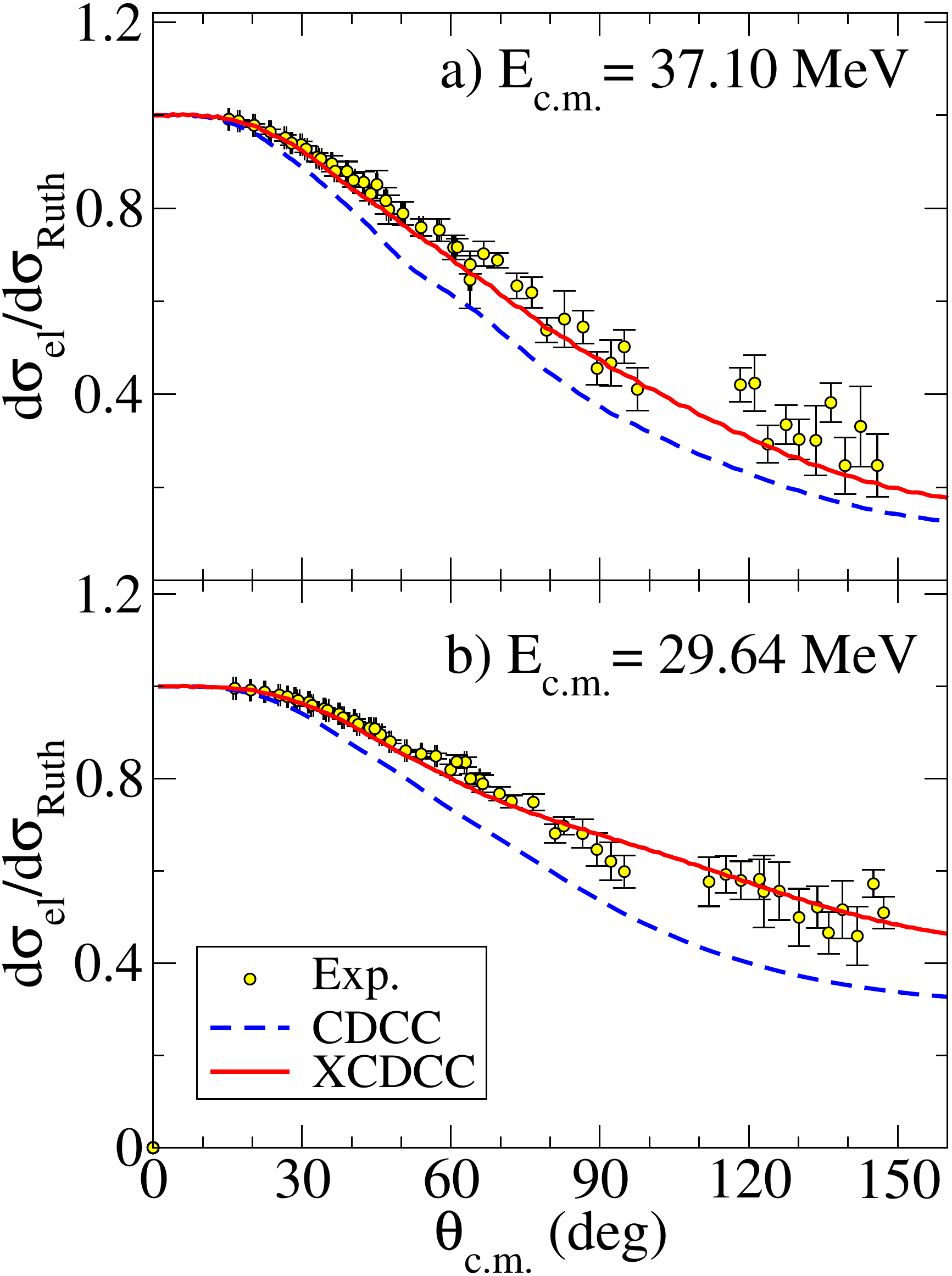}} \par}
\end{minipage}
\begin{minipage}[c]{.32\textwidth}
{\par \resizebox*{0.88\textwidth}{!}
{\includegraphics{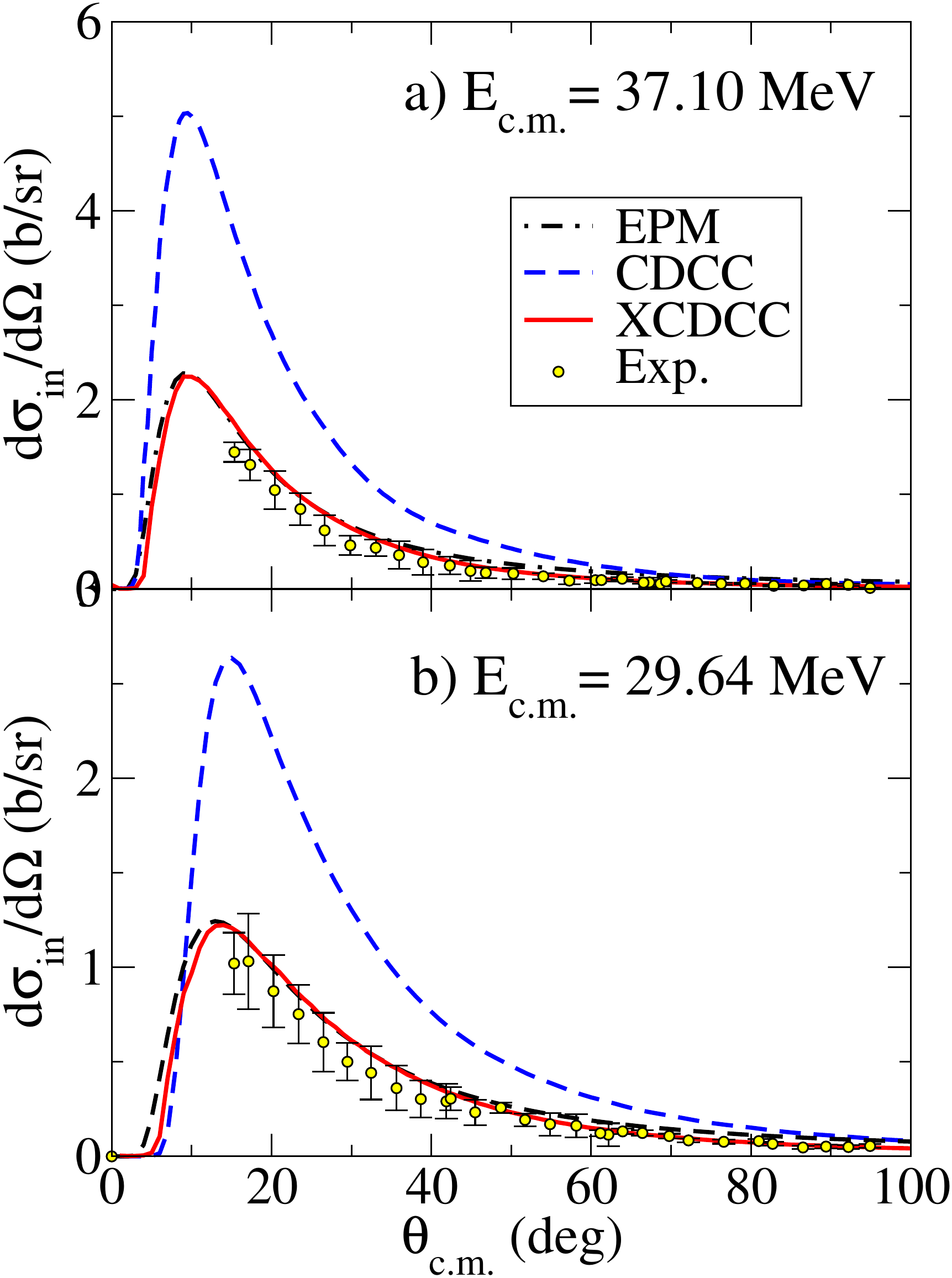}} \par}
\end{minipage}
\begin{minipage}[c]{.32\textwidth}
{\par \resizebox*{0.88\textwidth}{!}
{\includegraphics{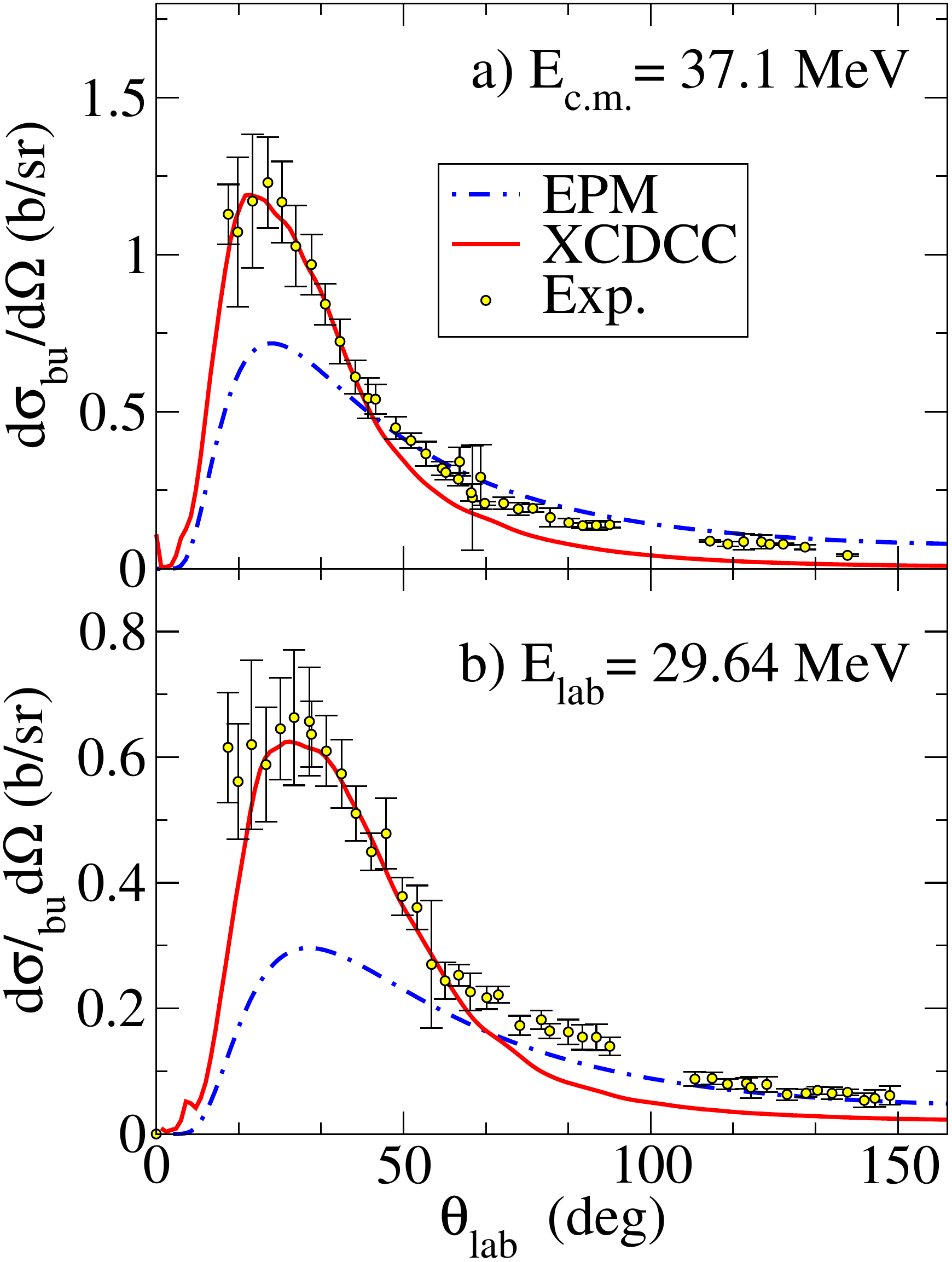}} \par}
\end{minipage}
\caption{\label{fig:be11au} Elastic (left), inelastic (middle) and breakup (right) cross sections for the reaction $^{11}$Be+$^{197}$Au at near barrier energies ($V_b \sim 40$~MeV). Experimental data are compared with CDCC, CDCC and first-order semiclassical pure $E1$ calculations (labeled EPM). Adapted from ref.~\cite{Pes17}.}
\end{center}
\end{figure}

The importance of the deformation on the structure of the projectile is clearly evidenced in the elastic and inelastic scattering of $^{11}$Be on $^{197}$Au at energies around and below the Coulomb barrier \cite{Pes17}, shown in Fig.~\ref{fig:be11au}. XCDCC calculations based on the particle-plus-rotor model of Ref.~\cite{Tar03} (solid lines) are able to reproduce simultaneously the elastic, inelastic and breakup angular distributions. On the other hand, standard CDCC calculations using single-particle wave functions fail to describe the elastic and inelastic data, even describing well the breakup (dashed lines). This is due to the overestimation of the $B(E1)$ connecting the ground state with the bound excited state \cite{Pes17}.

A  simpler DWBA, no-recoil  version of the formalism (XDWBA) has been also proposed in  Refs.~\cite{Cre11,Mor12}. An application of this formalism to the $^{11}$Be+$^{12}$C reaction at 69~MeV/u showed that the core excitation mechanism may interfere with the single-particle excitation mechanism, producing a conspicuous effect on the interference pattern of the resonant breakup angular distributions \cite{Mor12b}.

\subsubsection{Three-body observables}

Within the CDCC and XCDCC reaction formalisms, the breakup is treated as an
excitation of the projectile to the continuum so the theoretical cross sections are described in terms of
the c.m.\ scattering angle of the projectile and the relative energy of the constituents,
using two-body kinematics. For comparison with experimental data, it is
useful to have also the cross sections in terms of the angle and energy of the projectile fragments, since these quantities are more directly connected with the actual outcome of the measurements. 

In the case of the standard CDCC framework,  fivefold fully exclusive cross sections were already derived and presented by several authors \cite{Ise86,Tos01}. The method was recently generalized in Ref.~\cite{Die17} to the case of XCDCC. We briefly review the main formulas of the latter noting that the case without core excitation is recovered when the a single core state is considered. For simplicity, we ignore the target spin. 

In XCDCC, after solving the coupled-channel equations, one obtains some discrete breakup  transition amplitudes $T_{M_0,M'}^{i,J_0,J'}(\theta_i,K_i)$, connecting an initial state $|J_0  M_0 \rangle$ with a three-body final state comprised by  the target (assumed to be structureless), the valence particle and the core, and some discrete value of the final c.m.\ momentum  $\vec{K}_i=\{\theta_i,K_i\}$. The first step of the formalism is to relate these discrete amplitudes with the  actual breakup scattering amplitudes, without continuum discretization, that we denote as   $T_{\mu \sigma; M_0}^{I s; J_0}(\vec{k},\vec{K})$, where  $\vec{k}$ is the internal  relative momentum. 
Formally, these breakup transition amplitudes can be written in integral form as:
\be
T_{\mu \sigma; M_0}^{I s; J_0}(\vec{k},\vec{K})=
\langle \phi_{\vec{k}; I \mu; s \sigma}^{(-)} 
e^{i \vec{K} \cdot \vec{R}} | U | \Psi_{J_0,M_0}(\vec{K}_0)\rangle , 
\label{T-matrix1}
\ee
where $U=U_{bA}(\vec{r},\vec{R},\xi)+U_{xA}(\vec{r},\vec{R})$ and $\phi_{\vec{k}; I \mu; s \sigma}^{(-)}$ are two-body  {\it exact} scattering wave functions of the $b+x$ system for a relative final momentum $\vec{k}$ and  given core and valence spins. 

In order to relate the discrete and continuous amplitudes, we 
approximate the exact wavefunction $\Psi_{J_0,M_0}$ in the equation above by its XCDCC counterpart and we introduce the approximate completeness relation in the truncated discrete basis $\{ \Phi^{(N)}_{i,J,M}; i=1,\ldots,N \}$: 
\begin{align}
T_{\mu \sigma; M_0}^{I s; J_0}(\vec{k},\vec{K}) & \simeq  \sum_{i,J',M'} \langle \phi_{\vec{k}; I \mu; s \sigma}^{(-)}| \Phi^{(N)}_{i,J',M'}\rangle  \langle \Phi^{(N)}_{i,J',M'}
e^{i \vec{K} \cdot \vec{R}} | U | \Psi_{J_0,M_0}^\mathrm{CDCC}(\vec{K}_0)\rangle  \nonumber \\
&= \sum_{i,J',M'} \langle \phi_{\vec{k}; I \mu; s \sigma}^{(-)}| \Phi^{(N)}_{i,J',M'}\rangle T_{M_0,M'}^{i,J_0,J'}(\vec{K}) ,
\label{T-matrix2}
\end{align}
where the transition matrix elements $T_{M_0,M'}^{i,J_0,J'}(\vec{K})$ 
are to be interpolated from the discrete ones $T_{M_0,M'}^{i,J_0,J'}(\theta_i,K_i)$.  
The overlaps between the final scattering states and the discrete states, $\langle \phi_{\vec{k}; I \mu; s \sigma}^{(-)}| \Phi^{(N)}_{i,J',M'}\rangle $, are given explicitly in Refs.~\cite{Tos01} and \cite{Die17} for bin and PS functions, respectively.

The transition amplitudes of Eq.~(\ref{T-matrix2}),
$T_{\mu \sigma; M_0}^{I s; J_0}(\vec{k},\vec{K})$,
contain the dynamics of the process for the coordinates
describing the relative and center of mass motion 
of the core and the valence particle. From these amplitudes one can
derive two-body observables for a fixed spin of the core, $I$,
the solid angles describing the orientations of $\vec{k}$ ($\Omega_k$) and $\vec{K}$ ($\Omega_K$),
as well as the relative energy between the valence
and the core, $E_\mathrm{rel}$.
These observables 
factorize into the transition matrix elements and a kinematical factor:
\begin{align}
\frac{d^3\sigma^{(I)}}{d\Omega_k d\Omega_K dE_\mathrm{rel}} = & 
     \frac{\mu_{bx} k_I}{(2\pi)^5 \hbar^6} \frac{K}{K_0} \frac{\mu_{aA}^2}{2J_0+1}  \sum_{\mu, \sigma, M_0} |T_{\mu \sigma; M_0}^{I s; J_0}(\vec{k},\vec{K})|^2 ,
\label{three-obsv-cm}
\end{align}
where $\mu_{bx}$ and $\mu_{aA}$ are the valence-core and projectile-target
reduced masses. 

The three-body observables, assuming the energy of the core is measured, are
given by \cite{Tos01}:
\begin{equation}
\frac{d^3\sigma^{(I)} }{d\Omega_b d\Omega_x dE_b}=
     \frac{2 \pi \mu_{aA}}{\hbar^2 K_0} \frac{1}{2J_0+1} 
 \sum_{\mu, \sigma, M_0} |T_{\mu \sigma; M_0}^{I s; J_0}(\vec{k},\vec{K})|^2               
      \rho(\Omega_b, \Omega_x, E_b) ,
\label{three-obsv}
\end{equation}
where the phase space term $\rho(\Omega_b, \Omega_x, E_b)$, i.e., the number of states per
 unit core energy interval at solid angles $\Omega_b$ and $\Omega_x$,
 takes the form \cite{Fuc82}:
\begin{eqnarray}
\rho(\Omega_b, \Omega_x, E_b)=
     \frac{m_b m_x \hbar k_b \hbar k_x}{(2\pi\hbar)^6}
 \left[\frac{m_A}{m_x+m_A+m_x(\vec{k}_b-\vec{K}_{tot})\cdot\vec{k}_x/k_x^2}\right] .
\label{phfac}
\end{eqnarray}
Here, the particle masses are given by $m_b$ (core), $m_x$ (valence), and $m_A$ (target)
while $\hbar \vec{k}_b$ and $\hbar \vec{k}_x$ are the core and valence particle momenta
 in the final state. The total momentum of the system corresponds to $\hbar \vec{K}_{tot}$ and
the connection with the momenta in Eq.~(\ref{T-matrix2}) is made through:
\be
\vec{K}=\vec{k}_b+\vec{k}_x-\frac{m_a}{M_{tot}}\vec{K}_{tot} ; \quad
\vec{k}=\frac{m_b}{m_a}\vec{k}_x-\frac{m_x}{m_a}\vec{k}_b
\ee
with $m_a=m_b+m_x$ and $M_{tot}=m_b+m_x+m_A$ the total masses of the projectile and the
three-body system, respectively.

\begin{figure}[!ht]
\begin{center}
\begin{minipage}[t]{.45\linewidth}
\begin{center}\includegraphics[width=0.8\columnwidth]{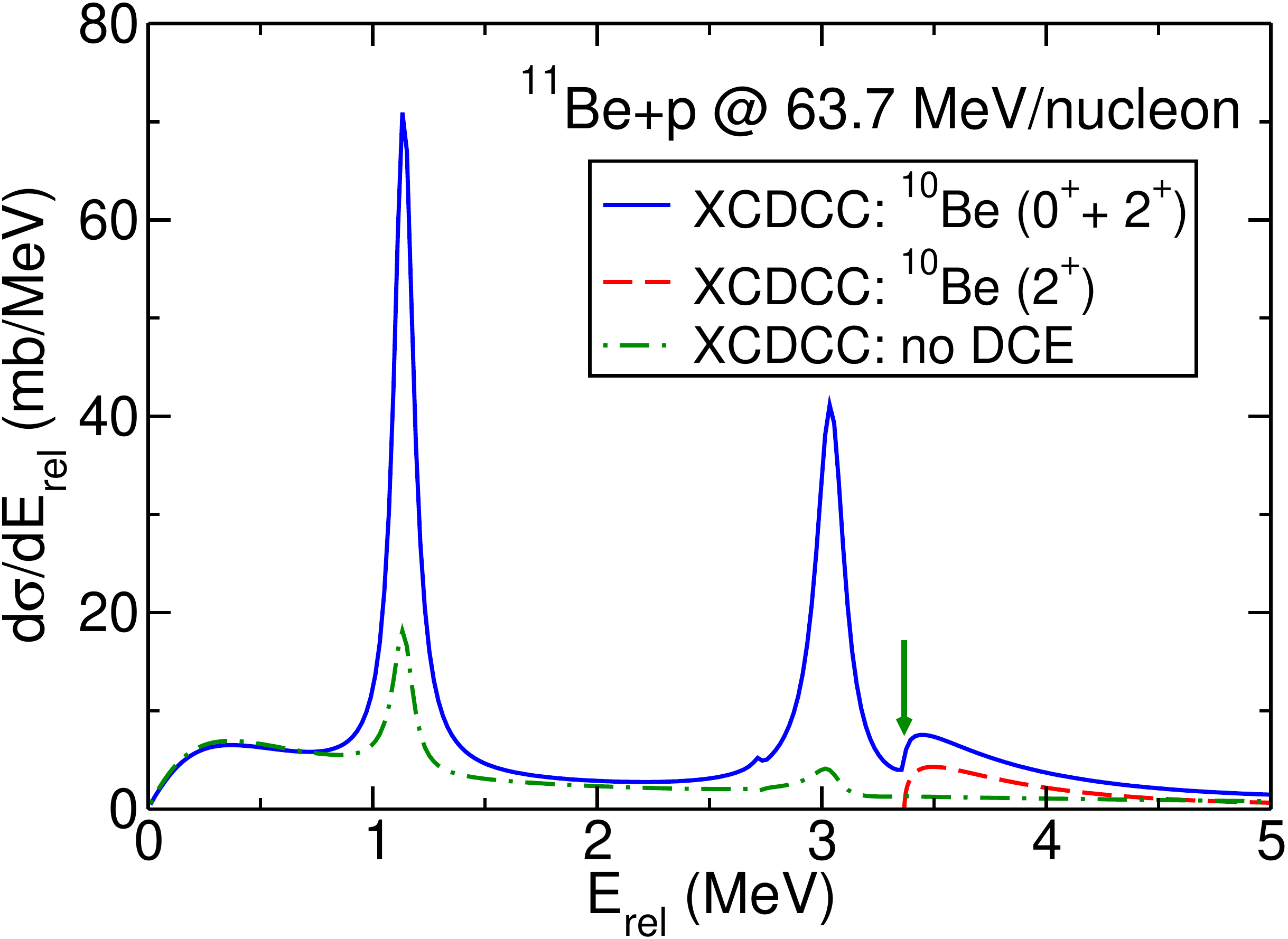} \end{center}
\end{minipage}
\begin{minipage}[t]{.45\linewidth}
\begin{center}\includegraphics[width=0.8\columnwidth]{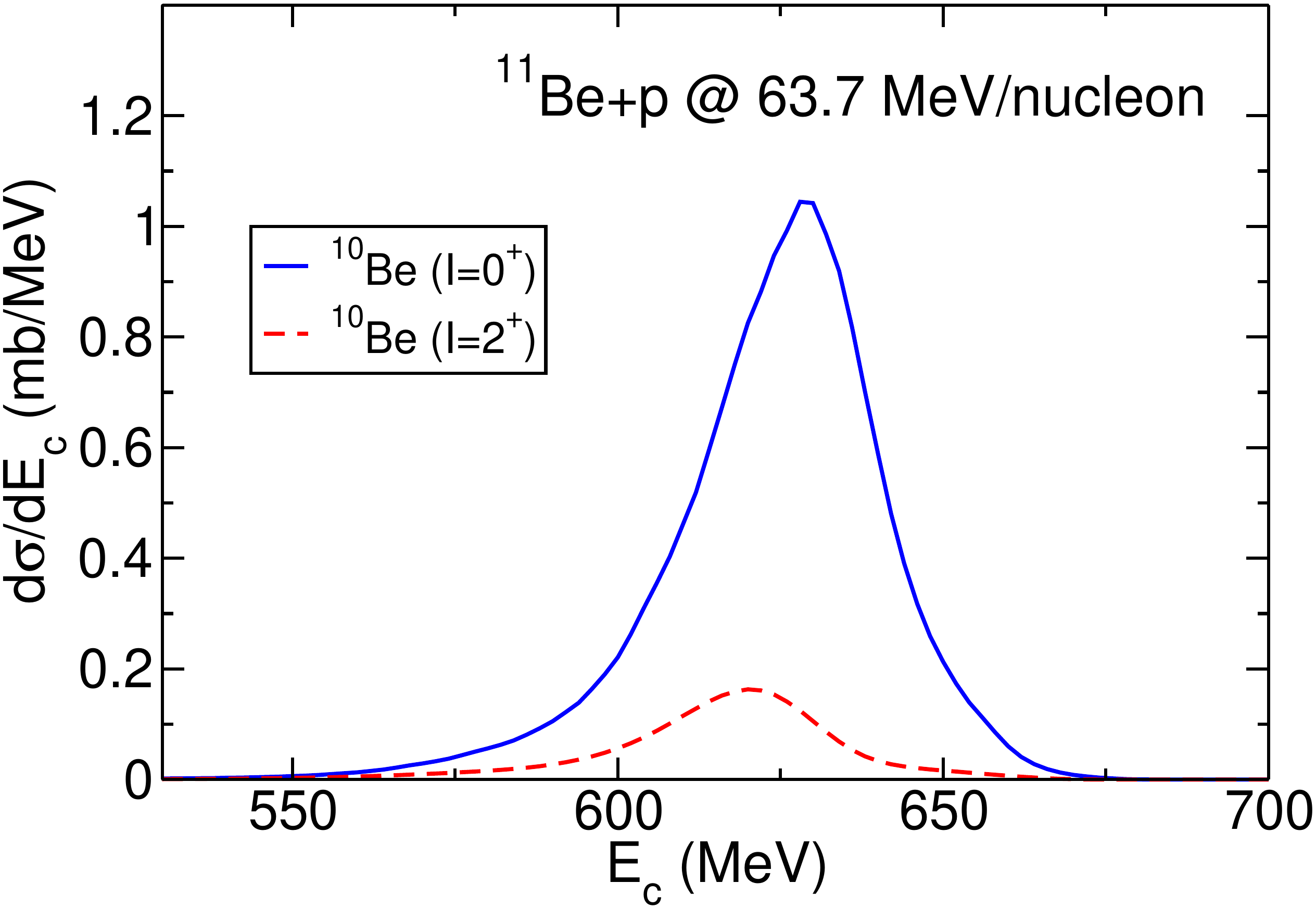}\end{center}
\end{minipage}
\caption{\label{fig:be11p_3b} Differential breakup cross  sections, with respect to the $n$-$^{10}$Be relative energy (a) and with respect to the final $^{10}$Be energy (b),  for the breakup of $^{11}$Be on protons at 63.7~MeV/nucleon. The contributions of the $^{10}$Be ground ($0^+$) and first excited ($2^+$) states are shown in each panel. In the left panel, the result of the calculation neglecting the $^{10}$Be excitation mechanism (labelled ``no DCE'') is also shown. The vertical arrow denotes the threshold for $^{10}$Be($2^+_1$)+n. Figure adapted from Ref.~\cite{Die17}.}
\end{center}
\end{figure}

An application of this formalism is presented in Fig.~\ref{fig:be11p_3b}, corresponding to the breakup of $^{11}$Be on a proton target at $E_p=63.7$~MeV/u. The left panel corresponds to the differential cross section with respect to the final $n$+$^{10}$Be relative energy and the right panel to the breakup with respect to the final $^{10}$Be energy (integrated in the $^{10}$Be and neutron angles). These observables were computed by  transforming the  XCDCC breakup transition amplitudes by means of Eqs.~(\ref{three-obsv-cm}) and (\ref{three-obsv}), and integrating over the unmeasured variables. The XCDCC calculations were performed with a  $^{11}$Be model including the $^{10}$Be ground ($0^+$) and first excited ($2^+$) states. The formalism allows to separate and quantify the contribution of these two states of $^{10}$Be. Note that, due to energy conservation, in the relative energy distribution the contribution coming from the $^{10}$Be($2^+$) state contributes only above the excitation energy of this state ($E_x=3.367$~MeV).

\subsubsection{Target excitations}
In addition to the excitations of the projectile constituents, excitations of the target nucleus may also take place and compete with the projectile breakup mechanism. Note that, within CDCC, the projectile breakup is treated as an inelastic excitation of the projectile to its continuum states and, thus, inclusion of target excitation amounts at including, simultaneously, projectile plus target excitations so their relative importance, and mutual influence, can be assessed. These target excitations can be treated with the collective models mentioned in Sec.~\ref{sec:collect}. It is worth noting that, within this three-body reaction model, target excitation arises naturally from the non-central part of the valence-target and core-target interactions.  To incorporate this effect, the  effective CDCC Hamiltonian must be now generalized as:  
\begin{equation}
\label{eq:Heff_tarx}
H= \hat{T}_{\vecr} + \hat{T}_{\vecR} + V_{bx}(\vecr_{bx}) + U_{bA}(\vecr_{bA},\xi_t) + U_{xA}(\vecr_{xA},\xi_t) + H_A(\xi_t),
\end{equation}
where $H_A(\xi_t)$ is the target internal Hamiltonian, and the $b-A$ and $x-A$ interactions depend now, in addition to the corresponding relative coordinate, on the target degrees of freedom (denoted as $\xi_t$).  
 Ideally, these $U_{xA}$ and $U_{bA}$ potentials should reproduce simultaneously  the elastic and inelastic scattering for the $x+A$ and $b+A$ systems, respectively. Once a specific model has been chosen for the target, its eigenstates  ($\Phi_{n,j_t}(\xi)$) can be computed. They are characterized by their angular momentum $j_t$ and the index $n$.

 In the presence of these target excitations, the CDCC wavefunction is now expanded in the generalized channel basis \cite{Mor12b,Die14,fresco} containing the target eigenstates 
\begin{align}
 \Psi^\mathrm{CDCC}_{c_0,J_T,M_T}(\vec{R},\vec{r},\xi)&=  \sum_{c} \frac{\chi_{c,c_0}^{J_T} (R)}{KR}
 \left\lbrace \left[i^L  Y_L(\hat{R}) \otimes \Phi_{i,j_p}(\vec{r}) \right]_J \otimes \Phi_{n,j_t}(\xi) \right\rbrace_{J_T,M_T},
 \label{eq:CDCC_tarx}
\end{align}
where $c$ denotes all the quantum numbers necessary to define the channel, i.e., $c=\lbrace L,i,j_p,J,j_t,n \rbrace$.

The explicit inclusion of target excitation was first done by the Kyushu group in the 1980s \cite{Yah86}, which considered the case of deuteron scattering. The motivation was to compare the roles of target-excitation and  deuteron breakup  in the elastic and inelastic scattering of deuterons. They applied the formalism to the $d$+$^{58}$Ni reaction at $E_d=22$ and 80~MeV, including the ground state and the first excited state of $^{58}$Ni ($2^+$) and found that, in this case, the deuteron breakup process is more important than the target-excitation.

\begin{figure}[!ht]
{\par\centering \resizebox*{0.55\columnwidth}{!}{\includegraphics{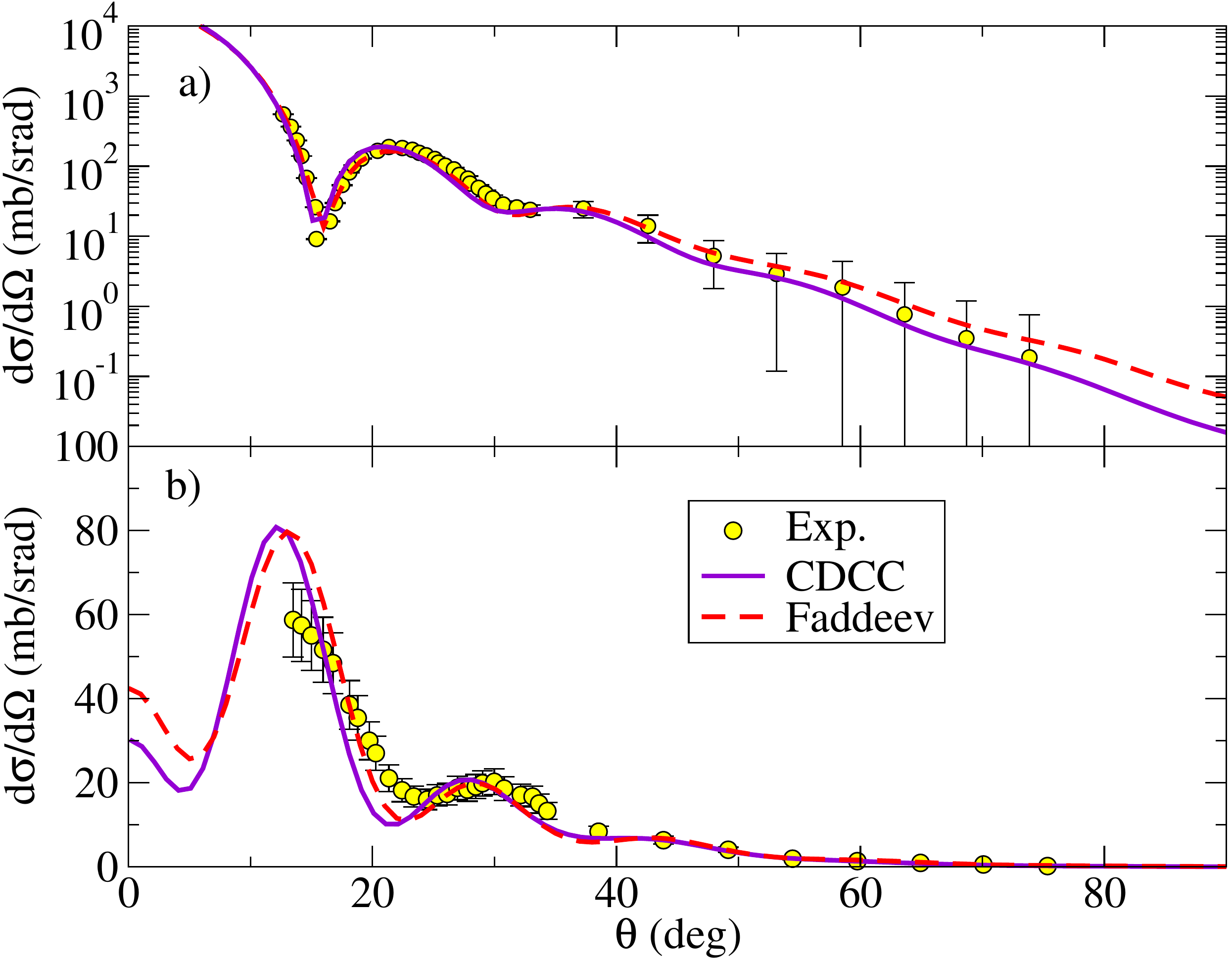}}\par}
 \caption{\label{fig:24Mg}Elastic (upper) and  $^{24}$Mg$(2^+)$ excitation (lower) differential angular cross sections for $d$+$^{24}$Mg $E_d=70$ ${\rm MeV}$. CDCC calculations are compared with Faddeev calculations from Ref.~\cite{Del16} and with the data of ref.~\cite{Kis76}. The calculations  employ CH89 \cite{CH89} parameterization for the $p$+$^{24}$Mg and $n$+$^{24}$Mg potentials and treat the target excitation with a collective model, assuming a deformation of $\beta=0.5$  for the $^{24}$Mg nucleus.  The plot is adapted from Ref.~\cite{Gom17a}.}
\end{figure}

Recently, the problem has been also addressed by some authors \cite{Chau11,Gom17a}. 
A recent application of the formalism is shown in Fig.~\ref{fig:24Mg}, which corresponds to the reaction $^{24}{\rm Mg}(d,d)^{24}{\rm Mg}^*$ at  $E_d=70$~MeV, including the ground and first excited states of $^{24}{\rm Mg}$, in addition to the deuteron breakup. The data are from Ref.~\cite{Kis76}. The target excitation was treated within the collective model, using a quadrupole deformation parameter of $\beta_2=0.5$. Also included are Faddeev calculations reported in Ref.~\cite{Del16}. Both calculations reproduce equally well the elastic differential cross section. The calculated  inelastic angular distributions are slightly out of phase with the data, but they agree well with each other, pointing to some inadequacy of the structure or potential inputs, which are the same in both reaction calculations. Although these inelastic cross section can be also well reproduced with standard DWBA calculations based on a deformed deuteron-target potential, it was shown in Ref. \cite{Del16} that the extracted deformation parameter obtained with the three-body approach is more consistent with that derived from nucleon-nucleus inelastic scattering.

\section{Inclusive breakup}
\label{sec:inclbu}
A common situation in many breakup experiments is that in which  only one of the fragments resulting from the projectile dissociation is detected experimentally. This is the case of reactions induced by neutron-halo nuclei in which, very often, only the heavy-charged fragment is detected. Considering for simplicity the case of a two-body projectile ($a=b+x$), and assuming that only the fragment $b$ is detected, the process can be schematically represented as  $A(a,b)X$, where $X$ denotes any possible final state of the $x+A$ system.

The evaluation of inclusive breakup reactions poses a challenging theoretical problem because many processes can in principle contribute to the $b$ singles cross section. When the two fragments $b$ and $x$ ``survive'' and the target remains in its ground state, the process is referred to as {\it elastic breakup} (denoted EBU hereafter), also called {\it diffraction dissociation}. 
Several theoretical methods have been developed for the evaluation of EBU cross sections for two- and three-body projectiles, including: the Faddeev formalism \cite{Fad60,deltuva2007,Del09b},  the distorted-wave Born  approximation (DWBA) \cite{Bau83}, the continuum-discretized coupled-channels  (CDCC) method  \cite{austern1987} and a variety of semiclassical approaches  \cite{bonaccorso1988,Typ94,Esb96,Kid94,Cap04}. 

In addition to EBU, the inclusive cross section will contain also nonelastic breakup (NEB) contributions in which the unobserved fragment $x$ interacts nonelastically with the target nucleus, such as target excitation, or $x$ capture by the target. If $x$ is a composite system, it also includes those processes in which this fragment is excited or even broken up. These processes are in fact effectively accounted for by the imaginary parts of the fragment-target potentials used in the CDCC effective Hamiltonian. 
After solving the CDCC equations, this gives rise to an {\it absorption cross section} which include, at least in an integrated way, these NEB contributions. 
 
In some applications, more detailed, differential cross sections for the NEB contribution are needed. An example is the so-called surrogate method  \cite{Esc12}, an indirect method for the evaluation of neutron-induced cross sections, whose application requires the knowledge of the incomplete fusion cross sections as a function of the excitation energy of the residual nucleus. This method will be briefly addressed in Sec.~\ref{sec:surrogate}. 

The problem of the calculation of detailed inclusive breakup cross sections attracted the attention of several groups in the 1980s, who tried to derive simple expressions to evaluate the cross section associated to these NEB processes \cite{Pam78,Bau80,Shy80,Uda81,Uda84,AV1981,Ich85}.  These models are based on a spectator-participant picture. The $b$ fragment, which  acts as spectator, is assumed to scatter elastically by the target nucleus and hence its relative motion with the target is represented by some optical potential $U_{bA}$. The fragment $x$, the participant, is allowed to interact in any possible way with the target. When $x$ scatters elastically by the target, we simply have the elastic breakup defined above. Conversely, NEB encompasses those processes in which $x$ scatters nonelastically by the target. The key idea of these models is the possibility of obtaining a closed-form expression for the sum over all these NEB processes. 
Interestingly, all these formulae  exhibit a common structure, given by
\begin{equation}
\label{eq:neb}
\left . \frac{d^2\sigma}{dE_b d\Omega_b} \right |_\mathrm{NEB} = -\frac{2}{\hbar v_{a}} \rho_b(E_b)
 \langle  \varphi_x  | W_x |  \varphi_x  \rangle   ,
\end{equation}
where  $\rho_b(E_b)=k_b \mu_{b} /[(2\pi)^3\hbar^2]$ is the density of states (with $\mu_b$ the reduced mass of $b+B$ and $k_b$ their relative wave number),  $W_x$ is the imaginary part of the optical potential $U_x$, which describes $x+A$ elastic scattering.
Expression (\ref{eq:neb}) can be regarded as a three-body version of the usual optical model theorem appearing in two-body scattering.
According to this expression, the NEB is the expectation value of the imaginary part of the $x-A$ system, evaluated with some suitable $x-A$ wavefunction $\varphi_x$ (the $x$-channel wavefunction hereafter). In two-body scattering, $\varphi_x$  is a scattering solution of the Schr\"odinger equation with the full optical potential $U_x$. In the three-body case considered here,  $\varphi_x$ should represent the $x-A$ relative motion compatible with the incoming boundary conditions, and subject to the constrain that fragment $b$ will be finally detected with a given momentum $\vec k_b$.

Although the pioneering inclusive breakup models agreed in the general form (\ref{eq:neb}) they differ in their prescription for the $x$-channel wavefunction. We focus here in the DWBA model of Ichimura, Austern and Vincent (IAV) \cite{IAV1985} and its three-body extension \cite{austern1987}. A detailed discussion of other NEB models has been given by Ichimura \cite{Ich90} as well as in more recent reviews \cite{Pot17,Gom21}.

\subsection{The Ichimura, Austern, Vincent (IAV) model} \label{sec:iav}
\begin{figure}[!ht]
\begin{center}
 {\centering \resizebox*{0.35\columnwidth}{!}{\includegraphics{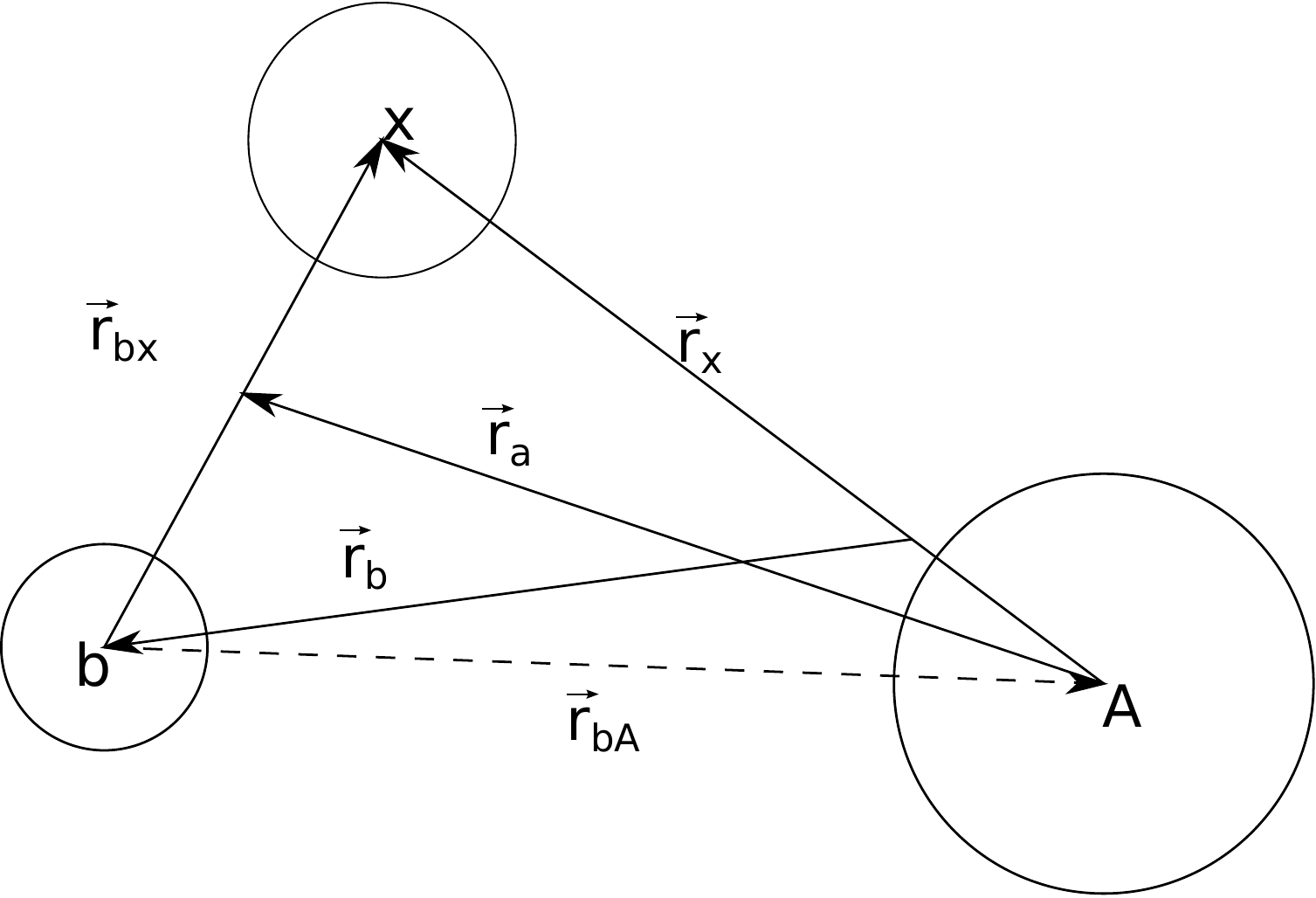}} \par}
\caption{\label{zrcoor} Relevant coordinates for the description of the $A(a,b)X$ reaction.}
\end{center}
\end{figure}

Ichimura, Austern and Vincent \cite{IAV1985} derived a DWBA formula for the inclusive breakup cross section based on the model Hamiltonian
 \begin{equation}
\label{eq:H3b}
H= T + V_{bx}(\vecr_{bx}) + U_{bA}(\vecr_{bA}) + H_A(\xi) + V_{xA}(\xi,\vecr_{x}) ,
\end{equation}
where $T$ is the total kinetic energy operator, $V_{bx}$ is the 
interaction binding the two clusters $b$ and $x$ in the projectile $a$,
 $H_{A}(\xi)$ is the Hamiltonian of the target 
nucleus (with $\xi$ denoting its internal coordinates) and $V_{xA}$ and 
$U_{bA}$ are the fragment--target interactions. The relevant coordinates 
are depicted in Fig.~\ref{zrcoor}. Note that the coordinate 
$\vecr_{b}$  connects the particle $b$ with the center of mass (c.m.) of 
the $x+A$ system. 
 
Using the post-form of the exact transition amplitude, the inclusive breakup cross section is given by
\begin{align}
\label{eq:tpost}
\frac{d^2\sigma}{d\Omega_b E_b }&  = \frac{2 \pi}{\hbar v_a} \rho(E_b) \sum_{c} |\langle \chi^{(-)}_{b} \Psi^{c,(-)}_{xA} |V_\mathrm{post}| \Psi^{(+)}  \rangle |^2  
  \delta(E-E_b-E^c) ,
\end{align}
where $V_\mathrm{post} \equiv V_{bx} + U_{bA}-U_{bB}$ is the post-form transition operator,  $\Psi^{(+)}$ is the system wavefunction with the incident wave in the $a+A$ channel,  and $\Psi^{c,(-)}_{xA}$ are the eigenstates of the $x+A$ system, with $c=0$ denoting the $x$ and $A$ ground states. Thus, for $c=0$ this expression gives the EBU part, whereas the terms $c \neq 0$ correspond to  NEB.   

In their original derivation, IAV worked out this expression to obtain a closed-form formula for the total inclusive breakup cross section, and then manipulated it to separate the elastic and nonelastic breakup contributions. We adopt here a different approach, first presented in Ref.~\cite{Jin15}. 

Within the assumed three-body model, and using the post-form representation, the total wave function of the 
system can be written in integral form as
\begin{align}
\Psi(\xi,\vecr_x,\vecr_{b}) & = \left[E^+ - T_{bB} - U_{bB}-H_{B} \right]^{-1}  V_\mathrm{post} \Psi(\xi,\vecr_x,\vecr_{b}) ,
\label{eq:Psi3b}
\end{align}
where $E^+=E+i\epsilon$, $\epsilon \rightarrow 0$, 
and $H_{B}$ is the Hamiltonian of the $x$+$A$ pair, given by
\begin{equation}
H_B(\xi,\vecr_x)= H_{A}(\xi) + T_{x} + V_{xA}(\xi,\vecr_{x}) .
\end{equation}

We consider now a specific final state of the detected particle $b$, 
characterized by a given final momentum of this fragment ($\vec{k}_b$). The motion of $b$ will be described by a distorted wave
with momentum ${\vec k}_b$, obtained as a solution of the single-channel equation
\begin{equation}
\left[  T_{bB}+ U_{bB}^\dagger-E_b\right]\chi_b^{(-)}(\vec{k}_b,\vecr_{b})=0 ,
\end{equation}
where $E_b$ is the kinetic energy associated with the $b+B$ relative motion.
 The wave function describing the motion of $x$ after the breakup, that  will be denoted as $Z_x(\xi,\vecr_x)$, can be 
obtained projecting the total wave function [Eq.~(\ref{eq:Psi3b})] onto this particular state of the $b$ particle, i.e.,
\begin{align}
 \zbx & \equiv \langle \vecr_x \chi_b^{(-)} | \Psi \rangle   =  \left[E^+ - E_b-H_{B} \right]^{-1} \langle \vecr_x\chi_b^{(-)}| V_\mathrm{post}|\Psi\rangle ,
\label{eq:zx_int}
\end{align}
which can be also written in differential form as
\begin{equation}
\label{eq:zeq}
\left[E^+ -E_b-H_{B} \right] \zbx=  \langle \vecr_x \chi_b^{(-)} | V_\mathrm{post}|\Psi\rangle .
\end{equation}
The source term of this equation involves the exact and hence unknown 
wave function $\Psi$ which, in actual calculations, must be approximated  by some calculable form. IAV adopted the  DWBA  factorized form
\begin{equation}
\label{eq:psi_dwba}
\Psi(\xi,\vecr_x,\vecr_{b}) \approx \phi^0_A(\xi) \phi_a(\vec{r}_{bx}) 
\chi^{(+)}_{a}(\vec{k}_a,\vec{r}_a) ,
\end{equation}
where $\phi_a(\vec{r}_{bx})$ is the projectile 
ground-state wave function and $\chi^{(+)}_{a}(\vec{k}_a,\vec{r}_a)$ is a distorted 
wave describing the $a+A$ motion in the incident channel. In practice,
the latter is commonly replaced by the solution of some optical potential describing  $a+A$ elastic scattering. 

 Using the  Feshbach projection formalism, the $\zbx$ function is decomposed as
\begin{equation}
\zbx = {\cal P} Z_x + {\cal Q} Z_x ,
\end{equation} 
where ${\cal P}$ is the projector operator onto the target ground state and 
${\cal Q}= 1 - {\cal P}$. Consequently, we can write ${\cal P} Z_x = \psix \phi^{0}_A(\xi)$, where the function $\psix$ is the seeked $x$-channel wavefunction that describes the $x+A$ relative 
motion when the target is in the ground state. It verifies the equation
\begin{equation}
\label{eq:inh_dwba}
(E^+_x - T_x - {\cal U}_x)  \psix = \langle \vecr_x \chi_b^{(-)}| V_{bx}|\Psi \rangle
\end{equation}
with $E_x=E-E_b$, 
and  ${\cal U}_x$ the formal optical model potential describing $x+A$ elastic scattering, i.e.,
\begin{equation}
{\cal U}_x =\langle \phi^0_A |V_{xA} + V_{xA} Q [E^+ - E_b - H_{QQ}]^{-1}  V_{xA}| \phi^0_A \rangle  ,
\end{equation}
where $H_{QQ} \equiv {\cal Q} H_{B} {\cal Q}$.  As usual, the formal potential ${\cal U}_x$ can be approximated by a simpler, energy-averaged  and possibly local potential with parameters adjusted to describe $a+A$ elastic scattering (denoted $U_x$ hereafter).   

To obtain the expression for the NEB component one makes use 
of the {\it coupled-channels optical theorem}, as formulated by 
Cotanch \cite{Cot10}, which can be regarded as a generalization of the usual optical theorem  to the multichannel case. 
If $\chi_i$ is the channel wave function and 
$W_i$ the diagonal imaginary part for this channel, the contribution to 
the absorption in this particular channel is given by \cite{Cot10}
\begin{equation}
\sigma^i_\mathrm{abs}=-\frac{2}{\hbar v_{el}} \langle \chi_i | W_i | \chi_i \rangle ,
\end{equation}
where $v_{el}$ is the projectile--target relative velocity in the 
incident (elastic) channel. 

We may use this result to calculate the NEB contribution by noting that 
the latter is nothing but the absorption occurring in the $x+A$ 
channel. The channel wave function is given by $\psix$, which is a 
solution of Eq.~(\ref{eq:inh_dwba}). Since this equation corresponds to a definite energy and direction of the $b$ particle, we consider the 
differential cross section corresponding to a range of the outgoing  
momenta of $b$,
\begin{equation}
d^2\sigma = -\frac{2}{\hbar v_{i}} \langle \varphi_x | W_x | \varphi_x \rangle   \, \frac{d\vec{k}_b}{2 \pi^3} .
\end{equation}
Transforming the element of momentum into 
energy and solid angle elements, we get the double differential cross section
\begin{equation}
\label{eq:iav}
\left . \frac{d^2\sigma}{dE_b d\Omega_b} \right |^\mathrm{IAV}_\mathrm{NEB} = -\frac{2}{\hbar v_{i}} \rho_b(E_b)  \langle \varphi_x | W_x | \varphi_x \rangle   .
\end{equation}
which, in combination with  with Eq.~(\ref{eq:inh_dwba}), provides the differential NEB cross section with respect to the angle and energy of the detected fragment $b$.

\subsection{The three-body model of Austern {\it et al.}}
Austern {\it et al.} \cite{austern1987}  proposed also the three-body approximation 
\begin{equation}
\label{eq:psi_3b}
\Psi(\xi,\vecr_x,\vecr_{b}) \approx \phi^0_A(\xi) \Psi^{\mathrm{3b}}(\vecr_x,\vecr_{b}) ,
\end{equation}
where $\Psi^{\mathrm{3b}}$ is  the solution of the three-body equation:
\begin{equation}
 [ \hat{T}_{aA} + \hat{T}_{bx}+ V_{bx} + U_{bA} + U_{xA} - E  ]  |\PsiTB \rangle =0 .
\end{equation}

Within this three-body approximation, the $x$-channel wavefunction is {\it simply} given by
\be
\label{eq:chiPsi}
\psixaus (\vecr_x) = \langle \vecr_x \chi_b^{(-)}  |  \PsiTB \rangle .
\ee
In spite of its apparent simplicity, this equation has the difficulty of requiring an accurate representation of the three-body wavefunction $\PsiTB$ in the full configuration space, since there is no natural cutoff in the integration variable $\vecr_b$. 

Alternatively, the three-body function $\psixaus (\vecr_x)$ can be obtained from an inhomogeneous equation similar to (\ref{eq:inh_dwba}), 
\begin{equation}
\label{eq:inh_3b}
(E^+_x - K_x - {\cal U}_x)  \psixaus = \langle \vecr_x \chi_b^{(-)}| V_\mathrm{post}| \Psi^{\mathrm{3b}^{(+)}} \rangle.
\end{equation}
Although this equation is a priori more difficult to implement than Eq.~(\ref{eq:chiPsi}), it has the advantage that the $\PsiTB$ function appears multiplied by $V_\mathrm{post}$, that will tend to emphasize small $b-x$ separations and hence one requires only an approximate three-body wavefunction accurate within that range.   This can be achieved, for instance, expanding $\PsiTB$  in terms of $b-x$ eigenstates, as done in the CDCC method, or in terms of Weinberg states \cite{Pan13}. The implementation of the method with CDCC wavefunctions is numerically challenging and the first calculation of this kind was reported only recently \cite{Jin19b}. 

\begin{figure}[!ht]
\begin{center}\includegraphics[width=0.78\columnwidth]{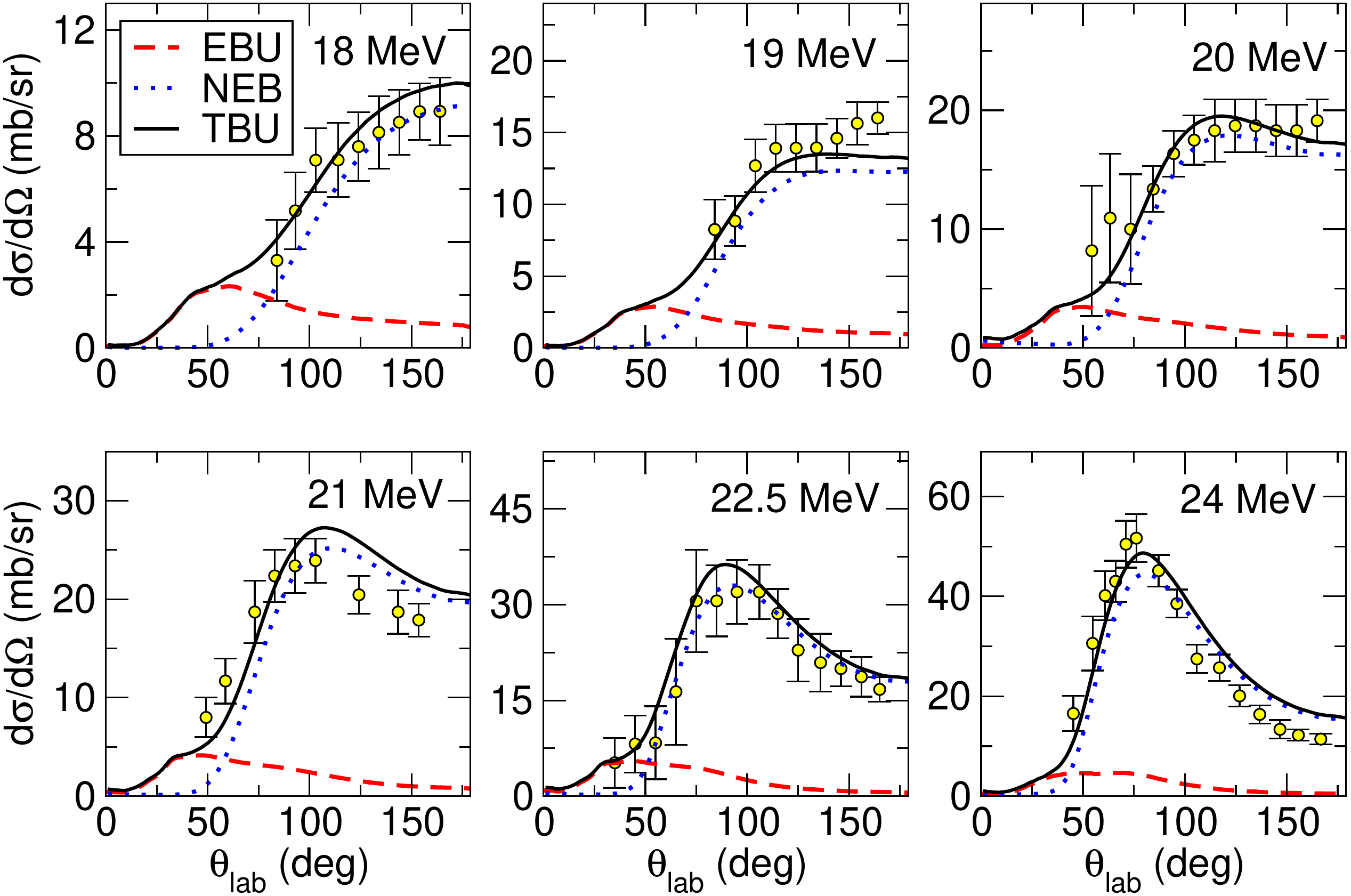} \end{center}
\caption{Angular distribution of $\alpha$ particles produced in the reaction $^{118}$Sn($^{6}$Li,$\alpha$)$X$ at the incident energies indicated by the labels. The dashed and dotted lines are the EBU and NEB contributions, and the solid line is their incoherent sum (TBU).}  
\label{fig:li6sn118} 
\end{figure}

The  IAV model has been recently revisited by several groups \cite{Car16,Jin15,Pot15} and its accuracy assessed  against experimental data with rather encouraging results \cite{Jin17,Pot17}.    As an example, in Fig.~\ref{fig:li6sn118} we show the application to the evaluation of the $\alpha$ production in the  $^6$Li+$^{118}$Sn reaction at the incident energies indicated by the labels (adapted from Ref.~\cite{Jin17}).  The EBU contribution (dashed line) was evaluated with the CDCC method whereas the NEB part (dotted line) was obtained with the IAV method.  Interestingly, one can see that the inclusive $\alpha$-yield is largely dominated by the NEB mechanism. The EBU is only important at small scattering angles (distant collisions, in a classical picture).  This dominance of the NEB over the EBU mechanism has been observed in other $^{6,7}$Li-induced reactions on a variety of energies and targets. The situation is qualitatively different for more weakly bound nuclei, such as the neutron halo nuclei $^{11}$Be, $^{11}$Li and the proton-halo  $^{8}$B. In all these cases, inclusive breakup cross sections corresponding to the production of the heavier projectile fragment (the {\it core}) shows a dominance of the EBU mechanism. 

\begin{figure}[!ht]
\begin{center}\includegraphics[width=0.35\columnwidth]{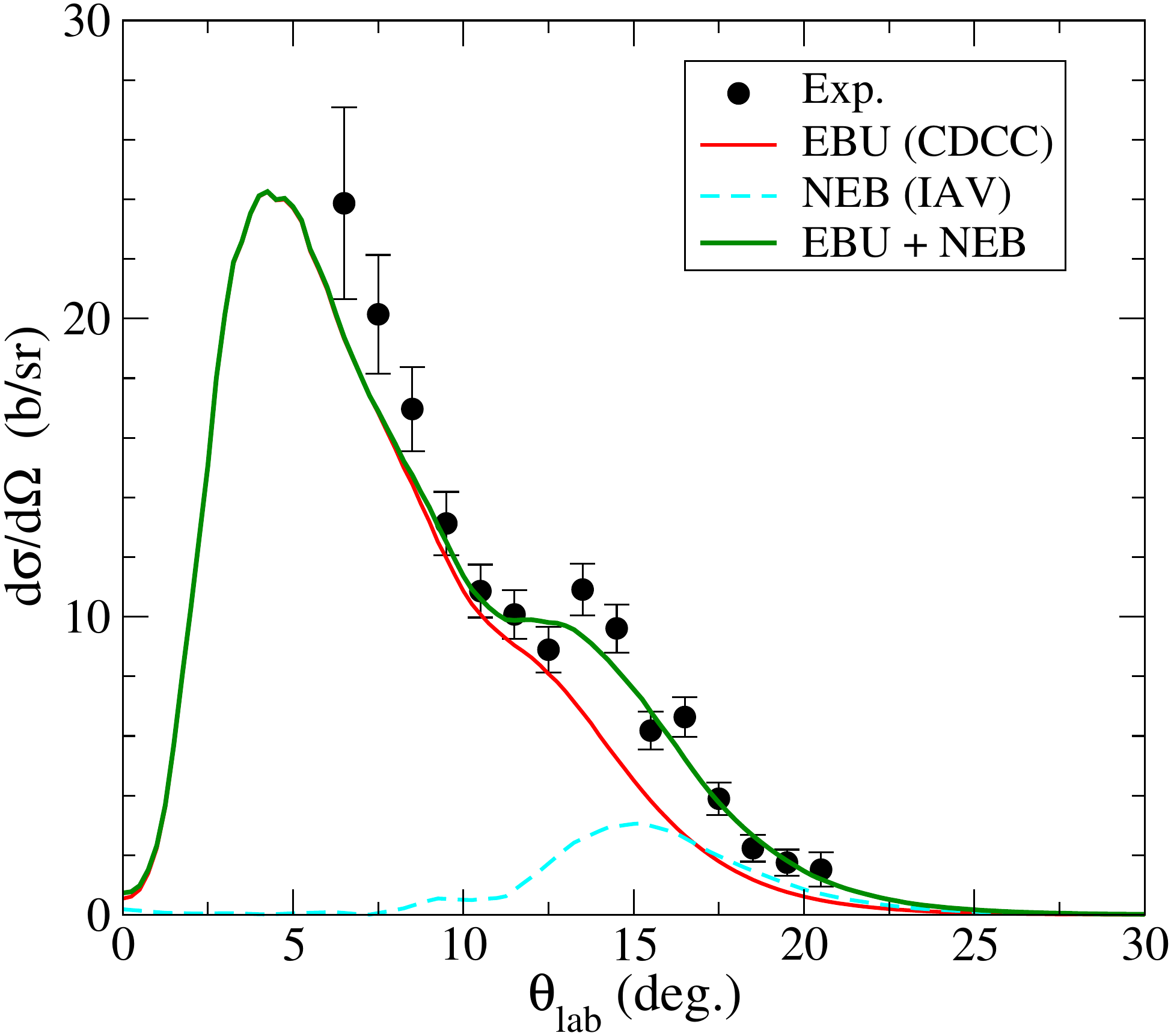} \quad
\includegraphics[width=0.45\columnwidth]{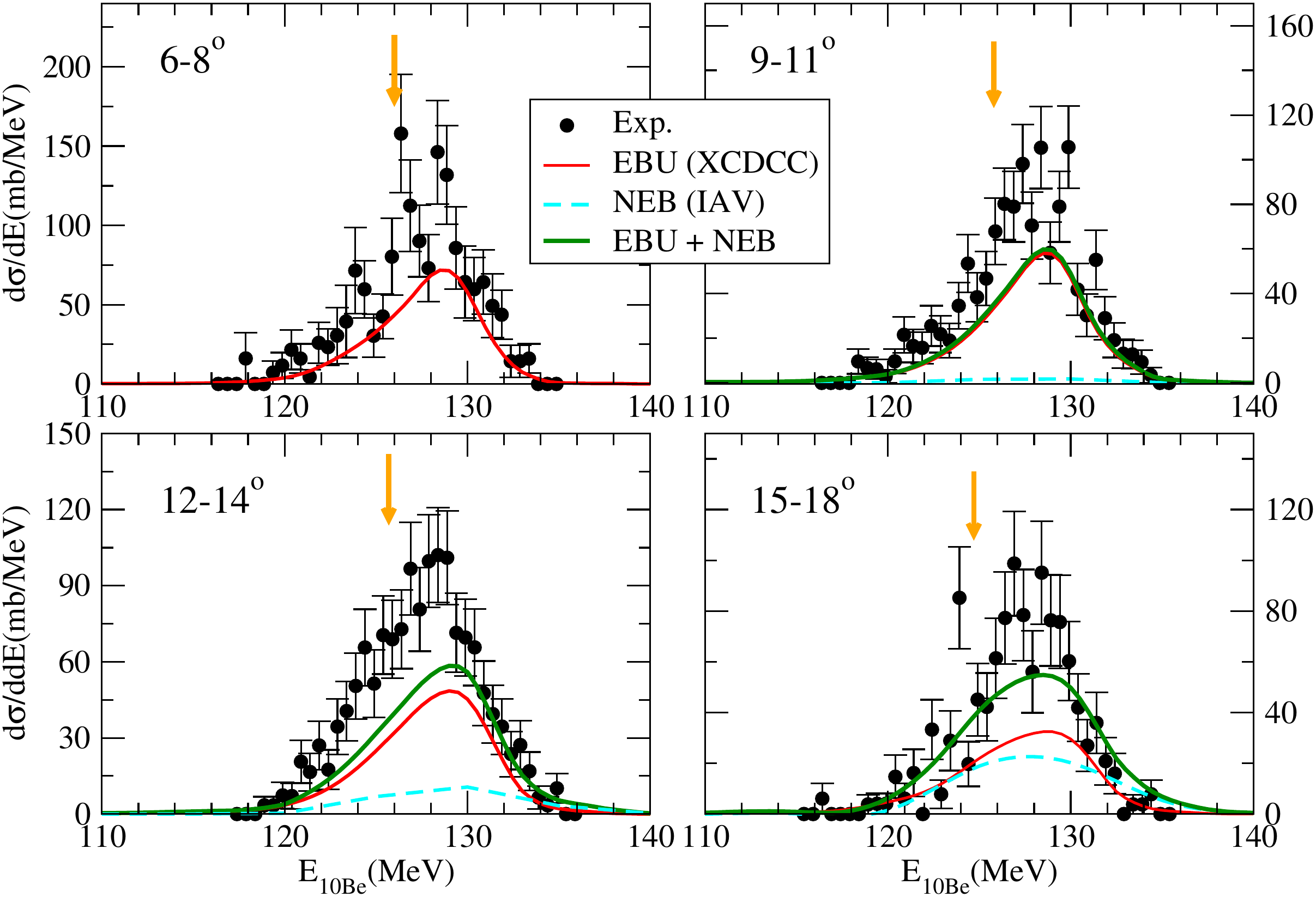}\end{center}
\caption{\label{fig:be11pb_e140} Angular (a) and energy(b) distribution of $^{10}$Be fragments produced in the reaction  $^{208}$Pb($^{11}$Be,$^{10}$Be)$X$ at the incident energy of 140 MeV. Experimental data are compared with the calculations for the EBU and NEB contributions, and their sum. Figure adapted from Ref.~\cite{Dua20}. } 
\end{figure}

An example is shown in Fig.~\ref{fig:be11pb_e140}, which shows the angular and energy distributions of $^{10}$Be fragments measured in the collision $^{11}$Be+$^{208}$Pb at 140~MeV \cite{Dua20}. The EBU and NEB contributions were calculated with the CDCC and IAV methods, respectively. It is seen that the EBU mechanism dominates for most angles and the NEB is only non-negligible in an angular region around the grazing angle. This dominance of the EBU in the case of halo nuclei can be interpreted as a consequence of the fact that, due to its large probability of being at large distances from the core, the halo particle(s) can be more easily removed from the projectile without distorting or exciting the core or the target, as compared to the case of a well bound nucleus. The orange arrow in the energy distributions correspond to $(10/11)E(\mathrm{^{11}Be})$, where $E(\mathrm{^{11}Be})$ is the expected $^{11}$Be kinetic energy calculated assuming a binary-like collision. The shift with the observed position of the energy peak is due to a Coulomb post-acceleration effect, as explained in Ref. \cite{Dua20}.

\section{Heavy-ion fusion reactions}
\subsection{Light systems and nuclear astrophysics}

In this section, we discuss fusion reactions. 
Fusion of light nuclei, such as $p+^7$Be and $\alpha+^{12}$C reactions, at deep subbarrier energies 
play an important role in nucleosynthesis \cite{bertulani2004,thompson2009}. For these reactions, even if the Coulomb 
barrier is overcome, 
it is unlikely that there exists a level of a compound nucleus at the same energy 
as the incident energy. Therefore, in order for fusion to take place, a di-nucleus system has to emit gamma rays 
to populate levels in the compound nucleus. In such radiative capture process, fusion reaction thus depends both on the 
tunnel dynamics of the Coulomb barrier and on the dynamics of electromagnetic transitions. This is the case also 
for the triple-alpha process to form the $^{12}$C nucleus \cite{ogata2009,akahori2015,LWT86,DB87,GDFJ2011,NNTB2012,ishikawa2013,NNT2013,SSD2016}. 
Since the operators of the electromagnetic transitions are long-ranged, fusion reactions may take place even outside 
the Coulomb barrier. Moreover, a few specific resonance states often play a decisive role in fusion of light nuclei. 

As the atomic number of a compound nucleus increases, the level density of the compound nucleus 
becomes large, and the compound nucleus may be formed even without emitting gamma-rays. 
We here discuss particularly 
the $^{12}$C+$^{12}$C fusion reaction, which plays an important role in several
astrophysical phenomena, such as the carbon burning in stellar evolution,
type Ia supernovae, and the X-ray superburst of an accreting neutron star.
It has been known that 
the fusion cross sections for the 
$^{12}$C+$^{12}$C reaction show many fine structures \cite{Spillane2007,Tumino2018,BMT2019,Tan2020,Fruet2020}. 
This is in contrast to the 
cross sections of the neighboring
systems, $^{12}$C+$^{13}$C and $^{13}$C+$^{13}$C,
which are more smooth functions of energy. 
The difference in the fusion cross sections for these three systems may be 
understood in terms of the level density of the compound nuclei, 
$^{24}$Mg, $^{25}$Mg and $^{26}$Mg\cite{Jiang13}.
An interesting observation is that 
the fusion cross sections for the $^{12}$C+$^{12}$C system match with the cross sections for 
the $^{12}$C+$^{13}$C and $^{13}$C+$^{13}$C systems at a few resonance energies, while 
the cross sections for the  $^{12}$C+$^{12}$C system are hindered as compared to the cross sections 
for the other systems at off-resonance energies \cite{Notani12,ZWD2020}. 

\begin{figure}[!tb]
\begin{center}
\includegraphics*[width=0.5\textwidth]{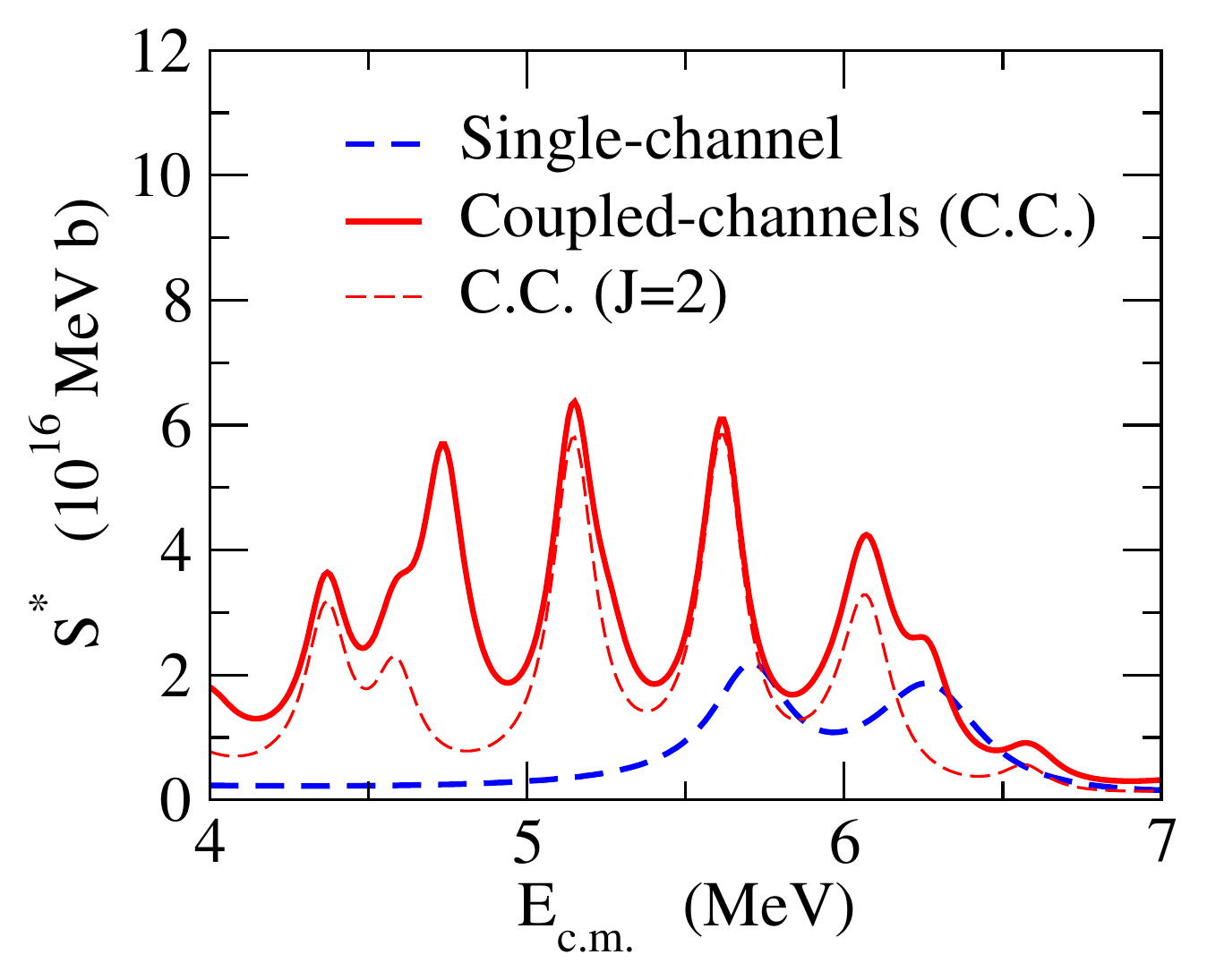}
\caption{
The modified astrophysical $S$-factor for the $^{12}$C+$^{12}$C reaction. 
The solid line is obtained by solving the coupled-channels equations with excitations to the first 
2$^+$ state in both the projectile and the target nuclei, and their mutual excitations. The contribution 
of $J=2$ is denoted by the thin dashed line, where $J$ denotes the spin of the compound nucleus. 
The dashed line shows the result of the single-channel calculation. 
The parameters for the coupled-channels calculation are given in Ref.~\cite{kondo1978}. 
}
\label{fig:12c12c}
\end{center}
\end{figure}

It has been known that the excitation of $^{12}$C to the first 2$^+$ state plays an important 
role in the $^{12}$C+$^{12}$C fusion reaction at subbarrier energies \cite{kondo1978,imanishi1968,imanishi1969,SGL1970,scheid1972}. 
Fig.~\ref{fig:12c12c} shows the result of the coupled-channels calculation which corresponds to that in Ref.~\cite{kondo1978}. 
Here, $S^*(E)$ is the modified astrophysical $S$-factor, defined as 
$S^*(E)=\sigma(E)E\,\exp(87.21/\sqrt{E}+0.46E)$,
where $\sigma(E)$ is the fusion cross section and the energy $E$ is measured in units
of MeV. The coupling to the first 2$^+$ state both in the projectile and in the target, as well as their mutual excitations, 
are taken into account. The solid and dashed lines show the result of the coupled-channels and the single-channel 
calculations, respectively. For the coupled-channels calculation, the contribution of $J=2$, $J$ being the spin of the 
compound nucleus, is also shown by the thin dashed line. 
One can  clearly see that the number of resonance peaks dramatically increases by 
taking account the channel coupling effects. This effect can be explained in terms of the double 
resonance mechanism \cite{SGL1970,scheid1972}. That is,  
the number of resonance peak increases when both the entrance channel (at the energy $E$ 
and the angular momentum $J$) and an excited channel 
(at the energy $E-\epsilon_2$ and the angular momentum $J-2$) have resonance states. 
This is, in a sense, similar to the Feshbach resonance, which has been intensively discussed 
in physics of cold atoms \cite{CGJT2010,PN2017}. 

See also Refs. \cite{ETJ2011,AD2013,taniguchi2021} for 
other recent coupled-channels calculations for this system. 

\subsection{Medium-heavy systems}

\subsubsection{Subbarrier enhancement of fusion cross sections}

\begin{figure}[!tb]
\begin{center}
\includegraphics*[width=0.9\textwidth]{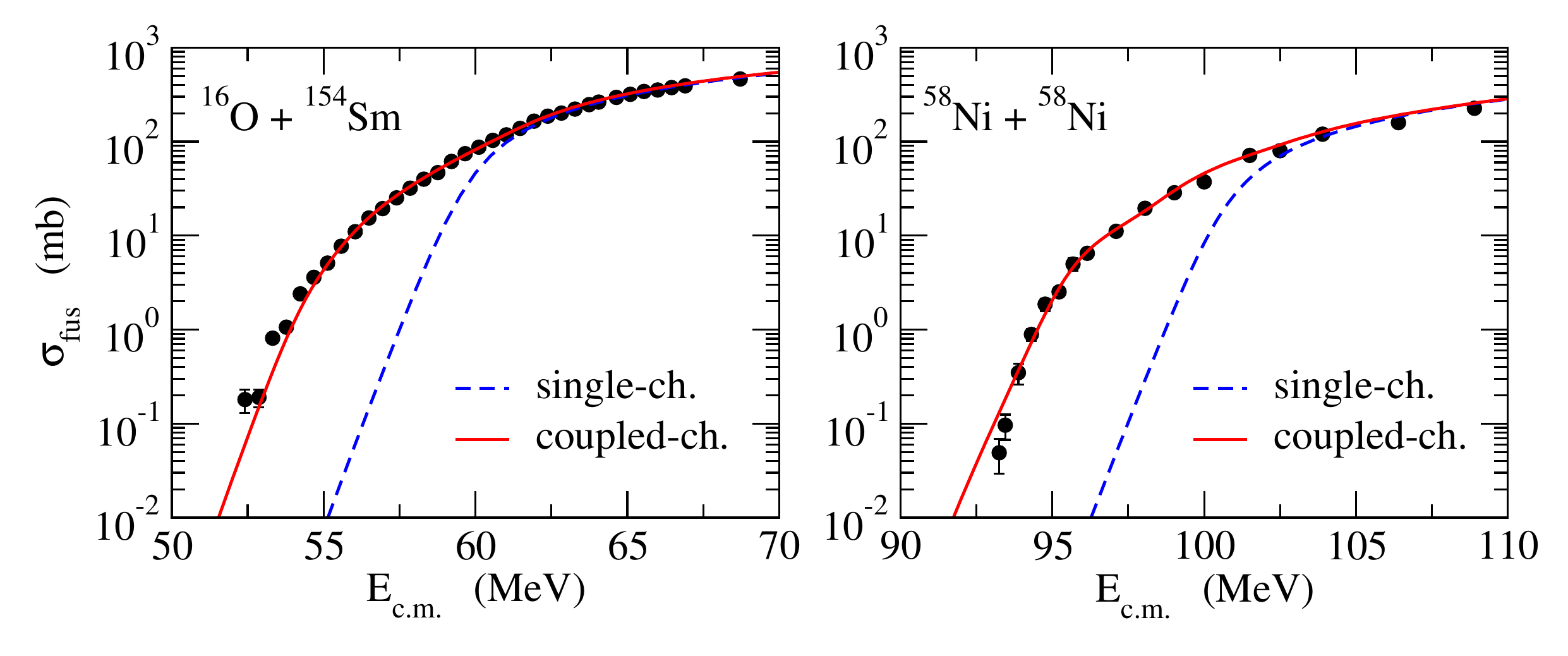}
\caption{Fusion cross sections for the $^{16}$O+$^{154}$Sm (the left panel) and the $^{58}$Ni+$^{58}$Ni (the 
right panel) systems. The dashed lines show the results of single-channel calculations, while the solid lines take 
into account the channel coupling effects. For the  $^{16}$O+$^{154}$Sm system, the excitation of 
the ground state rotational band in $^{154}$Sm is included. On the other hand, for the  $^{58}$Ni+$^{58}$Ni system, 
the quadrupole excitations up to the double phonon states are taken into account. The experimental data are 
taken from Refs. \cite{Leigh1995,beckerman1981}.
}
\label{fig:fus}
\end{center}
\end{figure}

In fusion of medium-heavy systems, the absorption inside the Coulomb barrier is strong and it is a good 
assumption that the compound nucleus is automatically formed once two nuclei touch with each other. 
In this circumstance, resonance structure in fusion cross sections cease to appear and the fusion dynamics 
is entirely governed by the penetration of the Coulomb barrier \cite{hagino2012}. We refer to Refs.~\cite{hagino2012,beckerman1985,beckerman1988,BT98,DHRS98,Back14,MS17} for previous reviews on heavy-ion fusion reactions 
in medium-heavy systems. 

A characteristic feature of fusion reactions in medium-heavy systems at energies around the Coulomb barrier 
is that fusion cross sections are significantly enhanced by the channel coupling effects. 
Figure \ref{fig:fus} shows the results of the coupled-channels calculations for the 
$^{16}$O+$^{154}$Sm (the left panel) \cite{hagino2012} and for the $^{58}$Ni+$^{58}$Ni (the 
right panel) \cite{HRD03,HY2015} as typical examples. For the $^{16}$O+$^{154}$Sm systems, the excitations of the 
ground state rotational band in $^{154}$Sm are taken into account while $^{16}$O is assumed to be inert. On the other 
hand,  for the $^{58}$Ni+$^{58}$Ni system, the excitations of the one and two quadrupole phonon states in both 
of the $^{58}$Ni nuclei 
are taken into account. In both systems, it is clearly seen that the fusion cross sections are 
significantly enhanced by several orders of magnitude 
at energies below the Coulomb barrier as compared to the 
prediction of a single-channel calculation (the dashed lines). The coupled-channels calculations well account for 
the enhancement of the fusion cross sections. 
 
\begin{figure}[!tb]
\begin{center}
\includegraphics*[width=0.5\textwidth]{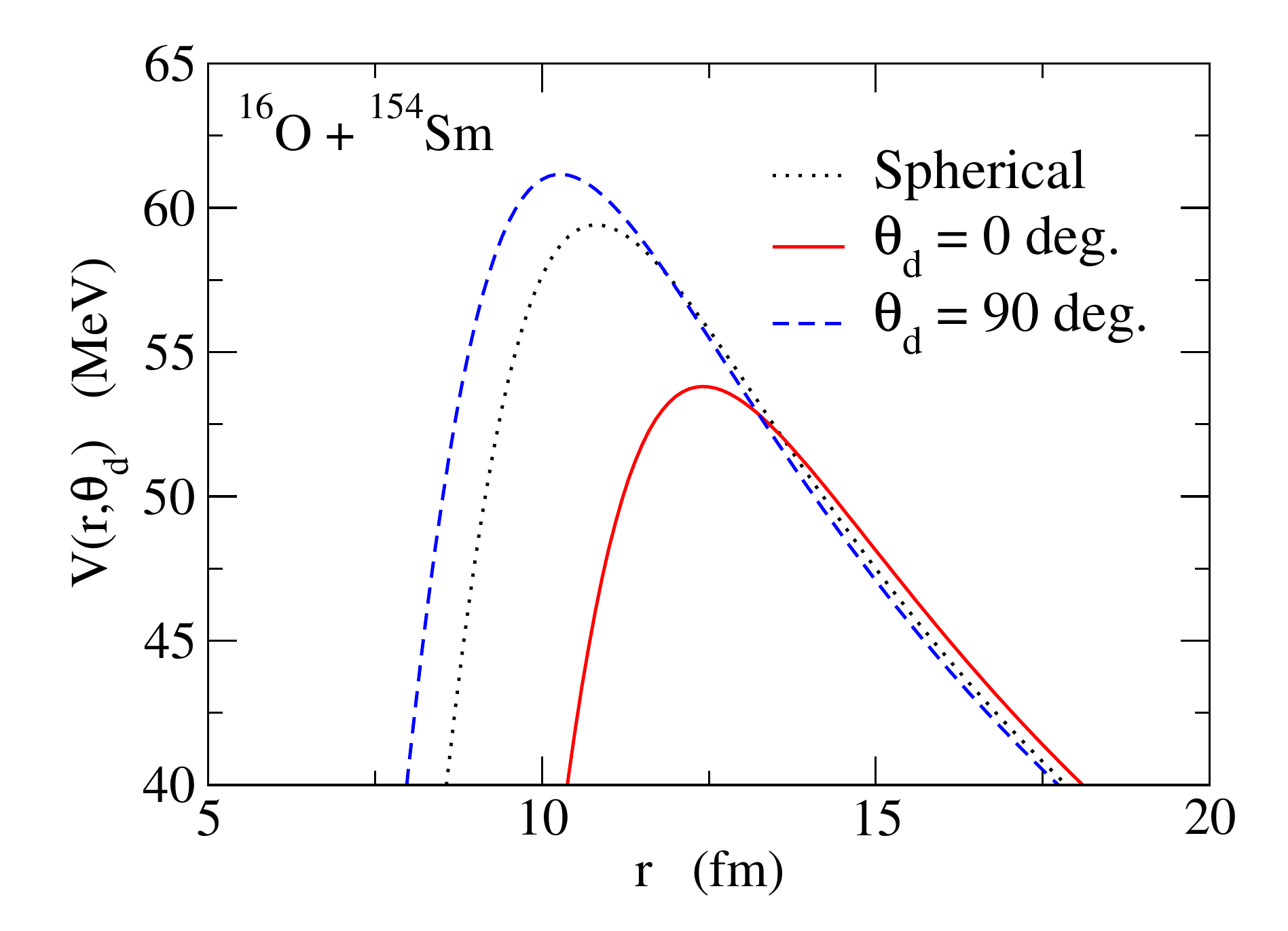}
\caption{
The dependence of the internucleus potential 
for the $^{16}$O+$^{154}$Sm system on the orientation angle $\theta_d$ of the 
deformed $^{154}$Sm target. The solid and the dashed lines show the potentials 
for  $\theta_d$= 0 and 90 degrees, respectively, for the quadrupole deformation 
parameter of $\beta_2=0.306$ and the hexadecapole deformation parameter 
of $\beta_4=0.05$. The dotted line shows the potential in the spherical limit.  
}
\label{fig:defpot}
\end{center}
\end{figure}

The subbarrier enhancement of fusion cross sections can be easily understood 
for the $^{16}$O+$^{154}$Sm system. Notice that the energy of the first 2$^+$ state, 
$\epsilon_2=89$ keV in the deformed $^{154}$Sm nucleus, 
is much smaller than the typical energy scale of fusion reactions 
characterized by the curvature of the Coulomb barrier \cite{hagino2012}, 
which is typically around 3.5 MeV. Neglecting the excitation energies of the 
ground state rotational band in $^{154}$Sm, the fusion cross sections can be 
calculated with Eq.~(\ref{eq:fus-orientation}) 
as a weighted average of contributions of different orientation angles. 
The solid and the dashed lines in Fig.~\ref{fig:defpot} show the internucleus potential 
for $\theta_d=0$ and $\pi/2$, respectively, with the deformation parameters 
of  $\beta_2=0.306$ and $\beta_4=0.05$. For prolate nuclei with $\theta_d=0$, 
the projectile nucleus approaches the 
target nucleus from the longer axis of the target, and thus 
the Coulomb barrier is lower than the potential in the spherical limit (see the dotted 
line), since the nuclear attraction acts from larger distances For $\theta_d=\pi/2$, 
the opposite happens and the Coulomb barrier is higher than that for the spherical case. 
The tunneling probability, and thus the fusion cross section, increases 
for $\theta_d=0$ and decreases for $\theta_d=\pi/2$. 
However, since the tunneling probability has an exponential dependence on the 
incident energy, the enhancement for  $\theta_d\sim 0$ leads to the main contribution 
to the fusion cross section, leading to the subbarrier enhancement of fusion cross 
sections. 
This also explains why the single-channel and the coupled-channels calculations 
lead to similar fusion cross sections at high energies, since the tunneling probability 
becomes close to unity irrespective of the deformation effect. 
  
For the $^{58}$Ni + $^{58}$Ni system, the interpretation is more complicated, since 
the geometrical interpretation cannot be applied as in the $^{16}$O+$^{154}$Sm 
system. However, 
as we have demonstrated in Fig.~\ref{fig:cc2ch}, the potential is dynamically 
altered by the channel coupling effects, which in general results in 
an enhancement of the penetrability at energies below the barrier. 
The subbarrier enhancement of fusion cross sections for this system 
reflects this effect. 

\subsubsection{Barrier distributions}

Equation (\ref{eq:fus-orientation}) for fusion cross sections with a deformed target 
can be symbolically expressed as a weighted sum of the contribution of eigen-channels 
$\alpha$ as  
\begin{equation}
\sigma_{\rm fus}(E)=\sum_\alpha w_\alpha\sigma_{\rm fus}^{(0)}(E; V_\alpha(R)), 
\end{equation}
where $w_\alpha$ are the weight factors and $\sigma_{\rm fus}^{(0)}(E; V_\alpha(R))$ 
is the single-channel fusion cross section with the potential $V_\alpha(R)$. 
For Eq.~(\ref{eq:fus-orientation}), the eigen-channel $\alpha$ corresponds to 
the orientation angle $\theta$ 
with the weight factor $w_\theta=2\pi\sin\theta$. Within the isocentrifugal 
approximation, this formula is exact when the 
excitation energy of the intrinsic state is zero, but the same formula holds 
to a good approximation even when the excitation energy is 
finite \cite{hagino2012,HTB1997}. In the latter case, the eigen-channel $\alpha$ can 
be defined as the eigenstates of $S^\dagger S$ \cite{hagino2012}, for which the 
potential $V_\alpha$ is obtained by diagonalizing 
$H_0(\xi)+V_{\rm coup}(\vec{R},\xi)$ in eq.~(\ref{eq:H3d}) at each $r$. 

\begin{figure}[!tb]
\begin{center}
\includegraphics*[width=0.5\textwidth]{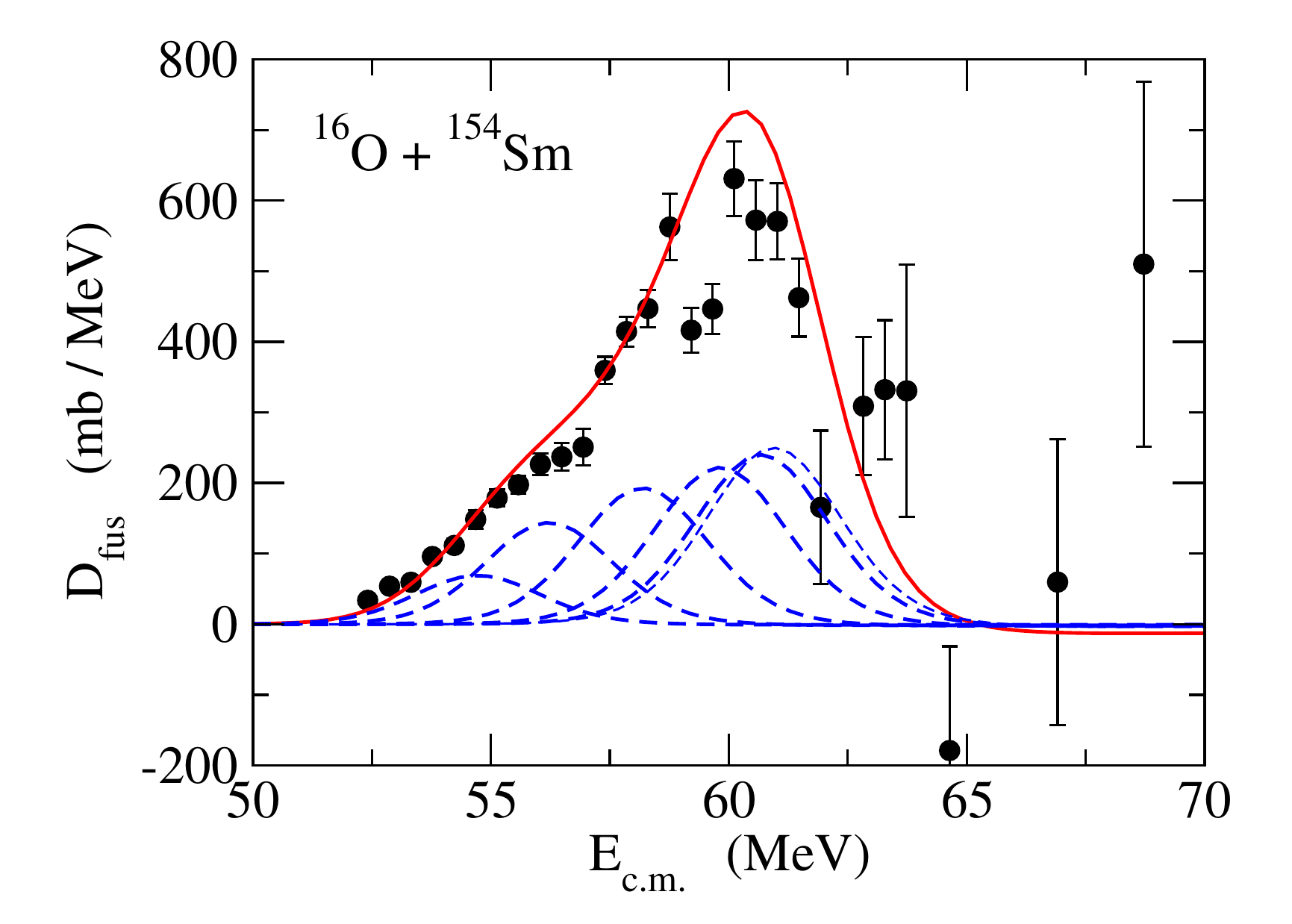}
\caption{
The fusion barrier distribution for the 
$^{16}$O+$^{154}$Sm system. 
The dashed lines show the contributions of different orientation angles 
of the deformed $^{154}$Sm, while the solid line denotes the sum of all the 
contributions. The experimental data are taken from Ref.~\cite{Leigh1995}.
}
\label{fig:154sm-bd}
\end{center}
\end{figure}

For a single-channel problem, one can derive a compact expression for fusion 
cross sections by approximating the Coulomb barrier by an inverse parabola 
\cite{hagino2012,BT98,Wong1973,RH2015,CGLH2020,TCH2017}. 
The resultant Wong formula \cite{Wong1973} reads
\begin{equation}
\sigma_{\rm fus}^{(0)}(E; V_\alpha(R))
=\frac{\hbar\Omega_\alpha}{2E}\,R_\alpha^2\,\ln\left[1+\exp\left(
\frac{2\pi}{\hbar\Omega_\alpha}(E-B_\alpha)\right)\right],
\end{equation}
where $B_\alpha$, $R_\alpha$, and $\hbar\Omega_\alpha$ are the height, the position, and 
the curvature of the Coulomb barrier for the potential $V_\alpha(R)$, respectively. 
Notice that the first energy derivative of $E\sigma_{\rm fus}^{(0)}$ from this 
formula is proportional to the $s$-wave penetrability of the Coulomb barrier 
with a parameter set of $(B,R_b,\hbar\Omega)$, 
\begin{equation}
\frac{d}{dE}[E\sigma_{\rm fus}^{(0)}(E)]
=\pi R_b^2\,P(E),
\end{equation}
with 
\begin{equation}
P(E)=\frac{1}{1+\exp\left[-\frac{2\pi}{\hbar\Omega}(E-B)\right]}.
\end{equation}
Since the first derivative of the penetrability $P(E)$ is a Gaussian-like function peaked at
$E=B$,  the second energy derivative of $E\sigma_{\rm fus}(E)$ shows in general multiple of 
peaks at $E=B_\alpha$ due to the channel coupling effects. 
This function thus provides a distribution of the barrier heights, 
and is referred to as the fusion barrier distribution $D_{\rm fus}$ \cite{DHRS98,RSS1991}, 
\begin{equation}
D_{\rm fus}(E)=\frac{d^2(E\sigma_{\rm fus}(E))}{dE^2}.
\end{equation}
The fusion barrier distribution has been extracted experimentally 
for several systems \cite{DHRS98,Leigh1995}. These barrier distribution have shown 
that the 
fusion barrier distribution is in general sensitive to the details of the channel 
couplings and thus 
provides a powerful tool to visualize the underlying dynamics of subbarrier fusion 
reactions. Fig.~\ref{fig:154sm-bd} shows the barrier distribution for the 
$^{16}$O+$^{154}$Sm system as an example. The figure also shows the barrier 
distribution obtained with the coupled-channels calculation (the solid line), 
for which the contribution 
of each orientation angle is shown by the dashed lines. 
One can see that the barrier distribution is structured, and the coupled-channels 
calculation well accounts for the shape of the barrier distribution. By analyzing the 
barrier distribution, one can infer the energies of 
the barrier height for the side ($\theta_d=\pi/2$) 
and for the tip ($\theta_d=0$) collisions. This has been utilized in fusion for 
superheavy elements, which we shall discuss in Sec.~5.4. 
We also mention that the Bayesian analysis may  be used to analyze fusion 
barrier distributions as well \cite{H2016}. 

A similar barrier distribution can be extracted using an excitation function of 
quasi-elastic scattering at backward angles. Here, the quasi-elastic barrier 
distribution is defined as the first energy derivative of the Rutherford ratio 
of quasi-elastic scattering \cite{Timmers1995}, 
\begin{equation}
D_{\rm qel}(E)=-\frac{d}{dE}\left(\frac{d\sigma_{\rm qel}(E)}{d\sigma_R(E)}\right), 
\end{equation}
where $d\sigma_{\rm qel}/d\Omega$ and  $d\sigma_R/d\Omega$ 
are quasi-elastic and the Rutherford cross sections, respectively. 
The quasi-elastic barrier distribution is defined at backward angles, corresponding 
to low partial waves in the semi-classical approximation, so that the quasi-elastic 
barrier distribution has a good correspondence to the fusion barrier distribution. 
A mapping onto the barrier distribution for the $s$-wave scattering can be achieved 
by introducing the effective energy defined by \cite{Timmers1995,HR2004}
\begin{equation}
E_{\rm eff}=2E\frac{\sin(\theta/2)}{1+\sin(\theta/2)}, 
\end{equation}
that is derived by subtracting from the total energy $E$ the centrifugal energy for the classical trajectory 
for  the scattering angle $\theta$. 
The quasi-elastic barrier distribution so obtained corresponds to the first energy 
derivative of the reflection probability of the Coulomb barrier, and is thus complementary 
to the fusion barrier distribution. In fact, it has been shown that the shape 
of the quasi-barrier distribution is similar to that of the fusion barrier distribution, 
even though the former is slightly less sensitive to the details of the couplings 
because of a low-energy tail in the distribution originated from the tail of the nuclear potential \cite{HR2004}. 

\begin{figure}[!tb]
\begin{center}
\includegraphics*[width=0.4\textwidth]{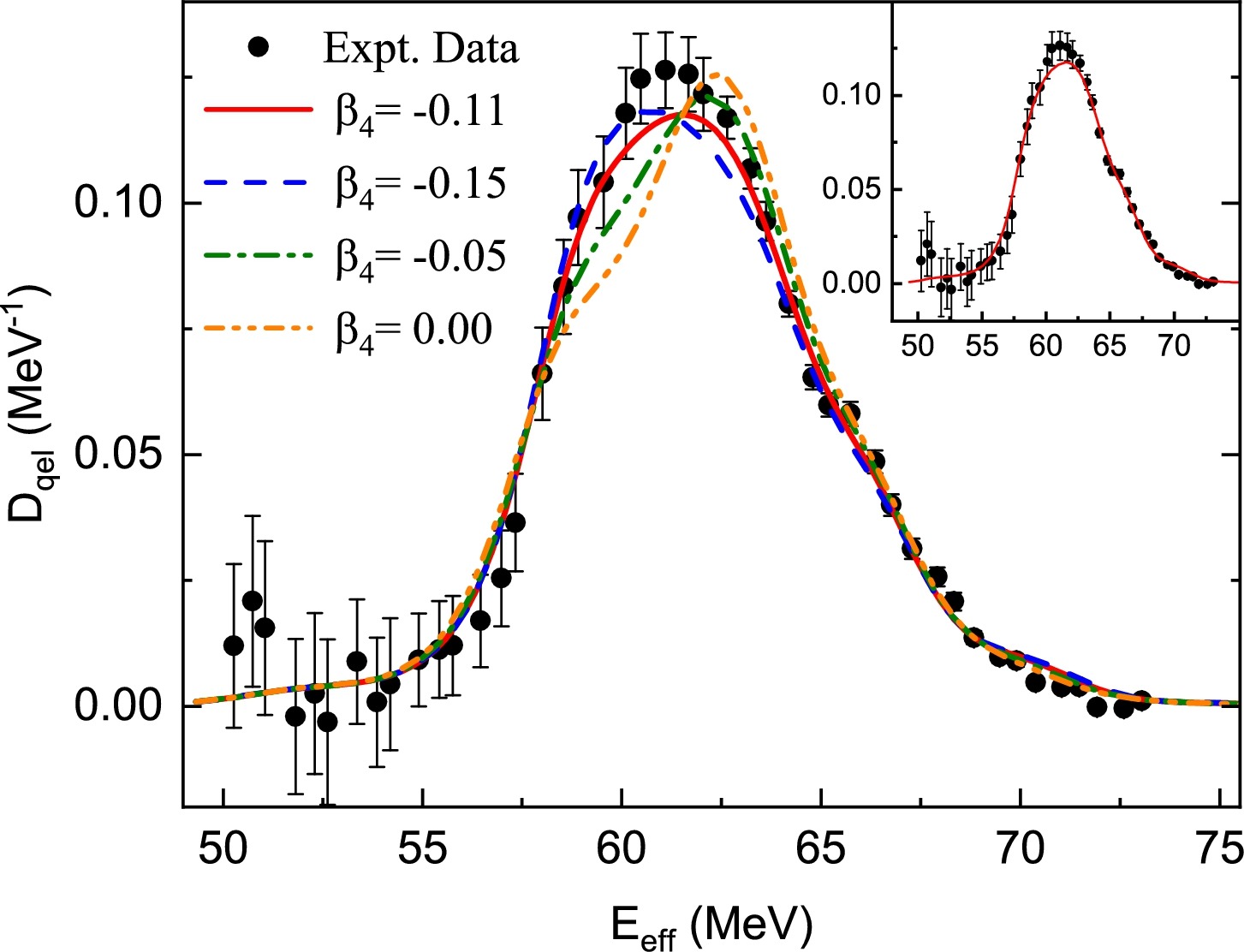}
\includegraphics*[width=0.4\textwidth]{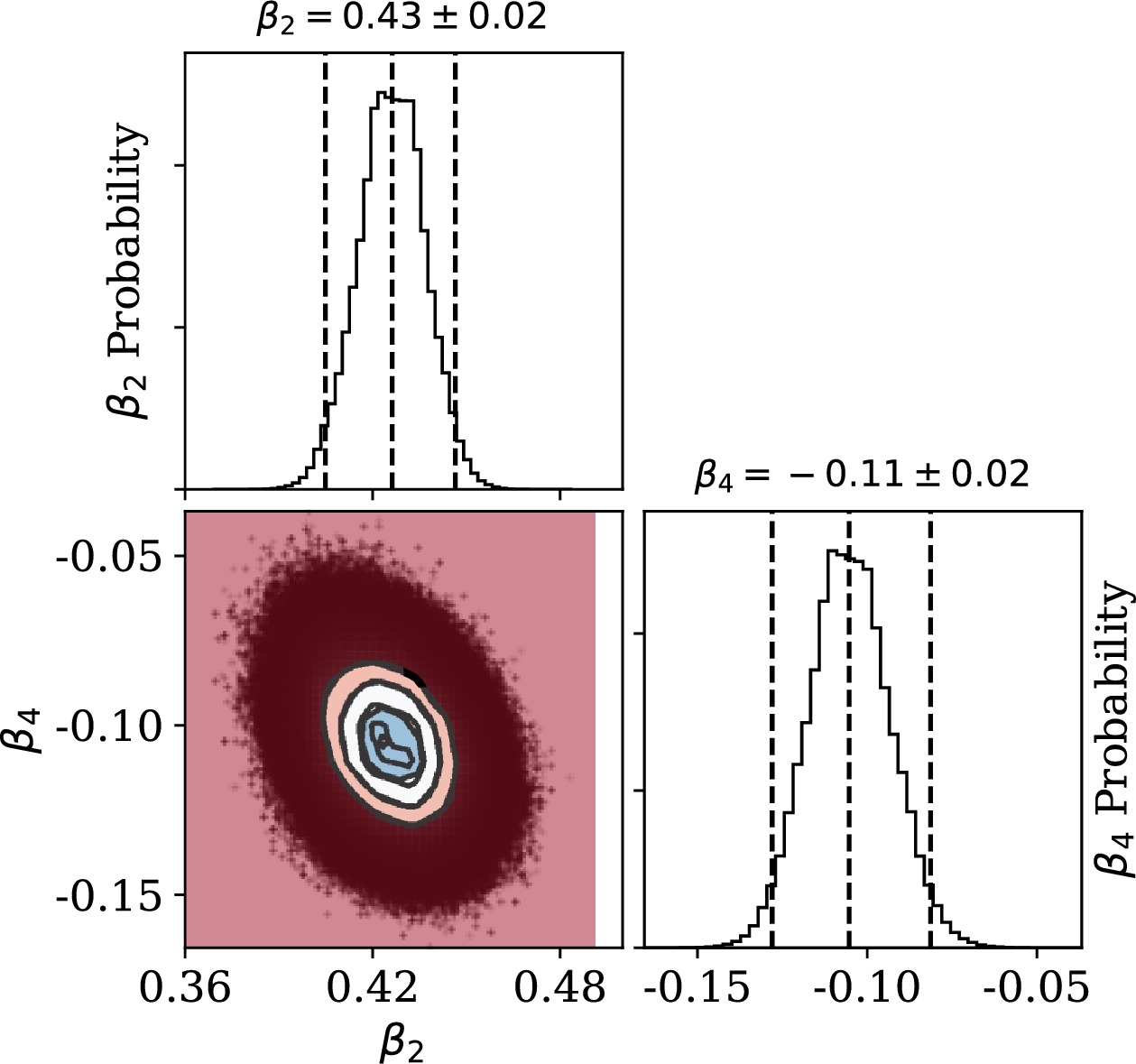}
\caption{The quasi-elastic barrier distribution for the $^{24}$Mg+$^{90}$Zr 
system and its comparison to the coupled-channels calculations with several 
values of the deformation parameters of the $^{24}$Mg nucleus. 
The right panel shows the posterior probability distribution from the Bayesian 
analysis. Taken from Ref. \cite{Gupta2020}. 
}
\label{fig:24mg-bd}
\end{center}
\end{figure}

Recently, the experimental 
quasi-elastic barrier distribution for the $^{24}$Mg+$^{90}$Zr system 
has been analyzed to extract the quadrupole and the hexadecapole deformation 
parameters of the 
$^{24}$Mg nucleus \cite{Gupta2020}. 
To this end, the authors of Ref.  \cite{Gupta2020} employed the Bayesian statistics 
and determined the values and the sign of the deformation parameters with the 
Markov-chain Monte-Carlo method. 
This is shown in Fig. \ref{fig:24mg-bd}. From this analysis, the hexadecapole 
deformation parameter of $^{24}$Mg, $\beta_4=-0.11\pm 0.02$, 
has been determined precisely for the first 
time. 

Incidentally,
the shape of the measured quasi-elastic barrier distribution has been shown 
significantly different between the $^{20}$Ne+$^{90}$Zr and  $^{20}$Ne+$^{92}$Zr 
systems \cite{Piasecki2009}.  While the barrier distribution for the $^{20}$Ne+$^{90}$Zr 
can be well reproduced by taking into account the rotational excitations of  $^{20}$Ne 
and the vibrational excitations of $^{90}$Zr, similar coupled-channels calculations 
do not reproduce the data for the $^{20}$Ne+$^{92}$Zr system, despite the fact that the 
couplings are well dominated by the rotational excitations of the 
strongly deformed $^{20}$Ne nucleus. See also Refs. \cite{Piasecki2012,Piasecki2012b,Piasecki2015,Piasecki2020}. 
The difference has been attributed to non-collective 
excitations in the target nuclei \cite{YHR2010,YHR2012,YHR2013,Piasecki2019}, 
originated from the two 
extra neutrons in $^{92}$Zr. 
 
\subsubsection{Deep subbarrier fusion hindrance}

At deep subbarrier energies, 
it has been systematically observed that 
experimental fusion cross sections fall off much steeper  
as compared to 
theoretical fusion cross sections based on the coupled-channels 
calculations \cite{Jiang2002,Back14,jiang2021}. 
Two theoretical models have been proposed to interpret this 
phenomenon, based either on the sudden approximation \cite{ME2006,ME2007} 
or on the adiabatic approximation \cite{IHI2007,IHI2009,Ichikawa2015}. 
In both models, the deep subbarrier fusion hindrance is attributed to the 
dynamics after two colliding nuclei 
touch with each other \cite{IHI2007b}. 
In the sudden model, one considers a shallow potential, for which the Pauli principle leads to a repulsive 
contribution at short distances \cite{SUGDH2017}. Fusion cross sections are then hindered at deep subbarrier 
energies as the contribution of 
high partial waves is suppressed. 
In contrast to the sudden model, the adiabatic model assumes that the internuclear potential is smoothly 
connected to a mono-nucleus potential, which is often described by the liquid drop model. 
After two nuclei touch with each other, it is not reasonable to expand the total wave function with the 
eigenstates of isolated nuclei, or at least it is not practical to do so as one would have to include a huge number 
of states. To take this into account, 
the adiabatic model assumes that 
the coupling is effectively damped after the touching \cite{IHI2007,IHI2009,Ichikawa2015}, 
which has been justified microscopically 
by the 
random phase approximation (RPA) with two-center shell model  wave functions \cite{IM2013,IM2015}.  
In the adiabatic model, the deep 
subbarrier hindrance is originated from a quenching of an enhancement of fusion cross sections 
due to the channel coupling effects. 

\begin{figure}[!tb]
\begin{center}
\includegraphics*[width=0.45\textwidth]{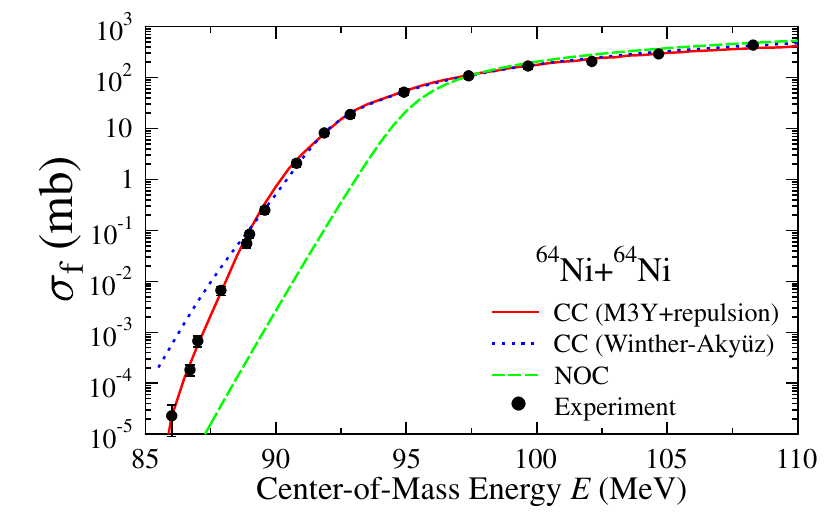}
\includegraphics*[width=0.45\textwidth]{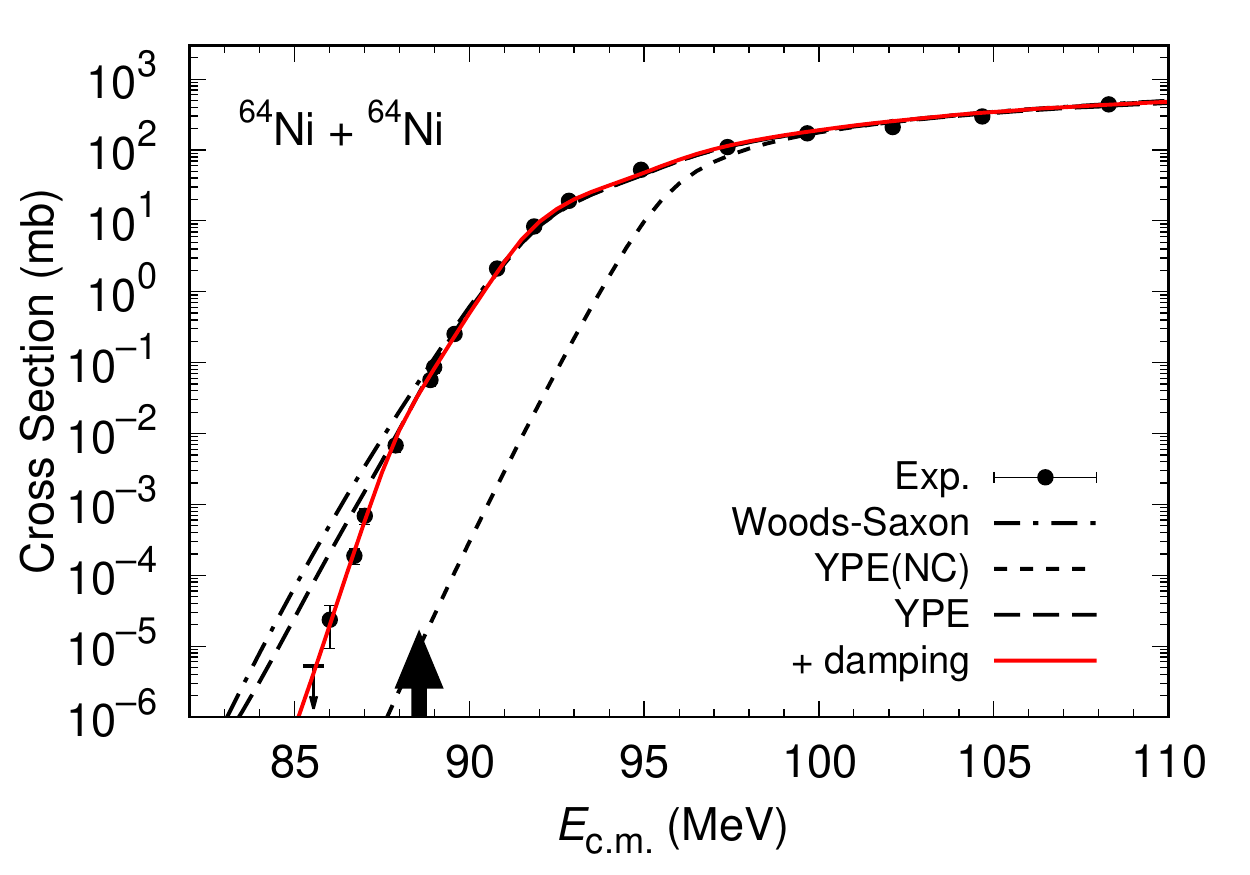}
\caption{The fusion cross sections for the $^{64}$Ni+$^{64}$Ni system 
obtained with the sudden model (the left panel) and the adiabatic model (the right panel). 
In both models, the standard coupled-channels calculations (the dotted line in the light panel and 
the dashed and the dot-dashed lines in the right panel) overestimate fusion cross sections 
at deep subbarrier energies. The arrow in the right panel indicates the threshold energy for 
the deep subbarrier fusion hindrance. Taken from Refs. \cite{ME2006,Ichikawa2015}. 
}
\label{fig:deep-subbarrier}
\end{center}
\end{figure}

Figure \ref{fig:deep-subbarrier} shows a comparison between the sudden (the left panel) 
and the adiabatic (the right panel) models for 
the fusion cross sections for the $^{64}$Ni+$^{64}$Ni system.  
The results of the standard coupled-channels calculations are indicated by the dotted line in the left 
panel and by the dashed and the dot-dashed lines in the right panel. Those calculations well account 
for the experimental fusion cross sections at energies around the Coulomb barrier, but they significantly 
overestimate the cross sections at deep subbarrier energies. By either introducing a repulsive potential in the sudden 
model or quenching the channel coupling effects in the adiabatic model, the deep subbarrier hindrance is 
well reproduced (see the solid lines). 
One can clearly see that both models work equally well. 
This indicates that 
fusion cross sections themselves do not distinguish the two models and 
other observables are needed to judge which model is more reasonable.
For instance, average angular momenta of a compound nucleus could be used for
this purpose \cite{Shrivastava2016}, even though it would be challenging to measure them directly 
at deep subbarrier energies.

\subsection{Semi-microscopic approach}

\begin{figure}[!ht]
\begin{center}
\includegraphics*[width=0.6\textwidth]{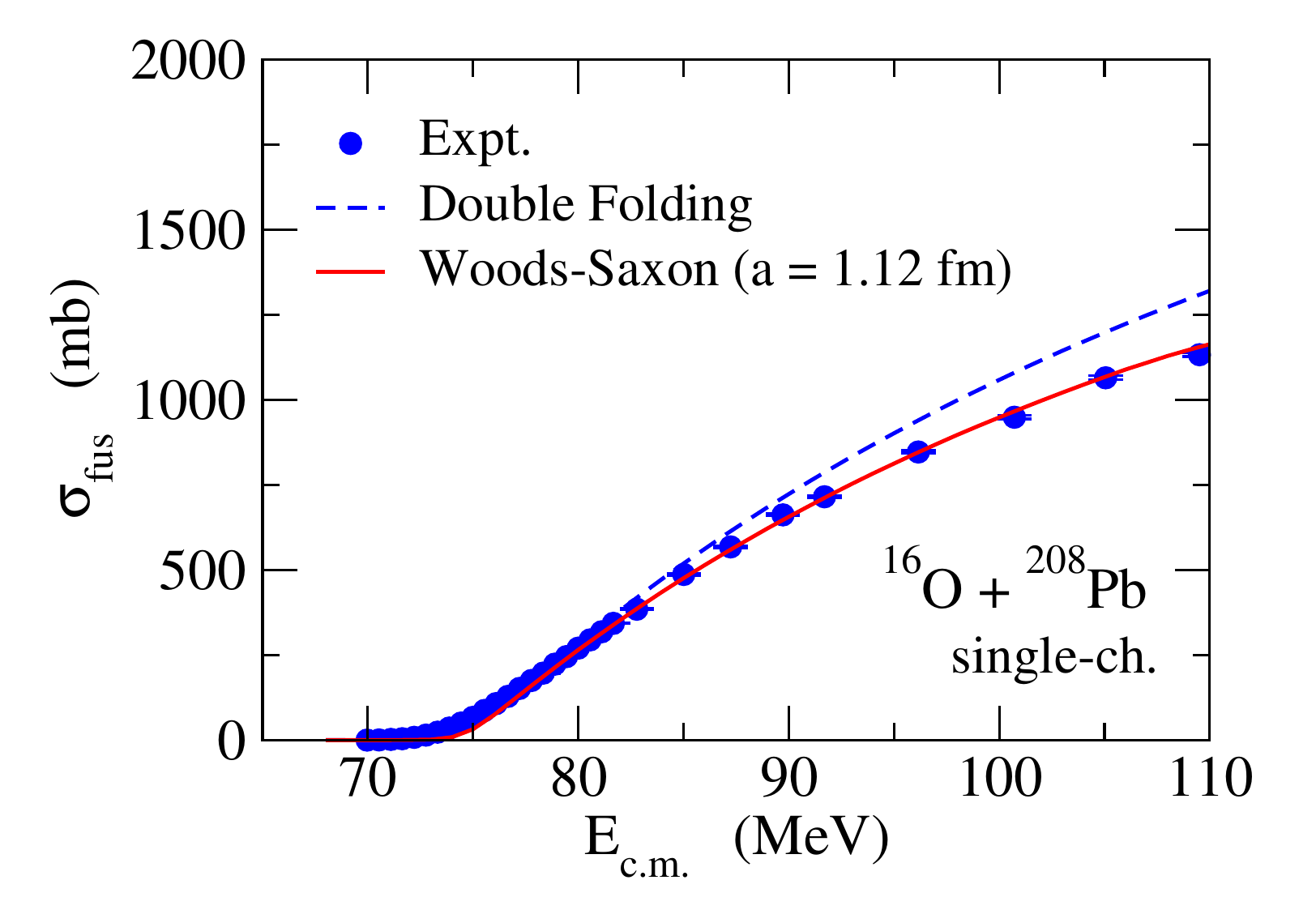}
\caption{The fusion cross sections  for the $^{16}$O+$^{208}$Pb system. 
The dashed and the solid lines are obtained with single-channel calculations with the double folding potential and with a Woods-Saxon potential, respectively. For the 
Woods-Saxon potential, the diffuseness parameter of $a=1.12$ fm is employed. 
The experimental data are taken from Ref. \cite{Morton1999}. 
}
\label{fig:16o208pb-folding}
\end{center}
\end{figure}

In Sec.~3, we have discussed the microscopic coupled-channels method and 
its application to inelastic scattering. Unfortunately, a similar approach does not 
work for heavy-ion fusion reactions. This is demonstrated in Fig.~\ref{fig:16o208pb-folding} 
(see also Refs.~\cite{HDG2002,DHD2007}). 
In the figure, the experimental fusion cross sections for the $^{16}$O+$^{208}$Pb 
system are compared with a single-channel calculation based on the double 
folding model with the density-dependent M3Y interaction (with the zero-range 
approximation for the exchange term) \cite{Khoa1995,BS1997}. One can see that 
the double folding model significantly overestimates the fusion cross sections at 
energies above the Coulomb barrier. This 
feature remains the same even if the channel coupling effects are taken 
into account \cite{DHD2007}. 
The double folding potential can be well fitted with a Woods-Saxon potential with 
the surface diffuseness parameter of $a\sim$ 0.63 fm. 
It has been known that the surface diffuseness parameter has to be phenomenologically increased 
to fit the experimental fusion cross 
sections (see the solid line in the figure) \cite{HRD03,Leigh1995,DHD2007,Newton2004,Newton2004b}, even though 
scattering data can be fitted with a surface diffuseness parameter that is consistent 
with a double folding potential \cite{Mukherjee2007,WHD2006,Gasques2007,Hinde2007,Evers2008}. 

The failure of the double folding model for fusion reactions may not be surprising, given 
that the model takes into account only a simple exchange term, that is, the knock-on 
exchange term in which the exchange operator is applied only to two interacting 
nucleons. From the point of view of many-body theories, such as the resonating 
group method (RGM), there are many other types in the exchange terms, which 
may play an important role especially in short distances relevant to fusion reactions
\cite{AH1982,AH1982b,AH1983,AH1983b}. Moreover, couplings to high-lying 
collective and non-collective states 
may also play a role in fusion reactions at energies above the barrier 
\cite{Newton2004b,Tokieda2020,Tokieda2020b}. 

Given this situation, a semi-microscopic coupled-channels approach has been 
developed \cite{HY2015,YH2016}. In this approach, 
the operator $\alpha_{\lambda\mu}$ in Eq.~(\ref{coupN}) is replaced by  (see Eq.~(\ref{multipole})), 
\begin{equation}
\alpha_{\lambda\mu}=
\frac{3eZ_T R_T^{\lambda}}{4\pi}\,Q_{\lambda\mu},
\end{equation}
where $Q_{\lambda\mu}$ is a many-body multipole operator
\begin{equation}
Q_{\lambda\mu}=\sum_ir_i^\lambda Y_{\lambda\mu}(\vec{r}_i).
\label{eq:q-microscopic}
\end{equation}
For the Coulomb coupling, $Q_{\lambda\mu}$ in Eq.(\ref{coupC}) is 
replaced by Eq.~(\ref{eq:q-microscopic}). 
Using many-body wave functions from microscopic nuclear structure calculations 
for $\varphi_{\alpha Im_I}(\xi)$ in Eq.~(\ref{eq:channelwf}), one can then construct 
the coupled-channels equations microscopically. This approach still uses a phenomenological 
potential $V_N(R)$, thus it is called the semi-microscopic approach. In this way, the difference 
in the coupling strengths between the nuclear and the Coulomb couplings can be automatically 
taken into account. In practical calculations, one may scale the coupling strengths to an empirical 
value known for the transition between the ground state and the first excited state for a given nucleus. 
Also one may scale the excitation energies to the experimental energy of the first excited state.  

\begin{figure}[!ht]
\begin{center}
\includegraphics*[width=1\textwidth]{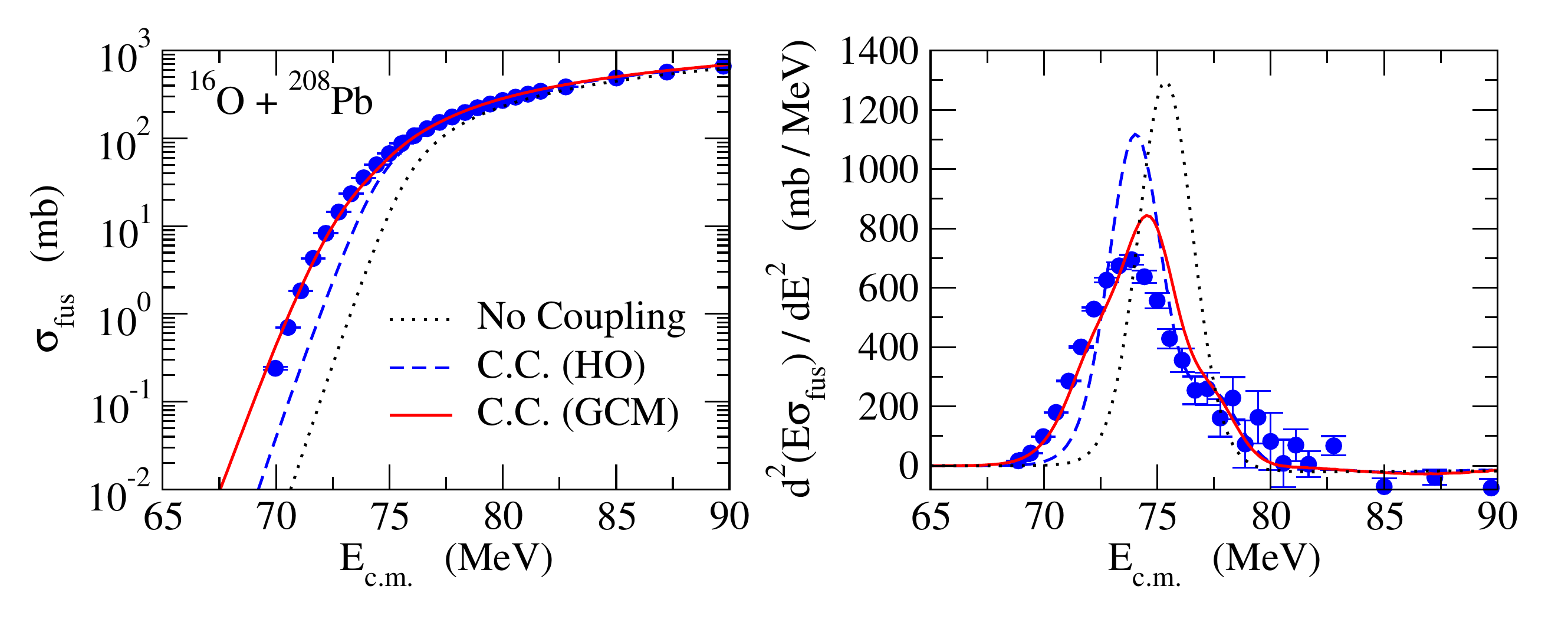}
\caption{The fusion cross sections (the left panel) and the 
fusion barrier distribution (the right panel) for the $^{16}$O+$^{208}$Pb system. 
The dotted lines show the result of the single-channel calculation. The dashed and 
the solid lines are obtained with the coupled-channels method with the vibrational 
couplings up to the double quadrupole and octupole phonon states in 
$^{208}$Pb. The dashed lines use the harmonic oscillator (HO) model, while 
the solid lines take into account the effects of anharmonicity with the generator 
coordinate method (GCM) based on the covariant density functional theory. 
The experimental data are taken from Ref. \cite{Morton1999}. 
}
\label{fig:16o208pb}
\end{center}
\end{figure}

Fig. \ref{fig:16o208pb} shows the results of the semi-microscopic approach for the $^{16}$O+$^{208}$Pb 
system\cite{YH2016}. To this end, the intrinsic states are constructed with the multi-reference covariant density functional 
theory, in which many Slater determinants are linearly superposed after the angular momentum and 
the particle number projections are applied. The left and the right panels show the fusion 
cross sections and the corresponding fusion barrier distribution, respectively. 
The dotted lines show the result of the single-channel calculation. On the other hand, the dashed lines are 
obtained by taking into account 
up to two octupole and quadrupole phonon states in the harmonic oscillator 
approximation. In this approximation, the octupole and the quadrupole phonons are treated as independent 
phonons. Such calculation underestimates the fusion cross sections, and overestimates the height 
of the main peak in the fusion barrier distribution \cite{Morton1999}. This problem is resolved by taking 
into account the anharmonicities of the octupole and the quadrupole phonons with the semi-microscopic 
coupled-channels approach, as is shown by the solid lines. 
It has been pointed out that the couplings between the octupole and the quadrupole phonons 
plays an important role in reproducing the experimental data \cite{YH2016}. 

The semi-microscopic coupled-channels approach can be combined with any microscopic nuclear 
structure calculation, such as shell model and (quasi-particle) random phase approximation. 
We mention that earlier studies with the interacting boson model (IBM) 
\cite{BBDK1992,BBK1993,BBK1994,BBK1994b,HTK1997,HKT1998,ZH2008} 
also belong to this approach. 

\subsection{Heavy and superheavy systems}

Superheavy elements, defined as the transactinide elements with $Z\geq 104$, have 
attracted a lot of attention in recent years \cite{HM2000,HHO2013,DHNO2015,HDS2021,Nazarewicz2018,Giuliani2019,H2019}. 
A standard way to synthesize those superheavy elements is to use heavy-ion 
fusion reactions at energies around the Coulomb barrier \cite{HM2000,HHO2013}. 
A characteristic in fusion of such massive systems is that the touching point appears 
outside the fission barrier, and thus the formation of a compound nucleus is not 
yet achieved even if the Coulomb barrier is overcome \cite{Swiatecki2005}. 
To form a compound nucleus, the inner fission barrier has to be overcome, but  
before that two nuclei may reseparate with a large probability after exchanging a few nucleons and energies. Such process is referred to as the quasi-fission process \cite{HDS2021}. 
Formation cross sections of a compound nucleus, that is, capture 
cross sections are then given as 
(see Eq.~(\ref{eq:sigma_abs})), 
\begin{equation}
\sigma_{\rm cap}(E)=\frac{\pi}{K^2}\sum_{J_T} (2J_T+1)
T_{J_T}(E)P_{\rm CN}(E,J_T).
\end{equation}
In this equation, 
$T_{J_T}(E)=
\left(1-\sum_c |S^{J_T}_{cc_0}|^2\right)$ is the penetrability of the Coulomb 
barrier, while $P_{\rm CN}(E,J_T)$ is the probability to overcome the inner barrier 
to form a compound nucleus. 
The Langevin approach has often been used to describe 
$P_{\rm CN}(E,J_T)$ \cite{ Swiatecki2005,ABGW2000,SKA2002,Swiatecki2003,AO2004,ZG2015}. 
$P_{\rm CN}(E,J)$ is almost one in light and medium-heavy systems, since the touching 
point is inside the inner barrier and thus it is a good approximation to assume that the compound 
nucleus is automatically formed once the touching configuration is reached. 
A transition to the massive regime is considered to occur when the charge product of the 
projectile and the target nuclei, $Z_PZ_T$, is around 1600-1800 \cite{Sham1984,SBKSA2011}. 

The compound nucleus in the superheavy region predominantly decays by fission. 
A complication is that the final products of the fusion-fission process often overlap 
with those of the quasi-fission process. It is thus difficult to separate them 
unambiguously, and a detection of fission fragments alone 
does not guarantee a formation of 
a compound nucleus.  For this reason, usually a formation of a compound nucleus is identified by measuring evaporation residues. Cross sections for the formation of 
evaporation residues are then given by, 
\begin{equation}
\sigma_{\rm ER}(E)=\frac{\pi}{K^2}\sum_{J_T} (2J_T+1)
T_{J_T}(E)P_{\rm CN}(E,J_T)W_{\rm suv}(E^*,J_T),
\label{eq:sigma_ER}
\end{equation}
where 
$E^*$ is the excitation energy of the compound nucleus and 
$W_{\rm suv}(E^*,J_T)$ is the survival probability of the compound nucleus 
against fission, which can be estimated using a statistical model. 

\begin{figure}[!ht]
\begin{center}
\includegraphics*[width=0.5\textwidth]{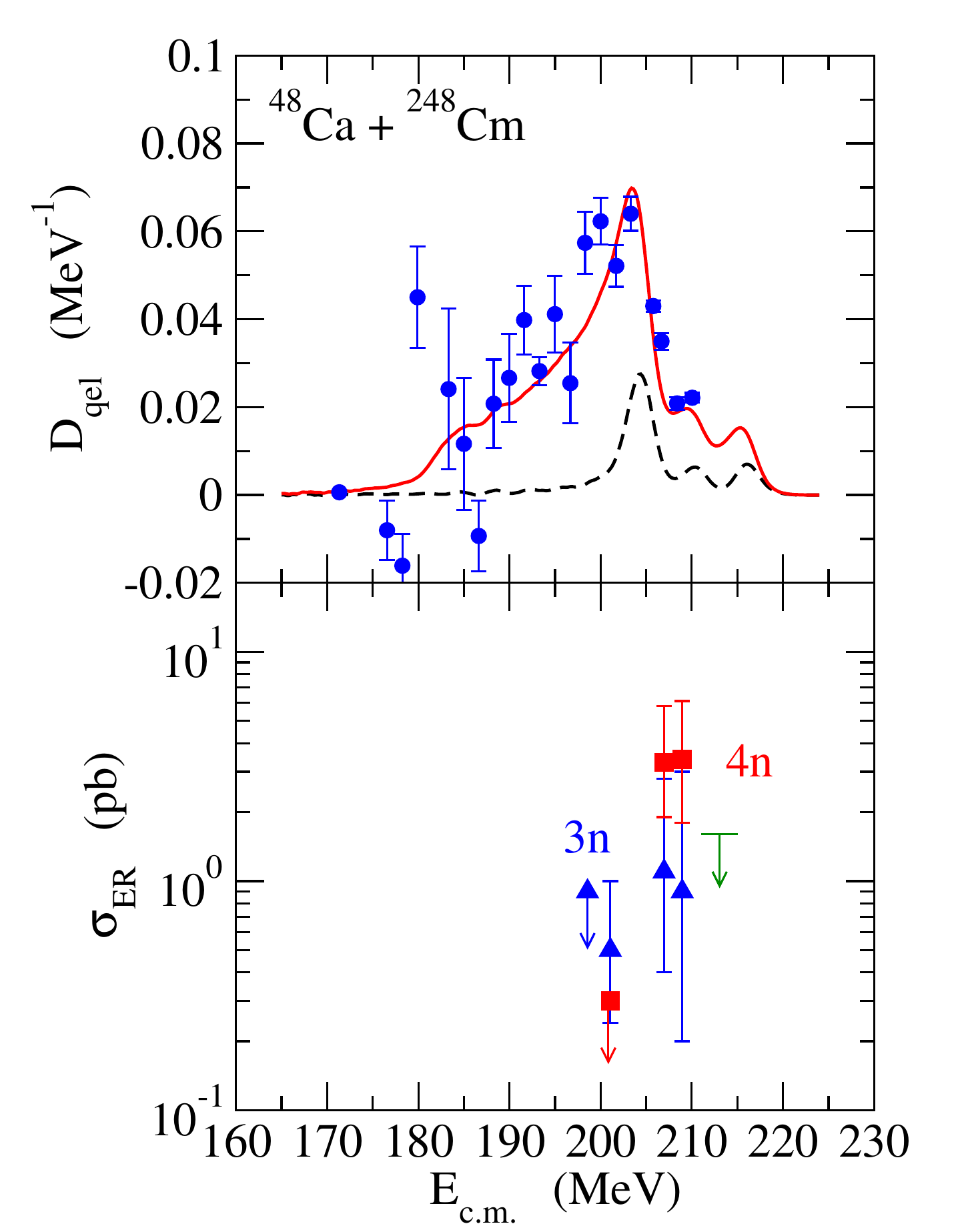}
\caption{(Upper panel) The barrier distribution for the 
capture process for the $^{48}$Ca+$^{248}$Cm system. 
The solid line is obtained with the coupled-channels calculation which 
includes the rotational excitation of $^{248}$Cm, the octuple phonon 
excitation of $^{48}$Ca, and a one-neutron transfer process. 
The dashed line shows the contribution of the side collision with $\theta_d=\pi/2$. 
The experimental data are taken from Ref. \cite{Tanaka2018}. 
(Lower panel) 
The evaporation residue cross sections for the same system. 
The experimental data are taken from Ref. \cite{Hofmann2012}. 
}
\label{fig:48ca238cm}
\end{center}
\end{figure}

Recently, the quasi-elastic barrier distribution has been successfully 
measured for several 
systems which are relevant to syntheses of superheavy elements 
\cite{Mitsuoka2007,Ntshangase2007,Tanaka2018,Tanaka2020}. 
Such barrier distribution is related to the capture process, whose probability 
is given by $T_{J_T}$ in Eq.~(\ref{eq:sigma_ER}).
As an example, the top panel of Fig. \ref{fig:48ca238cm} shows the experimental barrier distribution 
for the $^{48}$Ca+$^{248}$Cm system \cite{Tanaka2018} and its comparison to the 
coupled-channels calculations. In the coupled-channels calculation, the excitations 
of the rotational band in the deformed $^{248}$Cm nucleus, the octupole phonon 
excitation of $^{48}$Ca, as well as a one-neutron transfer reaction are taken 
into account. The coupled-channels calculation provides information on the energy 
region corresponding to the side collision, $\theta_d=\pi/2$ with respect to the 
beam direction, which is indicated by 
the dashed line in the figure. The measured evaporation residue cross sections 
$\sigma_{\rm ER}$  
for this system \cite{Hofmann2012} are shown in the lower panel of the figure. 
One can clearly see that the peak of $\sigma_{\rm ER}$ appears in the energy 
region corresponding to the side collision. 
This measurement of the capture barrier distribution thus provided  
the first experimental confirmation of the notion of compactness, 
that is, the side collision provides the dominant contribution to the compound nucleus 
formation since the touching configuration is compact such that the 
height of the inner barrier is effectively lowered \cite{Hinde1995}. 

\begin{figure}[!ht]
\begin{center}
\includegraphics*[width=0.5\textwidth]{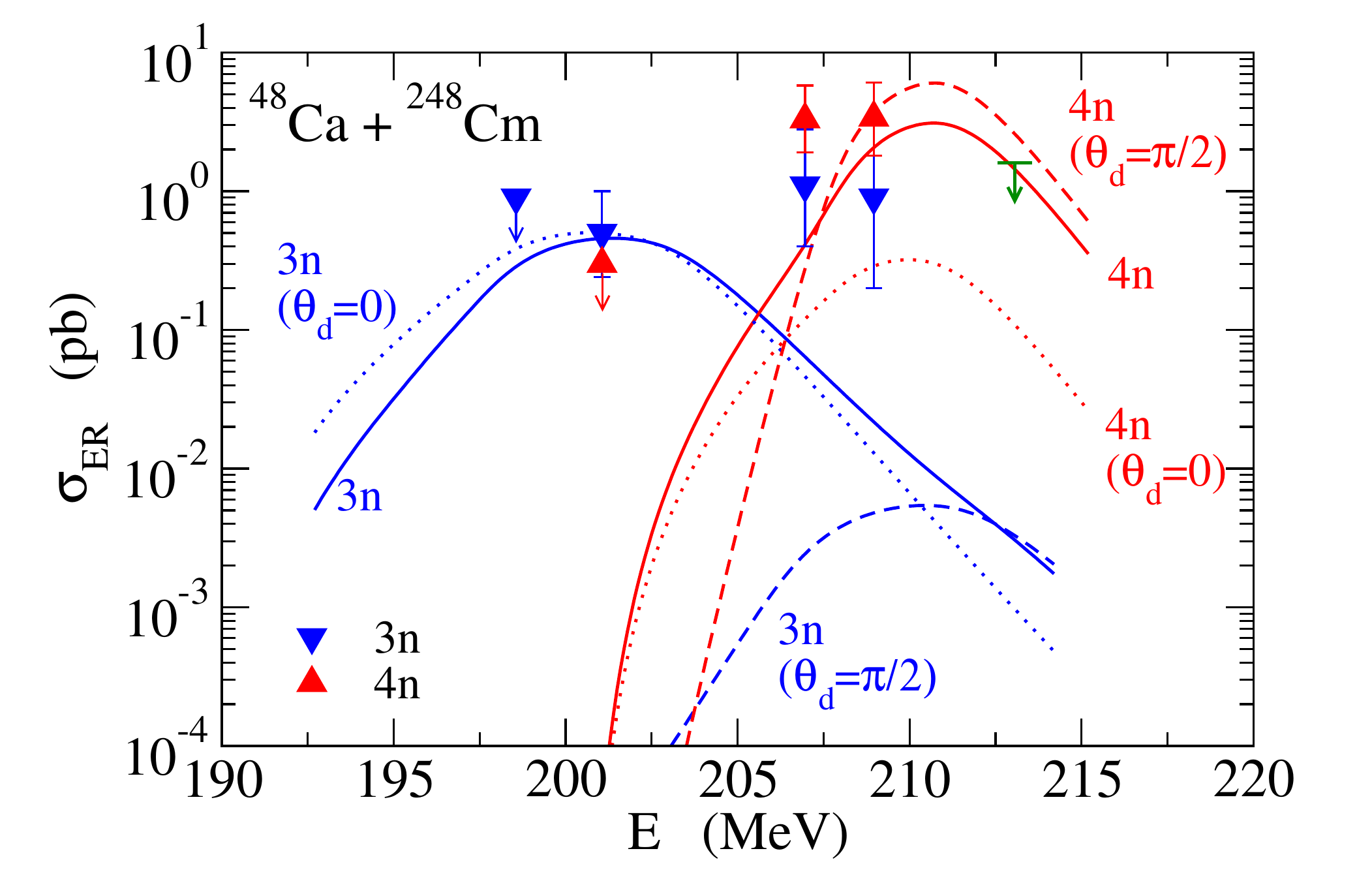}
\caption{
Evaporation residue cross sections for the $^{48}$Ca+$^{248}$Cm system 
obtained with the extended fusion-by-diffusion model taking into account the 
orientation angles of the deformed target nucleus (the solid lines). 
3$n$ and 4$n$ indicate the 3 neutron and 4 neutron evaporation channels, 
respectively. The dotted and the dashed lines show the contribution of the 
tip ($\theta_d=0$) and the side ($\theta_d=\pi/2$) collisions, respectively. 
 The experimental data are taken from Ref. \cite{Hofmann2012}. 
}
\label{fig:48ca238cm-fbd}
\end{center}
\end{figure}

Figure \ref{fig:48ca238cm-fbd} shows the role of orientation angle of $^{248}$Cm 
in the evaporation formation cross sections\cite{H2018}. 
To this end, the evaporation residue cross sections are computed as 
\begin{equation}
\sigma_{\rm ER}(E)=\frac{\pi}{K^2}\int^1_0d(\cos\theta_d) \sum_{J_T} (2J_T+1)
T_{J_T}(E;\theta_d)P_{\rm CN}(E,J_T;\theta_d)W_{\rm suv}(E^*,J_T),
\end{equation}
where $T_{J_T}(E;\theta_d)$ and $P_{\rm CN}(E,J_T;\theta_d)$ are the 
penetrability of the Coulomb barrier and the formation probability of the 
compound nucleus for a given orientation angle $\theta_d$, respectively. 
See also Ref. \cite{AHNC2012}. 
$P_{\rm CN}$ is estimated with the fusion-by-diffusion model 
\cite{Swiatecki2005,ABGW2000,Swiatecki2003,Cap2011} with 
an extension to introduce the angle dependence of the injection point. 
The injection point $s_{\rm inj}$ is the initial position to solve the diffusion equation to 
overcome the inner barrier, and is taken to be \cite{H2018}, 
\begin{equation}
s_{\rm inj}(\theta_d)=s_{\rm inj}^{(0)}+R_T\sum_\lambda \beta_{\lambda T}Y_{\lambda 0}(\theta_d), 
\end{equation}
where $s_{\rm inj}^{(0)}$ is the injection point in the spherical limit and 
$\beta_{\lambda T}$ is the deformation parameter. With $\beta_{2T}>0$, the injection 
point for $\theta_d=\pi/2$ is smaller than that for $\theta_d=0$. 
Since the effective barrier height is given by $B_{\rm eff}=V_{\rm fis}(s_{\rm sd})-V_{\rm fis}(s_{\rm inj})$, 
where $V_{\rm fis}$ is a fission barrier and $s_{\rm sd}$ is the saddle point, $B_{\rm eff}$ is lower 
for $\theta_d=\pi/2$ as compared to $B_{\rm eff}$ for $\theta_d=0$. In the figure, 
the dashed and the dotted lines show the contributions of the side ($\theta_d=\pi/2$) 
and the tip ($\theta_d=0$) collisions, respectively. For the 4 neutron evaporation 
channel ($4n$), which provides the most important evaporation channel in this 
system, one can clearly see that the side collision indeed provides the dominant 
contribution. 

The injection point has subsequently been determined 
more microscopically 
by using the time-dependent Hartree-Fock (TDHF) method \cite{SH2019}. 
Such method provides a hybrid approach which combines the TDHF with 
the Langevin approaches, which is suitable for applications to the region in which 
no experimental data is available. In fact, this approach has been applied to 
formation reactions of the superheavy element $Z=120$, revealing that the 
capture cross sections are not sensitive to a choice of the reaction systems  \cite{SH2019}.

\subsection{Fusion of weakly-bound nuclei}

Fusion of neutron-rich nuclei is important for nuclear astrophysics \cite{CH2008,Steiner2012} as well as 
for superheavy elements \cite{Loveland2007}. 
In addition, the reaction mechanism of fusion of neutron-rich nuclei is in itself 
intriguing, as several effects interplay with each other. 
Those effects include:  
i) the extended density distribution due to the weakly bound nature of 
neutron-rich nuclei, which results in a lowering of the 
Coulomb barrier \cite{TS1991}, 
ii)  the breakup process, which may hinder fusion cross sections since the 
lowering of the Coulomb barrier disappears; at the same time, it may also 
enhance fusion cross 
sections if couplings to a breakup channel 
dynamically lowers the Coulomb barrier \cite{DV1994,HVD00,DTT2002}, 
and iii) the transfer processes, for which 
the $Q$-value is positive for neutron-rich nuclei 
and fusion reaction are then significantly affected  \cite{CCS18,CKC2021}. 
In the case of reactions with weakly-bound nuclei, 
in addition to the complete fusion process,  
in which the whole projectile is absorbed by the target,  an important contribution to the fusion comes from the so-called incomplete 
fusion (ICF), in which only a part of the projectile is absorbed. The understanding and quantitative evaluation of  ICF has revealed as a very challenging theoretical problem.
We refer the reader to Refs.~\cite{canto2006,canto2015,CHU2021} for reviews as well as Refs. 
\cite{HPC92,TKS93} for earlier works based on the semi-classical approximation. 

\begin{figure}[!ht]
\begin{center}
\includegraphics*[width=0.5\textwidth]{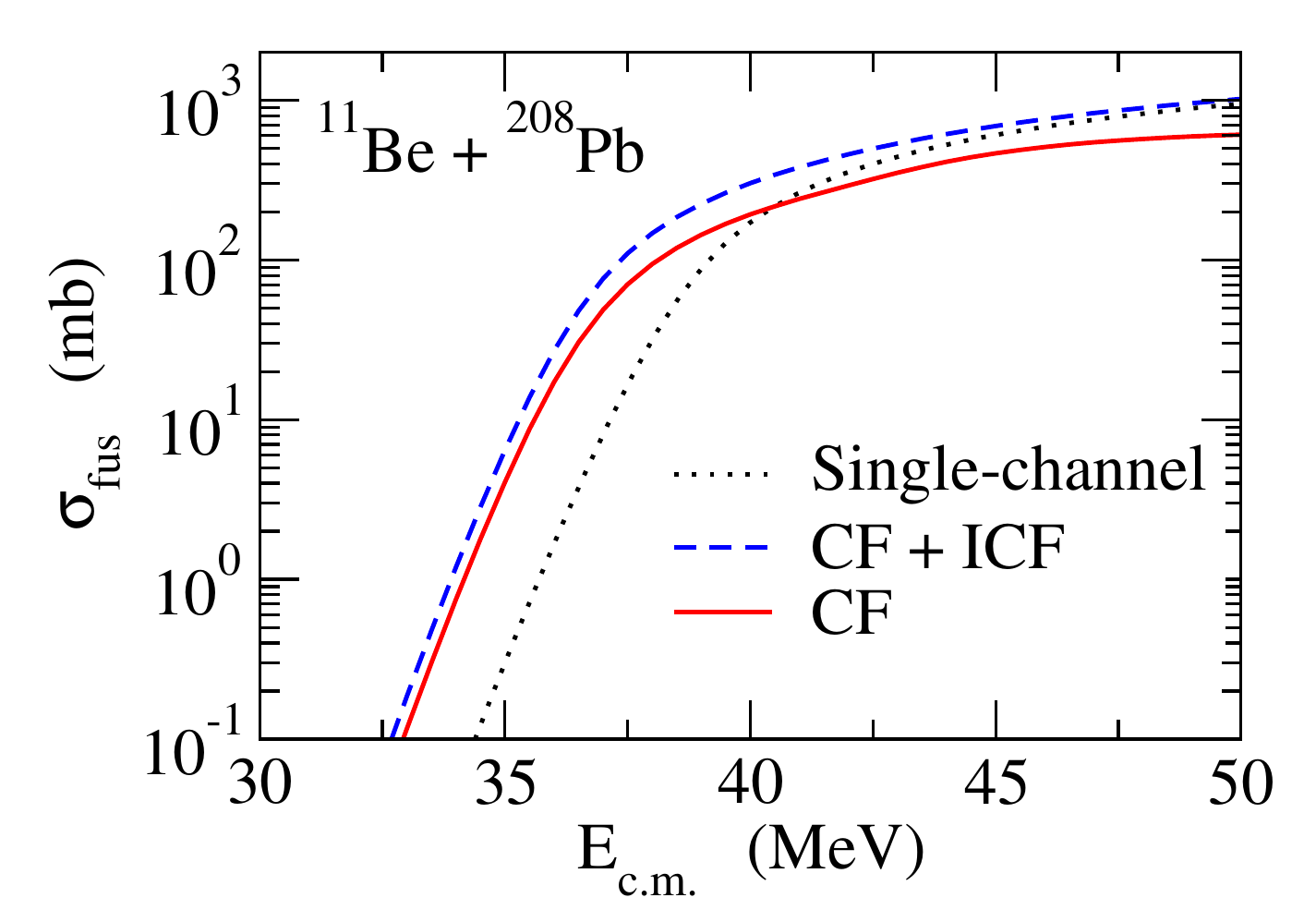}
\caption{Fusion cross sections for the $^{11}$Be+$^{208}$Pb system 
obtained in Ref. \cite{HVD00}. The dashed and the solid lines shows the total 
and the complete fusion cross sections, respectively. The dotted line denotes 
the result of a single-channel calculation without breakup of the projectile 
nucleus.
}
\label{fig:11be208pb}
\end{center}
\end{figure}

\subsubsection{The CF suppression phenomenon}
Experiments with light weakly-bound stable nuclei (such as $^{6,7}$Li or $^{9}$Be) have shown that the CF (conveniently defined  in these experiments as capture of the complete {\it charge} of the projectile) are suppressed by $\sim$20-30\% compared to the case of tightly bound nuclei \cite{Das99,Tri02,Das02,Das04,Muk06,Rat09,LFC15}.
Early analyses of these experiments tried to explain the phenomenon using coupled-channels calculations, including the coupling to low-lying excited states of the projectile and target \cite{Das99,Das02,Kum12,Zha14,Fan15}. Yet, these calculations  systematically failed to reproduce such suppression.  This failure has been attributed to the omission in these calculations of the breakup of the projectile; a scenario was suggested in which the weakly bound projectile breaks up prior to reaching the fusion barrier, with the subsequent reduction of probability of complete  capture. This interpretation is supported by the presence of large $\alpha$ yields as well as target-like residues  which are consistent with the capture of one of the fragment constituents  of the projectile, that is, ICF.   
To account for these observations, some authors have proposed a two-step scenario \cite{Das02,Dia07}  consisting on the elastic dissociation of the projectile followed by the capture of one of the fragments by the target.  However,  calculations based on a three-dimensional classical dynamical model \cite{Dia07}, which incorporates this two-step breakup-fusion mechanism, can only explain a small fraction of the observed CF suppression for $^{9}$Be \cite{Coo16} and $^{8}$Li \cite{Coo18} reactions. More encouraging results have been obtained with different methods based on the CDCC formalism, as described in the following subsections.

\subsubsection{Computation of fusion cross sections with the CDCC method}
Although the CDCC method was originally devised to evaluate the elastic and breakup observables, some attempts have been made to use this method to obtain complete and incomplete fusion cross sections. In Refs.~\cite{HVD00,DTT2002}, the CDCC approach was applied to fusion of 
a weakly-bound projectile. 
In these calculations, the CF was identified 
with the absorption from the entrance channel while the ICF with
the absorption from the breakup channels.    
Figure \ref{fig:11be208pb} shows the result of such calculation for the $^{11}$Be+$^{208}$Pb reaction \cite{HVD00}. 
The solid line shows 
the CF cross sections, while the dashed line shows the total 
fusion cross sections
(that is, a sum of complete and incomplete fusion cross sections). For comparison, 
the figure also shows the result of the single-channel calculation (dotted 
line). By comparing the solid and the dotted lines, one can see that 
the CF cross sections may be enhanced as compared to the fusion 
cross sections without the breakup process at energies below the Coulomb barrier, while 
they are suppressed above the barrier. 

\begin{figure}[!ht]
\begin{center}
\begin{minipage}{0.45\textwidth}
\includegraphics[width=0.7\columnwidth]{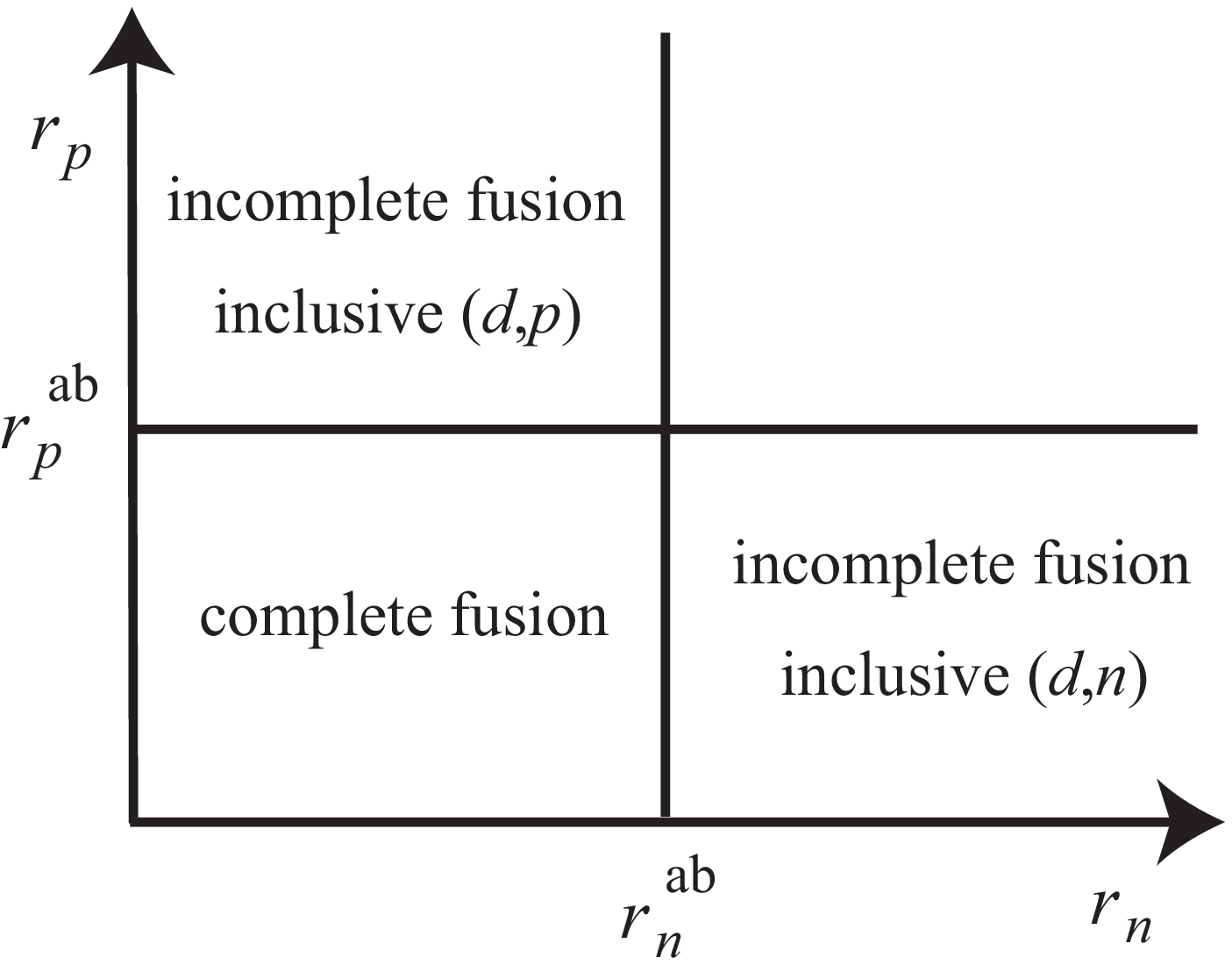}
\end{minipage}
\begin{minipage}{0.45\textwidth}
\includegraphics[width=0.7\columnwidth]{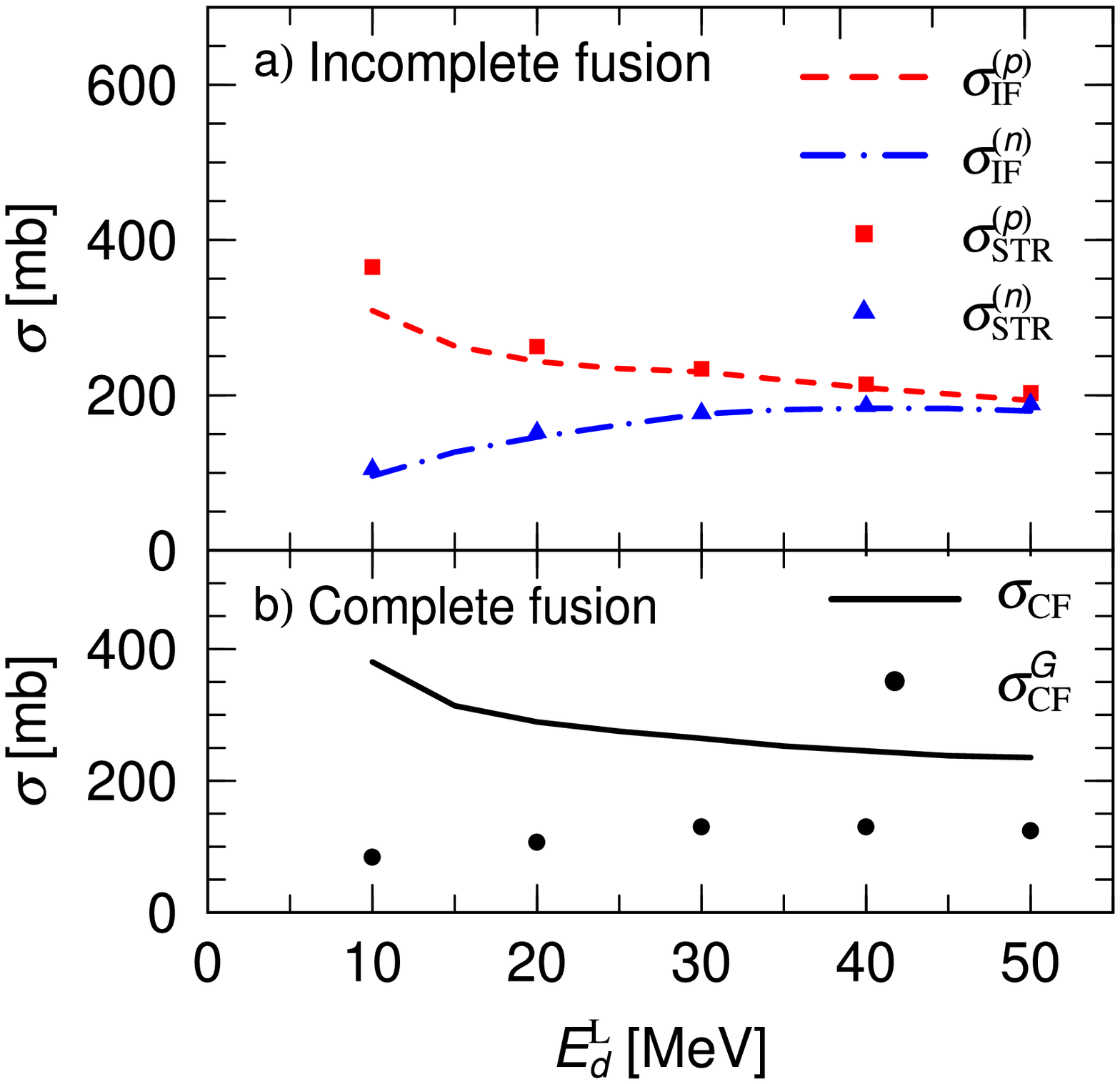}
\end{minipage}
\end{center}
\caption{\label{fig:hashimoto} Left: Schematic illustration of the four integration regions employed in the method of Hashimoto {\it et al.}~\cite{Hash09}. Right: 
a) Comparison of $\sigma_{\rm ICF}^{(p)}$ (short-dashed line)
and $\sigma_{\rm ICF}^{(n)}$ (dash-dotted line) obtained in \cite{Hash09}  with $\sigma_{\rm STR}^{(p)}$ (squares)
and $\sigma_{\rm STR}^{(n)}$ (triangles)
given by the Glauber model~\cite{Ye09} \
b) Complete fusion cross sections calculated with
the Glauber model (dots) and with the method of \cite{Hash09}
(solid line).} 
\end{figure}

A limitation of this method is that it can only be  applied to  projectiles composed by a heavy charged fragment and a light uncharged one (such as $^{11}$Be), as it relies on the assumption that the center of mass of the projectile is close to that
of the heavy fragment, and far from the light one.   Thus, it cannot
be used for projectiles like $^{6,7}$Li, that break up into two fragments of comparable masses. 
To overcome this difficulty, Hashimoto {\it et al.}~\cite{Hash09}  proposed an alternative approach based also on the CDCC method, and applied it to the case of deuteron scattering.  Their idea  is to transform the CDCC wavefunction from its natural coordinates 
\{$\vecr$, $\vecR$\} to the coordinates \{$\vecr_p$, $\vecr_n$\}, 
\begin{equation}
|\Psi(\vecr,\vecR) |^2 \, d\vecr d\vecR =  |\tilde{\Psi}(\vecr_p,\vecr_n) |^2  d\vecr_p  d\vecr_n    \, ,
\end{equation}
and then associate the CF and ICF cross sections with the absorption taking place in different regions of the $\{ r_p , r_n \}$ space, as  illustrated in the left panel of Fig.~\ref{fig:hashimoto}. The distances $r^\mathrm{ab}_p$ and $r^\mathrm{ab}_n$ denote the absorption radii for the proton and neutron, such that for $r_p > r^\mathrm{ab}_p$ ($r_n > r^\mathrm{ab}_n$) the proton (neutron) absorption becomes negligible. The  CF is identified with the absorption taking place when both the proton and neutron are inside their respective absorption radii, i.e.,
\begin{eqnarray}
\sigma_{\rm CF}=
  \frac{1}{|N|^2}\frac{K}{E}
  \int_{r_p<r_p^{\rm ab}}  d\vecr_p
  \int_{r_n<r_n^{\rm ab}}  d\vecr_n
  |\tilde{\Psi}(\vecr_p,\vecr_n) |^2
  \{ W_p(\vecr_p)+W_n(\vecr_n) \},
\label{eq-CFX}
\end{eqnarray}
where $K$ is the incident wave number,  $E$ the energy and  $N$ is a normalization constant.  
Similarly, the ICF cross section is obtained from the absorption occurring in the region where one of the two fragments is inside its absorption radius while the other is outside it, i.e.,
\begin{eqnarray}
\sigma_{\rm ICF}^{(p)}=
  \frac{1}{|N|^2}\frac{K}{E}
  \int_{r_p<r_p^{\rm ab}}  d\vecr_p
  \int_{r_n>r_n^{\rm ab}}  d\vecr_n
  |\tilde{\Psi}(\vecr_p,\vecr_n) |^2
  W_p(\vecr_p),
\label{eq-IFXp}
\end{eqnarray}
\begin{eqnarray}
\sigma_{\rm ICF}^{(n)}=
  \frac{1}{|N|^2}\frac{K}{E}
  \int_{r_p>r_p^{\rm ab}}  d\vecr_p
  \int_{r_n<r_n^{\rm ab}}  d\vecr_n
  |\tilde{\Psi}(\vecr_p,\vecr_n) |^2
  W_n(\vecr_n) .
\label{eq-IFXn}
\end{eqnarray}

The method was applied to the reaction $d$+$^{7}$Li \cite{Hash09}. Although no comparison with experimental data was attempted, the authors compared their results with those computed with the Glauber calculations of Ref.~\cite{Ye09}.  This comparison is shown in the right panel of Fig.~\ref{fig:hashimoto} with the upper and lower panels corresponding respectively to the ICF and CF cross sections as a function of the deuteron incident energy. The agreement for the ICF is very satisfactory for energies as low as $E_d$=10~MeV. Since the Glauber model is a high-energy approximation, the agreement at these relatively low energies is somewhat unexpected. By contrast, for the CF part the Glauber method gives significantly smaller cross sections. A discussion of these results is provided in Ref.~\cite{Ye09}.

Another procedure to extract the CF and ICF cross section from the CDCC method was proposed by Parkar and co-workers \cite{Par16}. The central idea of their method is to perform a series of CDCC calculations with different choices of the fragment-target potentials. In particular, to obtain the  ICF cross section for the capture of a given fragment, they perform a CDCC calculation using short-range imaginary potential for the interaction of that fragment and the target, and a real potential for the other fragment. Using this method, these authors have been able to obtain a reasonable account of the CF, ICF and TF cross sections of reactions induced by $^{6,7}$Li projectiles \cite{Par16,Par18}, although with some arbitrary choice of the short-range of the fusion potential.

More recently, the authors of Refs.~\cite{Ran20,Cor20} have proposed an alternative method which requires a single CDCC calculation to compute the TF, CF and ICF cross sections. In this CDCC calculation the fragment-target interactions are also modeled with optical potentials with a short-ranged imaginary part. For a two-body projectile, the total fusion cross section is  then computed as:
\begin{equation} 
\sigma_\mathrm{TF} = \frac{1}{|N|^2}\ \frac{K}{E}\ 
\  \left\langle\ \Psi^{(+)}\, \left| \, W^{(1)} + W^{(2)}\,  \right| \Psi^{(+)}\  \right\rangle \, ,
\label{sigTF-1}
\end{equation}
where $W^{(1,2)}$ represent the imaginary parts of the fragment-target interactions. 
This CDCC wavefunction can be split into bound ($\Psi^{B}$) and continuum ($\Psi^{C}$) components:
\begin{equation} 
{\Psi}^{(+)}({\vecR},\vecr) = {\rm \Psi}^{B}({\vecR},{\vecr})  + {\Psi}^{C}({\vecR},{\vecr}) ,
\end{equation}
where ${\Psi}^{B}$ and ${\Psi}^{C}$ are respectively its components in the bound and bin subspaces. They are given by the channel expansions
\begin{eqnarray} 
{\Psi}^{B}({\vecR},{\vecr}) &=& \sum_{\beta\, \in\, B} \  \chi_{\beta}({\bf R})  \phi_\beta ({\bf r}) \label{PsiB-1}\\
{\Psi}^{C}({\vecR},{\bf r}) &=& \sum_{\gamma\, \in\,  C} \  \chi_{\gamma}({\vecR})  \phi_\gamma ({\vecr}) , 
\label{PsiC-1}
\end{eqnarray}
where $\phi_\beta$ and $\phi_\gamma$ are respectively bound and unbound states of the projectile, and $\chi_\beta$ and $\chi_\gamma$ are
the corresponding wave function describing the projectile-target relative motion.

Assuming that matrix-elements of the imaginary potentials connecting bound channels to bins are negligible,  Eq.~(\ref{sigTF-1}) can  be put in the form
\begin{equation} 
\sigma_\mathrm{TF} = \sigma_\mathrm{TF}^{B} \ +\  \sigma_\mathrm{TF}^{C},
\end{equation}
with
\begin{eqnarray} 
 \sigma_\mathrm{TF}^{B} &=&  \frac{1}{|N|^2} \frac{K}{E} \sum_{\beta, \beta^\prime\, \in\, {B}} \
 \left\langle \chi_\beta \left|  W_{\beta \beta^\prime}^{(1)}+ W_{\beta \beta^\prime}^{(2)}\,
 \right|   \chi_{\beta^\prime} \right\rangle \label{TF-B}\\
 \sigma_{TF}^{C} &=&  \frac{1}{|N|^2} \frac{K}{E} \sum_{\gamma,\gamma^\prime \in\,{C}} \
 \left\langle \chi_\gamma \left|  W_{\gamma \gamma^\prime}^{(1)} + W_{\gamma \gamma^\prime}^{(2)}\,
 \right|   \chi_{\gamma^\prime} \right\rangle .
 \label{TF-C}
\end{eqnarray}
where $W_{\alpha \alpha^\prime}^{(i)} =\left(\phi_{\alpha} \left| {W}^{(i)} \right| \phi_{\alpha^\prime} \right),$
with $\alpha, \alpha^\prime$ standing for either $\beta, \beta^\prime$ or $\gamma, \gamma^\prime$, are the matrix-elements of the imaginary potentials.\\

Then, by performing an angular momentum expansion of the wave functions and the imaginary potentials,  Eqs.~(\ref{TF-B}) and
(\ref{TF-C}) become
\begin{eqnarray}
\sigma_{TF}^{B} &=& \frac{\pi}{K^2}\,\sum_{J_T} (2J_T+1)\ \mathcal{P}_{B}^{TF}(J_T)  \label{TF-B1} \\
\sigma_{TF}^{C} &=& \frac{\pi}{K^2}\,\sum_{J_T} (2J_T+1)\ \mathcal{P}_{C}^{TF}(J_T) \label{TF-C1},
\end{eqnarray}
with
\begin{eqnarray}
\mathcal{P}_{B}^{TF}(J_T)  &=& \mathcal{P}_{B}^{(1)}(J_T) +  \mathcal{P}_{B}^{(2)}(J_T) \label{P_TF-B} \\
\mathcal{P}_{C}^{TF}(J_T)  &=& \mathcal{P}_{C}^{(1)}(J_T) +  \mathcal{P}_{C}^{(2)}(J_T) \label{P_TF-C} .
\end{eqnarray}
where $\mathcal{P}_{B}^{(i)}(J_T)$ and $\mathcal{P}_{C}^{(i)}(J_T)$ are the probabilities of absorption of fragment $c_i$ in
bound channels and in the continuum, respectively, resulting from the contributions of ${W}^{(i)}$ to the TF cross section.

In terms of these probabilities,  the authors of Ref.~\cite{Ran20,Cor20} introduce  the ICF probabilities 
\begin{eqnarray}
 {\mathcal P}^{ICF1}(J_T)  &=& {\mathcal P}^{(1)}_{C} (J_T) \times \left[\ 1 - {\mathcal P}^{(2)}_{C} (J_T)\ \right]   \label{PICF1}\\
{\mathcal P}^{ICF2}(J_T)  &=&  {\mathcal P}^{(2)}_{C} (J_T) \times \left[\ 1 - {\mathcal P}^{(1)}_{C} (J_T)\ \right]   \label{PICF2},
\end{eqnarray}
and the  {\it sequential complete fusion} probability 
\begin{equation}
{\mathcal P}^\mathrm{SCF}(J_T)  =  2\  {\mathcal P}^{(1)}_{C} (J_T) \times  {\mathcal P}^{(2)}_{C} (J_T).
 \label{TSCF}
\end{equation}

In terms of the introduced probabilities, the following fusion cross sections are defined:
\begin{itemize}
    \item Direct complete fusion (CF):
 \begin{equation}
 \sigma_{DCF} = \sigma_{TF}^{B} ,
 \label{DCF}
 \end{equation}
  which describes the simultaneous capture of the two fragments. 
  
  \item Sequential complete fusion (SCF):
   \begin{equation}
\sigma_{SCF} = \frac{\pi}{K^2}\,\sum_{J_T} (2J_T+1)\ \mathcal{P}^{SCF}(J_T) .
\label{SCF}
\end{equation}

\item  ICF of fragment $c_{i}$ (ICFi) 
\begin{equation}
\sigma_{ICFi} = \frac{\pi}{K^2}\,\sum_{J_T} (2J_T+1)\ \mathcal{P}^{ICFi}(J_T) .
\label{sigICFi}
\end{equation}

\end{itemize}

In this formalism, the  CF, ICF and TF cross sections are then given by
\begin{eqnarray}
\sigma_\mathrm{CF} &=& \sigma_{DCF}\,+\,\sigma_{SCF},  \label{sigCF} \\
\sigma_\mathrm{ICF} &=&\sigma_{ICF1} \,+\, \sigma_{ICF2}, \label{sigICF}\\
\sigma_{TF} &=& \sigma_{CF} + \sigma_{ICF} . \label{sigmaTF-sum} 
\end{eqnarray}

\begin{figure}
\begin{center}
\begin{minipage}{0.45\textwidth}
\includegraphics[width=0.7\columnwidth]{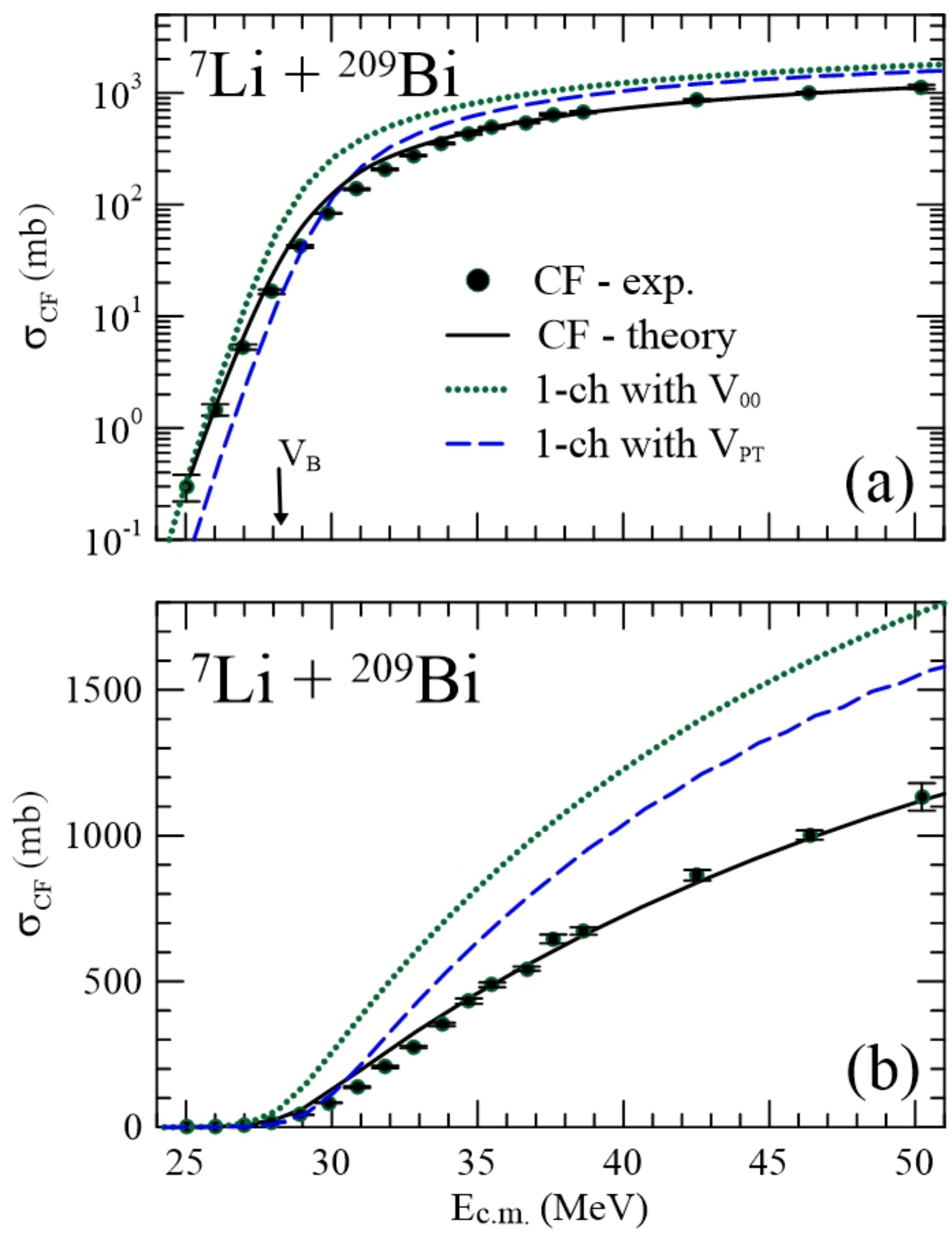}
\end{minipage}
\begin{minipage}{0.45\textwidth}
\includegraphics[width=0.7\columnwidth]{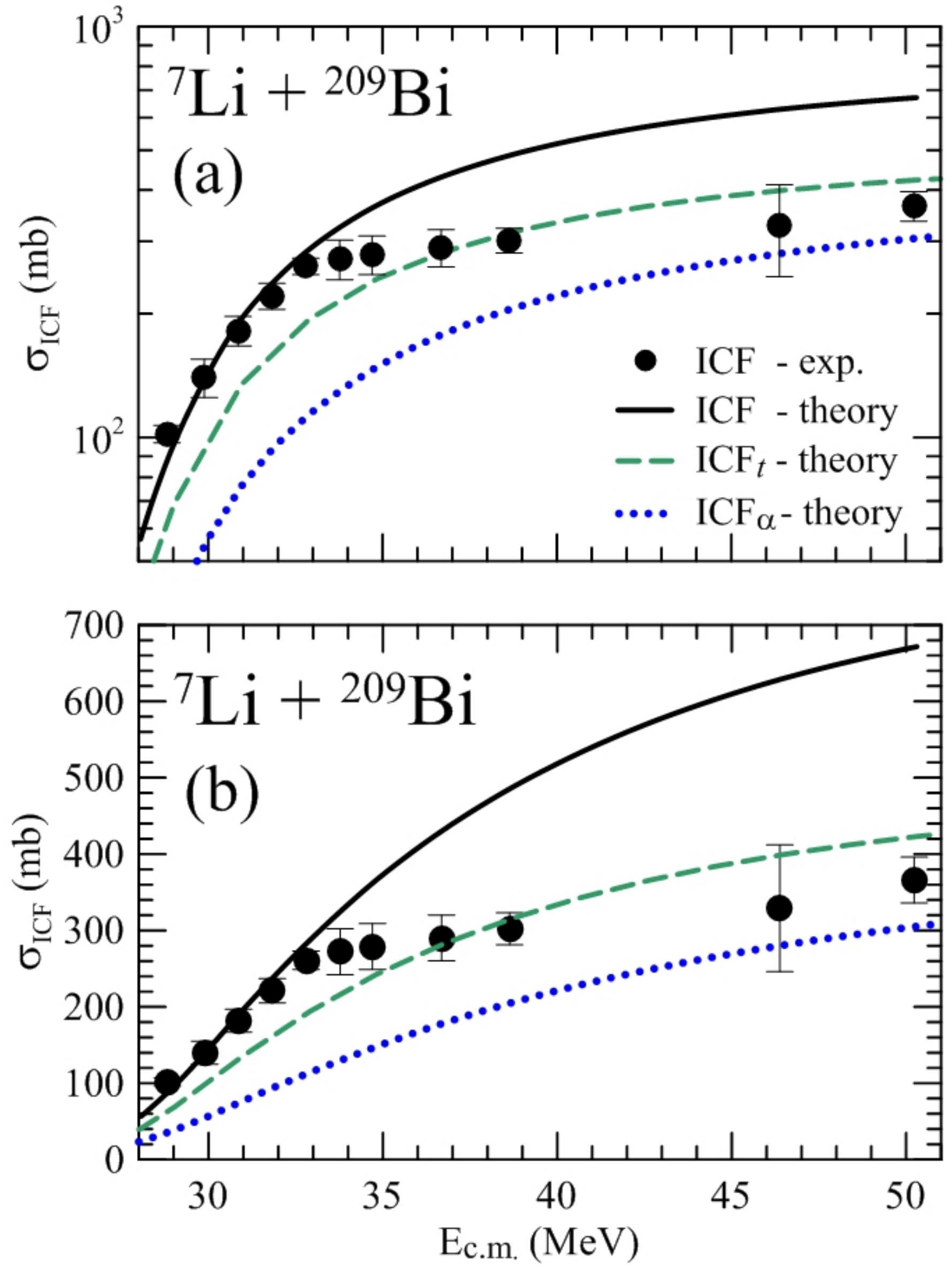}
\end{minipage}
\end{center}
\caption{\label{fig:li7bi_canto} CF (left) and ICF (right) cross sections for $^{7}$Li+$^{209}$Bi. Experimental data from Refs.~\cite{Das02,Das04} are compared with the calculations of Ref.~\cite{Cor20}. Figure reproduced from this reference with permission from APS.} 
\end{figure}

An application of this model is shown in Fig.~\ref{fig:li7bi_canto}, where CF (left panel) and ICF (right panel) data for the reaction $^{7}$Li+$^{209}$Bi \cite{Das02,Das04} are compared with the predictions of the model. The upper and lower panels display the same results in logarithmic and linear scale, respectively. For the CF, the agreement with the data is very satisfactory, both above and below the barrier (indicated by the arrow). For the ICF, the separate contributions for triton capture and $\alpha$ capture are shown, with the former  giving a much larger cross section. The calculations are found to reproduce well the data up to $E_\mathrm{c.m.} \approx 34$~MeV, but overestimate them at higher energies. A detailed discussion of these results can be found in Ref.~\cite{Cor20}.

\subsubsection{Evaluation of CF and ICF cross sections with the IAV model}

As discussed in Sec.~4, the IAV model is intended to provide the total inclusive cross section corresponding to the detection of the $b$ fragment in  reactions of the form $A(a,b)X$. This results from the fact that the imaginary part appearing in the expectation value of Eq.~(\ref{eq:iav}) accounts in principle for {\it all} processes in which the participant fragment $x$ interacts nonelastically with the target nucleus. This will include the ICF cross section, but also other NEB processes not associated with the formation of a compound nucleus of the $x+A$ system, such as target excitation.  The isolation of the ICF cross section from the total NEB cross section is indeed not a non-trivial problem. An intuitive approach consists in identifying the ICF with the absorption due to a short-ranged imaginary potential. This strategy has been however barely explored so far within the IAV model. 

\begin{figure}
\begin{center}
\includegraphics[width=0.85\columnwidth]{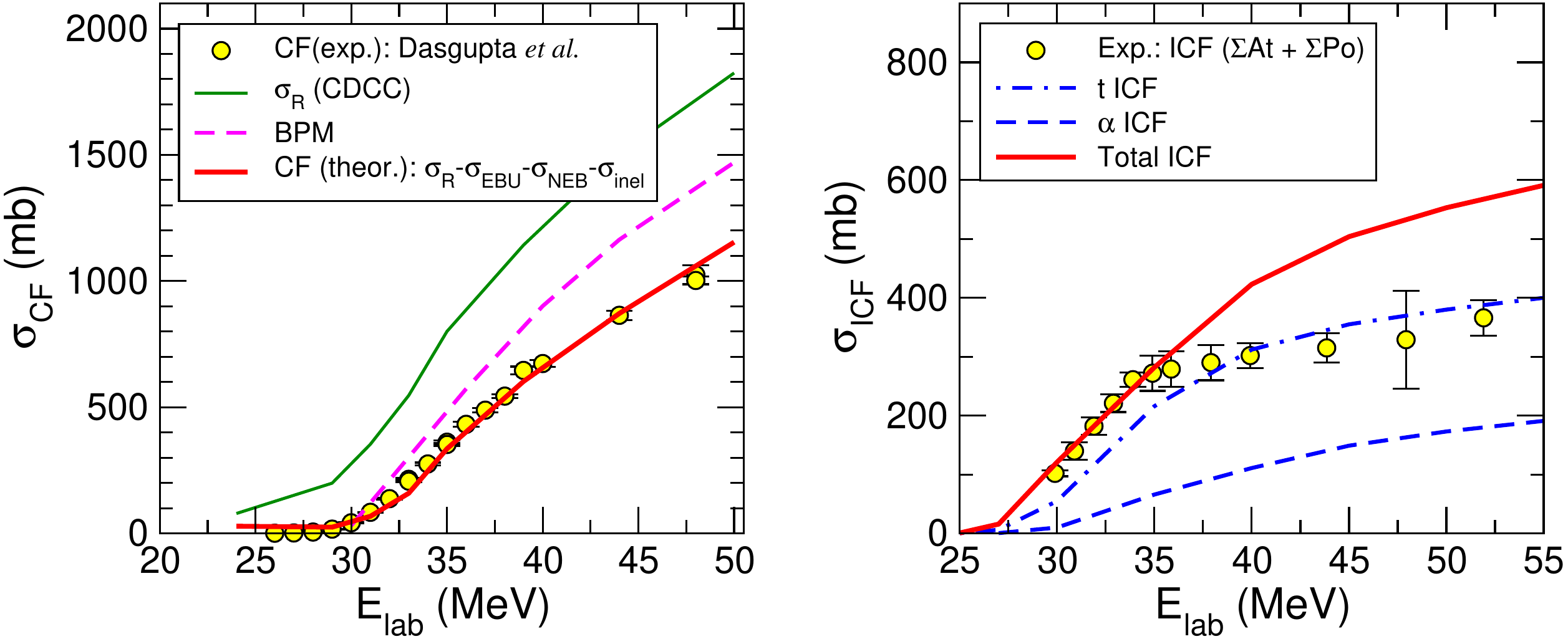}
\end{center}
\caption{\label{fig:li7bi_iav} CF (left) and ICF (right) cross sections for $^{7}$Li+$^{209}$Bi. Experimental data from \cite{Das02,Das04} are compared with the calculations based on the IAV model. In the CF plot, we include also the reaction cross section (from a CDCC calculation) and the fusion computed with the barrier penetration model (BPM).} 
\end{figure}

A tentative application of this idea is shown in the right panel of  Fig.~\ref{fig:li7bi_iav}, corresponding to the reaction  $^{7}$Li+$^{209}$Bi. The symbols correspond to the ICF data of Dasgupta {\it et al.}~\cite{Das02,Das04}. For the calculations we show the individual contributions to the ICF cross section, namely,  $\alpha$-ICF (i.e., $\alpha$ absorbed) and $t$-ICF ($t$-absorbed) as well as their sum.  To compute the $\alpha$-ICF ($t$-ICF), the imaginary part of the $\alpha$+$^{209}$Bi ($t$+$^{209}$Bi) system was replaced by a short-range imaginary potential of Woods-Saxon form and parameters $W_0=50$~MeV, $r_i=1.0$~fm, $a=0.2$~fm.  The results are very similar to those reported  in Ref.~\cite{Cor20} and shown in Fig.~\ref{fig:li7bi_canto}, in which the authors made use of the absorption and survival probabilities extracted from the CDCC calculations. Further calculations are needed to elucidate the usefulness and applicability of the IAV model to evaluate ICF cross sections. 

The IAV model has been also used to infer CF cross sections of weakly bound nuclei \cite{Jin19a,Jin19b}. The idea of this method is to decompose the reaction cross section as follows    
 \begin{equation}
 \label{eq:decomp}
 \sigma_R \approx    \sigma_\mathrm{CF} + \sigma_\mathrm{inel} + \sigma_\mathrm{EBU} +
 \sigma^{(b)}_\mathrm{NEB} + \sigma^{(x)}_\mathrm{NEB} .
 \end{equation}
In this expression, $\sigma_\mathrm{inel}$ corresponds to the excitation of the projectile and/or target without dissociation (i.e., inelastic scattering). The terms $\sigma_\mathrm{EBU}$ and $\sigma^{(b,x)}_\mathrm{NEB}$ correspond to the EBU and NEB contributions. In the latter, one distinguishes the cases in which either the fragment $x$ or $b$ interacts nonelastically with the target whereas the other scatters elastically.  
A successful determination of the CF cross section from the decomposition (\ref{eq:decomp}) requires that all other quantities involved in this formula can be evaluated accurately. The pure inelastic scattering cross sections ($\sigma_\mathrm{inel}$) are standardly computed by means of coupled-channels calculations including low-lying collective excitations of the projectile and target (see Sec.~\ref{sec:collect}).  The EBU part  can be accurately calculated, for instance, using the continuum-discretized coupled-channels (CDCC) method discussed 
in Sec. 3.   Finally, the NEB contributions can be evaluated with the IAV model, at least for projectiles with a developed two-boy structure, such as  $^{6,7}$Li.

An application of the method to the $^{7}$Li+$^{209}$Bi reaction is shown in the left panel of Fig.~\ref{fig:li7bi_iav}, adapted from  Ref.~\cite{Jin19a}. The circles are the CF data from Ref.~\cite{Das04}, the solid green line is the reaction cross section obtained from a CDCC calculation and the solid red line the calculated CF cross section inferred from Eq.~(\ref{eq:decomp}) assuming a two-body model ($\alpha+t$) for $^{7}$Li. For comparison, a single-channel barrier penetration model (BPM) calculation (dashed lined) is shown. As is seen, the data are largely suppressed with respect to the BPM. By contrast, the CF extracted from Eq.~(\ref{eq:decomp}) explains very well the data.  As discussed in Ref.~\cite{Jin19a}, the reduction with respect to the BPM is found to be mostly due to the competition due to the $^{209}$Bi($^{7}$Li, $\alpha$)X channel, which includes, among others, the $t$-ICF channel (see the right panel).  

\subsubsection{Origin of the ICF large cross sections}
\begin{figure}[!ht]
\begin{center}
 {\centering \resizebox*{0.5\columnwidth}{!}{\includegraphics{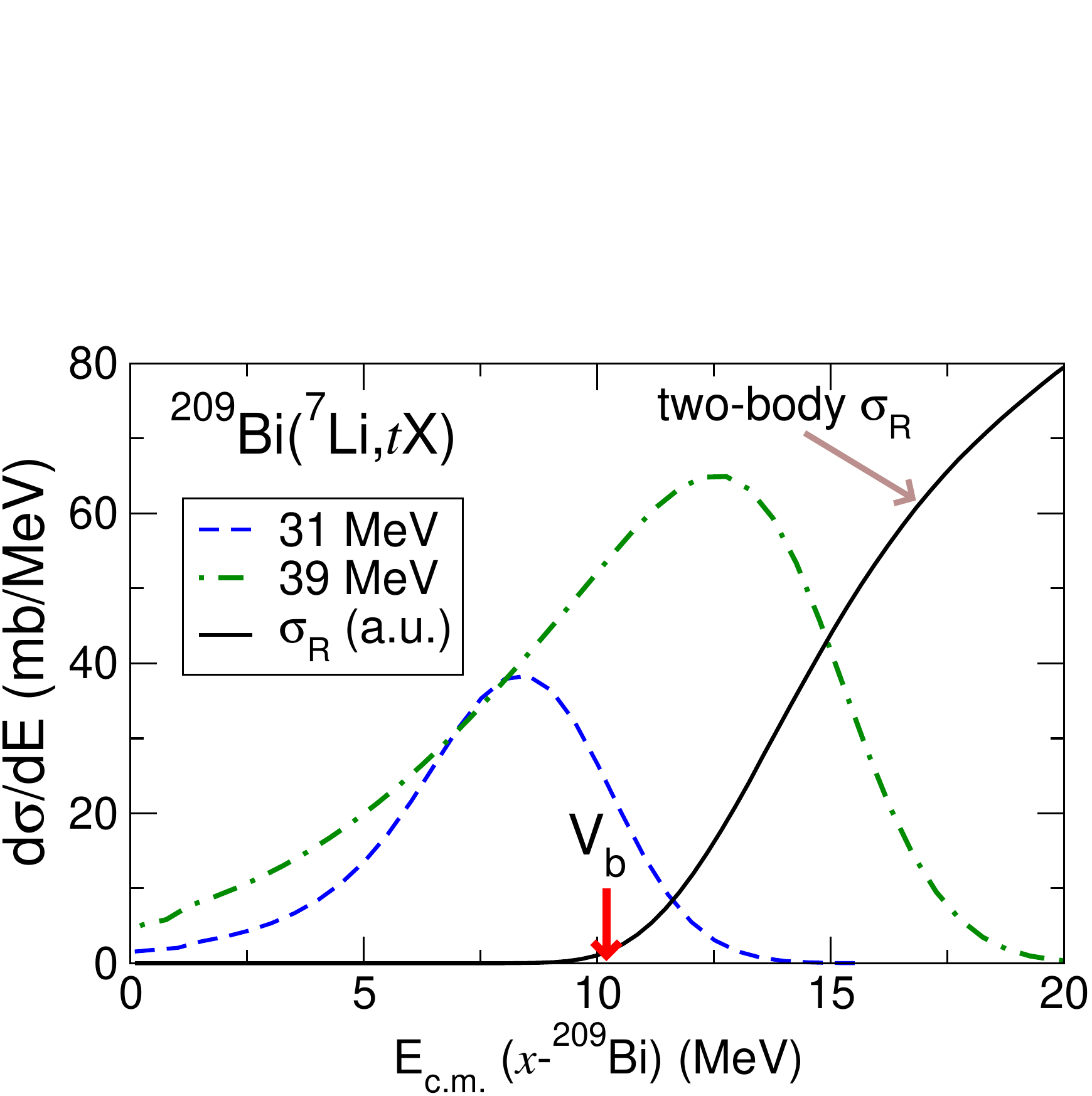}} \par}
\caption{\label{fig:thm} Illustration of the Trojan Horse mechanism for the  $^{209}$Bi($^{7}$Li,\,$\alpha$$X$)  reaction. The broken lines are the three-body cross sections ($^{7}$Li,\,$\alpha$$X$) at the  incident energies indicated by the labels. The black solid lines are the two-body reaction cross sections (in arbitrary units). The nominal position of the barrier for the $t$+$^{209}$Bi reactions is indicated by the arrow. }
\end{center}
\end{figure}

As shown in Refs. \cite{Jin15,Jin17,Jin19a}, the elastic dissociation (EBU) mechanism yields a  comparatively small contribution to the total inclusive breakup cross section in reactions induced by weakly bound nuclei  such as deuterons or $^{6,7}$Li. Consequently, the widely accepted hypothesis that the CF suppression arises as due to the two-step mechanism, consisting on the elastic dissociation of the projectile followed by the capture of one of the produced fragments does not seem to be supported by these calculations. In fact, IAV calculations performed for the $^{7}$Li+$^{209}$Bi reaction using the three-body version of the IAV method, Eq.~(\ref{eq:inh_3b}), in which the $^{7}$Li continuum is explicitly included, did not find significant differences with respect to the DWBA IAV calculations \cite{Jin19b} for the calculated $\alpha$ production yields. In fact, the calculations of Ref.~\cite{Jin19a} suggest that the large $\alpha$ yields observed in $^{6,7}$Li reactions can be explained in terms of a {\it Trojan horse mechanism}. 
The idea is that, for a three-body reaction of the form $a+A$, with $a=b + x$, a particular channel of the form $a+A \to b+ c +C$ will be enhanced  
with respect to the free, two-body reaction $x+A \to c + C$ due to the fact that the $a+A$ system is above its Coulomb barrier. Loosely speaking, the $x$ particle is brought inside its Coulomb barrier by the heavier particle $a$.
 The Trojan Horse method has become a standard tool in nuclear astrophysics as an indirect way of obtaining information of low-energy charged-particle induced reactions by means of three-body reactions (see e.g.~\cite{Spi11}) and its formal aspects can be found elsewhere \cite{Typ03}. We illustrate in Fig.~\ref{fig:thm} the phenomenon for the $^{7}$Li+$^{209}$Bi at hand. For that  we compare the reaction cross sections for the {\it two-body} reaction $t$+$^{209}${Bi}, as a function of the center-of-mass energy for each system, with the {\it three-body} cross sections  $^{209}$Bi($^{7}$Li,\, $\alpha$$X$)  for two different $^{7}$Li incident energies. The vertical arrow indicates the position of the Coulomb barrier for the $t$+$^{209}${Bi} system. As expected, the reaction cross section for the two-body reaction drops very quickly as the energy decreases and approaches the Coulomb barrier. By contrast, the three-body cross sections remain very large, even at energies well below  their nominal barrier. These results provide a natural explanation of the large  $\alpha$  yields observed experimentally and confirmed by the IAV model.

 The picture that emerges from these calculations is the following. The weakly-bound projectile $a$ overcomes the $a+A$ Coulomb barrier, bringing also the $x$ fragment inside its Coulomb barrier via the just described Trojan Horse mechanism. This triggers the non-elastic processes between $x$ and $A$ which give rise to the large variety of emerging fragments observed experimentally and, in turn, to the suppression of CF. The present results add numerical support to the suggestion put forward by Cook {\it et al.}~\cite{Coo18}.

\subsubsection{Application to surrogate reactions}
\label{sec:surrogate}
The gradual improvement in the models oriented to the computation of ICF cross sections has driven their applicability of the so-called {\it surrogate method} (SRM). This method provides an indirect way of evaluating compound-nucleus cross sections in reactions for which the direct measurement is difficult or even not possible. An example is the extraction of neutron-induced cross sections of the form $(n,\chi)$, where $\chi$ is a given decay channel product ($\gamma$, fission fragment, etc). Following the Bohr hypothesis, it is customary assumed that in these reactions the formation and decay of a compound nucleus take place independently of each other. In order to obtain information on the decay of the compound nucleus ($B^*$) occurring in the reaction of interest ($a + A \to B^* \to c + C$), one uses the alternative (surrogate) reaction $a +A \to b+ B^*$ that involves a projectile-target combination ($a+A$) that is experimentally more accessible. 

We consider for definiteness the extraction of $(n,\chi)$ cross sections from a $(d,pX)$ surrogate reaction.  Compound-nuclear reactions are properly described in the Hauser-Feshbach formalism, which considers the  conservation of angular momentum $J$ and parity $\pi$.  The cross section for the ``desired'' reaction $A(n,\chi)$ is given by
\begin{eqnarray}
\sigma_{(n,\chi)}(E_{n}) &=& \sum_{J_T,\pi}  \sigma_{J_T,\pi}^{CN}(E_{ex},J_T,\pi) \;\; G_{\chi}^{CN}(E_{ex},J_T,\pi) \; ,
\label{eq:DesReact}
\end {eqnarray}
\noindent
where $ \sigma_{J_T,\pi}^{CN}(E_{ex},J_T,\pi)$ is the cross section for the CN formation and  $G_{\chi}^{CN}(E_{ex},J_T,\pi)$ the branching ratio for the decay in the channel $\chi$. Note that the excitation energy  $E_{ex}$ of the compound nucleus $B^*$ is related to the center-of-mass energy $E_a$ in the entrance channel via the energy needed for separating $a$ from $B$: $E_a=E-S_a(B)$.  
The objective of the surrogate method is to experimentally determine or constrain the decay probabilities $G_{\chi}^{CN}(E_{ex},J_T,\pi)$, which are often difficult to calculate accurately.

In the surrogate reaction,  the same CN nucleus $B^*$ is formed and the decay product of interest ($\chi$) is measured in coincidence with the outgoing particle $b$. The probability for this process can be written as:
\begin{equation}
P_{S,\chi}(E_{ex}) = \sum_{J_T,\pi} F_{S}^{CN}(E_{ex},J_T,\pi) \; G_{\chi}^{CN}(E_{ex},J_T,\pi) \; ,
\label{Eq:SurReact}
\end{equation}
where the subscript $S$ denotes the specific surrogate reaction (in this case, $A(d,p)B^*$), $F_{S}^{CN}(E_{ex},J_T,\pi)$ is the probability of forming $B^*$ in this surrogate reaction (with specific values of $E_{ex}$, $J$ and $\pi$) and $G_{\chi}^{CN}(E_{ex},J_T,\pi)$ is the same branching ratio appearing in the direct reaction (\ref{eq:DesReact}). The probability $P_{S, \chi}(E_{ex})$ can be obtained experimentally as the ratio between  the number of coincidences between the $b$ particle and the decay particle $\chi$, $N_{S,\chi}$, and the total number of surrogate events, $N_{S}$, i.e.:
\begin{equation}
P^{exp}_{S,\chi}(E_{ex}) = \frac{N_{S,\chi}}{N_{S} \, \epsilon_\chi} \; ,
\label{Eq:CoincProb}
\end{equation}
where $\epsilon_\chi$ is the efficiency of detecting the exit-channel $\chi$ for the reactions in which $b$ is detected. 

Ideally, if a reliable prediction of $F_{S}^{CN}(E_{ex},J_T,\pi)$ is possible, with an accurate determination of $P^{exp}_{S,\chi}(E_{ex})$ for a range of energies and angles of $b$, one might be able to extract  $G_{\chi}^{CN}(E_{ex},J_T,\pi)$ which can be then used to calculate the desired cross section by means of (\ref{eq:DesReact}). In practice, this approach is not always feasible due to the lack of some of this required information and the approach has relied on additional approximations. In particular, most practical applications have made use of the so-called ``Weisskopf-Ewing   approximation'', which assumes that the branching ratios $ G_{\chi}^{CN}(E_{ex},J_T,\pi)$ are independent of the angular momentum and spin, giving rise to the simplified cross section:
\begin{eqnarray}
\sigma_{(n,\chi)}(E_{n}) &=& \sigma^{CN}(E_{ex}) \; G_{\chi}^{CN}(E_{ex}) \; ,
\label{eq:WE}
\end {eqnarray}
where $\sigma^{CN}(E_{ex})$ is to be understood as the CN cross section summed over all possible $J_T,\pi$ values. Applying the same approximation to the surrogate reaction, and using $\sum_{J_T,\pi} F_{S}^{CN}(E_{ex},J_T,\pi)=1$ we have
\begin{equation}
P_{S,\chi}(E_{ex}) = G_{\chi}^{CN}(E_{ex}) \; ,
\label{Eq:SurReact_WE}
\end{equation}
allowing the determination of the desired cross section as
\begin{eqnarray}
\sigma_{(n,\chi)}(E_{n}) &=& \sigma^{CN}(E_{ex}) \; P_{S,\chi}(E_{ex}) \; ,
\label{eq:WE_2}
\end {eqnarray}
which avoids the need of the probabilities $F_{S}^{CN}(E_{ex},J_T,\pi)$.

\begin{figure}[!ht]
\begin{center}
 {\centering \resizebox*{0.5\columnwidth}{!}{\includegraphics{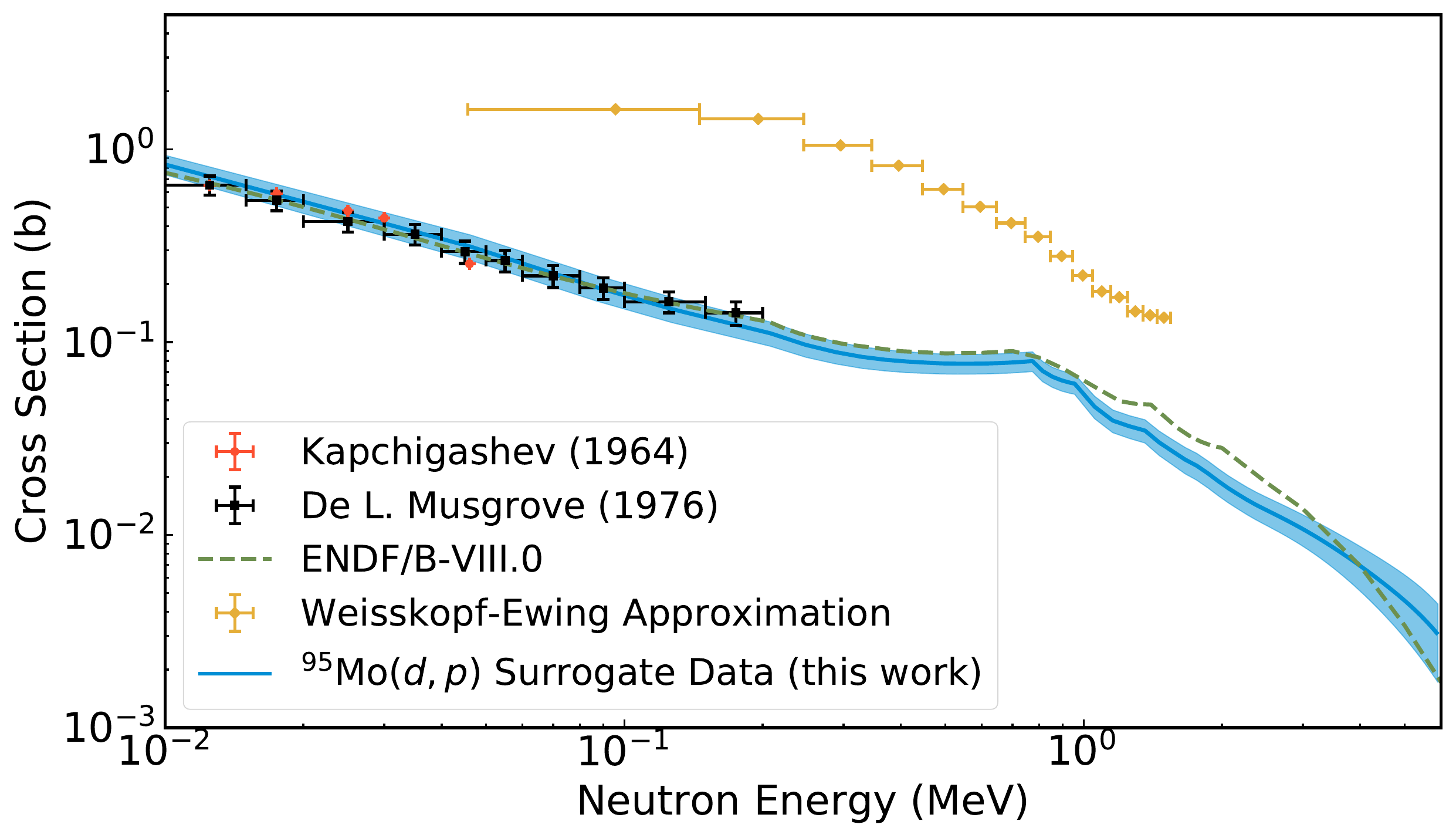}} \par}
\caption{ \label{fig:surrogate_mo95} Application of the surrogate method (SRM) to  the $^{95}$Mo$(n,\gamma)$ reaction. The solid blue region is the $(n,\gamma)$ cross section obtained from the SRM, the red circles and black squares are the  direct measurements \cite{Abr08,Mus76}. The uncertainty due to experimental data and fitting error is indicated by the shaded band. The result obtained using the Weisskopf-Ewing   approximation is also shown (gold diamonds). Quoted from Ref. \cite{Rat19}.}
\end{center}
\end{figure}

One of the few cases in which the idealized SRM approach has been successfully applied is shown in Fig.~\ref{fig:surrogate_mo95}, taken from Ref.~\cite{Rat19}. In this case, the important required probability $F_{S}^{CN}(E_{ex},J_T,\pi)$ was provided by the IAV model. The extracted $(n,\gamma$) cross section is in excellent agreement with the direct measurement (black and red squares). As shown also in this figure, the result using Weisskopf-Ewing approximation departs significantly from the direct measurement.

\section{Summary}
The coupled-channels method has been a standard tool to analyze and 
interpret experimental data for nuclear reactions. Unlike the distorted wave 
Born approximation (DWBA), it treats couplings to excited states in the colliding 
nuclei to all orders in the coupling strengths, 
and thus it automatically describes the interplay among several nuclear reaction 
processes.  
Moreover, a microscopic optical potential can also be formulated 
via the coupled-channels approach. 
In this paper, we have reviewed the current status and the recent developments of 
the coupled-channels approach to nuclear reactions. 

It has been known well that the channel coupling effects play an important role in 
heavy-ion fusion reactions of medium-heavy systems 
at energies around the Coulomb barrier. There, the fusion cross sections are 
largely enhanced due to the channel coupling effects as compared to the prediction 
of single-channel calculations. 
An interesting observation is that the standard coupled-channels calculations, which 
well account for fusion cross sections around the Coulomb barrier, overestimates fusion 
cross sections at deep subbarrier energies. To explain this, a model has been proposed 
in which the channel coupling effects are effectively quenched after  two nuclei 
touch with each other. 

The channel coupling effects also play an important role in nuclear astrophysics and 
in fusion for superheavy elements. For instance, it has been shown that 
the number of resonance peaks 
increase significantly in the $^{12}$C+$^{12}$C fusion reaction due to the coupling 
to the first 2$^+$ state in $^{12}$C. For superheavy elements, it has been 
pointed out that the channel coupling effects play a crucial role in 
determining the optimum energy for fusion with a deformed target nucleus. 

Conventionally, the coupled-channels approach has been formulated with 
the macroscopic collective models. Such treatment has limitations, since one 
often has 
to assume the harmonic oscillator for vibrations  and the rigid rotor for 
rotational motions. In this regard, the (semi-)microscopic coupled-channels 
approach has been developed. In this approach, excited states of the colliding 
nuclei are described with microscopic nuclear structure calculations. The internucleus 
potentials may also be constructed microscopically based on the double folding 
model. This method has been applied both to inelastic scattering and to subbarrier 
fusion reactions. In particular, for inelastic scattering, this approach has been 
successfully applied to extract information on cluster structure in light nuclei. 

In weakly-bound nuclei, the breakup process plays an important role. To describe such process, the coupled-channels approach has to be extended to include excitations to the continuum spectrum. For this purpose, the continuum-discretized coupled-channels (CDCC) method has been widely used.  Although the method was originally devised to describe deuteron elastic and breakup reactions, it was soon extended to other weakly-bound systems and observables, including reactions induced by other weakly bound stable (e.g. $^{6,7}$Li) as well as two- and three-body halo nuclei (such as $^{11}$Be, $^{19}$C, $^{6}$He and $^{11}$Li) impinging on protons and composite target nuclei. Another advance has been the inclusion of core degrees of freedom and the use of microscopic fragment-target interactions. Additionally, the method has been successfully employed in transfer (e.g. $(d,p)$), proton-induced knockout reactions of the form $(p,pn)$ and $(p,2p)$ and  fusion  reactions of weakly bound nuclei at near-barrier energies, which have been found to be affected by the breakup channels. In recent years, several models based on the CDCC method have been proposed to the evaluation of incomplete fusion cross sections, which has remained a challenging and elusive problem for many years. The revival of the inclusive breakup models first proposed in the 1980s, provides also a promising tool to the evaluation of incomplete fusion cross sections. 

In this way, the coupled-channels approach has played a central role in 
nuclear reaction physics. However, at the same time, 
there still remain many theoretical 
challenges. For instance, it has still been difficult to include breakup, transfer, 
and inelastic excitations simultaneously. This is particularly the case with 
pair transfer and multi-nucleon transfer channels, which is relevant to reactions of 
exotic nuclei with neutron skins. Moreover, it has also been a theoretical 
challenge to develop a many-body theory to describe reaction processes involving 
high excitations, in which the concepts of diffusion and friction are relevant. 
This is particularly important  for heavy-ion deep inelastic collisions and fusion for superheavy nuclei. Even though some of these issues may need to wait for 
a development of powerful computers, several works are now in progress towards 
these goals. 

Another important remark is that some of the developments described in this 
paper, such as CDCC with core and target excitations, the four-body CDCC calculations, nonelastic breakup (IAV), and multiple scattering methods, are not currently available in a form of publicly available codes. It would be advisable that these codes become available in future once the formalisms will have been consolidated and tested, so future generations can employ these formalisms and developments. 

\bigskip

\noindent
{\bf Acknowledgements} 

We thank all our collaborators 
for discussions and fruitful collaborations. 
This work was 
supported in part by JSPS KAKENHI
Grant Numbers JP19K03861, JP21H00120, JP17740148, JP22740169, JP22560820, JP25400255, JP16K05352, and in part by the COREnet program of RCNP, Osaka University. 
A.M.M.\ is supported by the Spanish Ministerio de Ciencia e Innovaci\'on y Universidades under project PID2020-114687GB-I00 and by the Consejería de Econom\'ia, Conocimiento, Empresas y Universidad (Junta de Andaluc\'ia) under project  P20\_01247.

\begin{appendix}
\section{Details of the microscopic coupled-channels method}
\label{secappa}

In this appendix, we describe the MCC framework for direct reactions, elastic and inelastic scattering in particular, based on the MST. Some formulae appearing below are essentially the same as those in Sec.~\ref{sec:cca}. Nevertheless, we intend to reformulate them with explicit indices to show unambiguously the representation of the structural input, i.e., transition densities of Eq.~(\ref{3trden}), and its implementation in the MCC reaction framework. 

\subsection{Multiple scattering theory}
\label{secappa1}

The basis of the MCC method discussed in this article is the MST developed by Foldy~\cite{foldy1945}, Watson~\cite{watson1953}, and Kerman, Mc-Manus, and Thaler~\cite{kerman1959}. Below we briefly recapitulate the MST. Let us consider an NA scattering process. The Sch\"{o}dinger equation is written by%
\begin{equation}
\left(  T_{\mathrm{NA}}+\sum_{i}v_{i}+H_{\mathrm{A}}-E\right)  \Psi^{\left(
+\right)  }=0, \label{3sc}%
\end{equation}
where $T_{\mathrm{NA}}$\ is the kinetic energy operator regarding the NA relative motion, $v_{i}$ is the NN bare interaction between the incident nucleon and the $i$th nucleon in A, $H_{\mathrm{A}}$ is the internal Hamiltonian of A, $E$ is the total energy of the system. For simplicity, we omit many-nucleon forces in Eq.~(\ref{3sc}), and the Coulomb interaction is disregarded in the formulae shown below. The total wave function is denoted by $\Psi^{\left(  +\right)  }$, which contains the complete information on the reaction system. Although it is represented only implicitly, $\Psi^{\left(+\right)  }$\ involves components corresponding to all possible boundary conditions.

As the starting point of the MST, we restrict ourselves to consider the process whose boundary condition is expressed by the following LS equation.%
\begin{equation}
\bar{\Psi}^{\left(  +\right)  }=\phi+G_{0}^{\left(  +\right)  }\sum_{i}%
v_{i}\bar{\Psi}^{\left(  +\right)  }, \label{3ls}%
\end{equation}
where $\phi$ is the plane wave and 
$G_0^{(+)}$ is the Green function given by
\begin{equation}
G_{0}^{\left(  +\right)  }=\frac{1}{E-T_{\mathrm{NA}}-H_{\mathrm{A}%
}+i\varepsilon}.
\end{equation}
It is then understood that one can impose the boundary condition
\begin{equation}
\bar{\Psi}^{\left(  +\right)  }\rightarrow\phi+\sum_{c}f_{c}\frac{e^{iKR}}{R}
\label{3bc}%
\end{equation}
when the displacement $R$ of the incident nucleon from the c.m. of A is sufficiently larger than the range $R_{\mathrm{N}}$ of the strong interaction. The scattering amplitude is denoted by $f_{c}$ with $c$ being the channel index. The important point is that $\bar{\Psi}^{\left(+\right)  }$ is a subset of $\Psi^{\left(  +\right)  }$ because of the restriction of the boundary condition. For instance, $\bar{\Psi}^{\left(+\right)  }$ does not explicitly include the rearrangement channels.

The transition matrix corresponding to Eq.~(\ref{3ls}) is given by%
\begin{equation}
T=\sum_{i}\Lambda_{i},
\end{equation}%
with
\begin{equation}
\Lambda_{i}\equiv v_{i}+v_{i}G_{0}^{\left(  +\right)  }\sum_{j}v_{j}%
+v_{i}G_{0}^{\left(  +\right)  }\sum_{j}v_{j}G_{0}^{\left(  +\right)  }%
\sum_{k}v_{k}+... \label{3lambda}%
\end{equation}
where $i,$ $j,$ $k,$ $...$ are the indices of the nucleons in A. The essence of the MST is the resummation of the series in Eq.~(\ref{3lambda}) as follows. First, we rewrite Eq.~(\ref{3lambda}) as%
\begin{equation}
\Lambda_{i}=v_{i}+v_{i}G_{0}^{\left(  +\right)  }\Lambda_{i}+v_{i}%
G_{0}^{\left(  +\right)  }\sum_{j\neq i}\Lambda_{j},
\end{equation}
which reads%
\begin{equation}
\Lambda_{i}=\frac{1}{1-v_{i}G_{0}^{\left(  +\right)  }}v_{i}+\frac{1}%
{1-v_{i}G_{0}^{\left(  +\right)  }}v_{i}G_{0}^{\left(  +\right)  }\sum_{j\neq
i}\Lambda_{j}.
\end{equation}
By introducing the effective interaction $\tau_{i}$ defined through%
\begin{equation}
\tau_{i}=v_{i}+v_{i}G_{0}^{\left(  +\right)  }\tau_{i}, \label{3effint}%
\end{equation}
$\Lambda_{i}$ is written by%
\begin{equation}
\Lambda_{i}=\tau_{i}+\tau_{i}G_{0}^{\left(  +\right)  }\sum_{j\neq i}%
\Lambda_{j}. \label{3lambda2}%
\end{equation}
One may find that Eq.~(\ref{3lambda2}) is a series of multiple scattering with respect to $\tau_{i}$, whereas Eq.~(\ref{3lambda}) to $v_{i}$. Because $\tau_{i}$ takes into account infinite number of interacting processes between the incident nucleon and the $i$th nucleon in A, it is much easier to handle than $v_{i}$ as $\tau_{i}$ has no repulsive core.

The transition matrix with the resummation reads%
\begin{equation}
T=\sum_{i}\tau_{i}+\sum_{i}\tau_{i}G_{0}^{\left(  +\right)  }\sum_{j\neq
i}\tau_{j}+\sum_{i}\tau_{i}G_{0}^{\left(  +\right)  }\sum_{j\neq i}\tau
_{j}G_{0}^{\left(  +\right)  }\sum_{k\neq j}\tau_{k}+...
\end{equation}
If the wave function of A to be operated by this $T$ is antisymmetrized regarding the exchange of each nucleon-pair inside A, we obtain%
\begin{equation}
T=\frac{A}{A-1}\bar{T}, \label{3tbar}%
\end{equation}%
\begin{equation}
\bar{T}=\bar{\tau}+\bar{\tau}G_{0}^{\left(  +\right)  }\bar{T},
\end{equation}%
\begin{equation}
\bar{\tau}=\frac{A-1}{A}\tau, \label{3taubar}%
\end{equation}
where $A$ is the number of nucleons in A; we have dropped the index $i$ of $\tau$. One can interpret this result of the MST as a rewriting of Eq.~(\ref{3sc}) to%
\begin{equation}
\left[  T_{\mathrm{NA}}+A\bar{\tau}+H_{\mathrm{A}}-E\right]  \bar{\Psi
}^{\left(  +\right)  }=0 \label{3sc2}%
\end{equation}
with the modification of Eq.~(\ref{3tbar}) to the resulting transition matrix $\bar{T}$. Equation (\ref{3sc2}) makes sense, however, only for the scattering process governed by the asymptotic condition of Eq.~(\ref{3bc}).

\subsection{Coupled-channel equations based on the MST}
\label{secappa2}

Let us concentrate for a while on the NA elastic and inelastic processes. For simplicity, we do not take the intrinsic spin of the incident nucleon into account. We first expand the total wave function $\bar{\Psi}^{\left(+\right)}$ in terms of the eigenstates $\Phi_{nIm_{I}}$ of $H_{\mathrm{A}}$, where $n$, $I$, $m_{I}$ are the energy index, the total spin, its third component, respectively, of A; the parity of A is understood to be specified but not shown in the subscript of $\Phi$. Here, it is assumed that $\Phi_{nIm_{I}}$ are either bound states or resonant states with narrow widths. When a broad resonance comes into play, it is not isolated well from continuum states near the resonant energy, and we need to consider both resonant and nonresonant continuum states on the same footing. Such cases are discussed in Sec.~\ref{sec33}.

To derive CC equations to be solved, we make the partial-wave expansion of $\bar{\Psi}^{\left(  +\right)  }$. The partial scattering wave characterized by the total angular momentum $J_T$ and its third component $M_{T}$ is defined by%
\begin{align}
\bar{\Psi}^{\left(  J_TM_{T}\right)  }_{n_0 I_0 L_0}\left(  \boldsymbol{R},\xi\right)   &
=\sum_{nIL}\frac{\chi_{nIL,n_0I_0L_0}^{\left(  J_T\right)  }\left(  K_{nI},R\right)  }%
{KR}\sum_{m_{I},m_L}\langle  Im_{I}Lm_L|J_TM_{T}\rangle  i^{P}\Phi_{nIm_{I}}\left(
\xi\right)  i^{L}Y_{Lm_L}(  \boldsymbol{\hat{R}} ) \nonumber\\
&  =\sum_{nIL}\frac{\chi_{nIL,n_0I_0L_0}^{\left(  J_T\right)  }\left(  K_{nI},R\right)
}{KR}\left[  i^{P}\Phi_{nI}\left(  \xi\right)  \otimes i^{L}Y_{L}(  \boldsymbol{\hat{R}} )  \right]  _{J_TM_{T}}. \label{3psijm}%
\end{align}
Here, $\boldsymbol{R}$ and $\xi$ are the NA relative coordinate and all internal coordinates of A, respectively, $P=0$ (1) for the positive (negative) parity state of A. $\Phi_{nIm_{I}}$ satisfies%
\begin{equation}
H_{\mathrm{A}}\Phi_{nIm_{I}}\left(  \xi\right)  =\epsilon_{nI}\Phi_{nIm_{I}%
}\left(  \xi\right)  ,
\end{equation}
where $\epsilon_{nI}$ is the eigenenergy. The NA scattering wave function is denoted by $\chi_{nIL,n_0I_0L_0}^{\left(  J_T\right)  }$, where $L$ is the relative orbital angular momentum. 
The NA relative wave number $K_{nI}$ is defined by%
\begin{equation}
K_{nI}=\frac{\sqrt{2\mu E_{nI}}}{\hbar},
\end{equation}
where $\mu$ is the NA reduced mass and%
\begin{equation}
E_{nI}=E-\epsilon_{nI}.
\end{equation}
In the present case, the channel is characterized by $n$, $I$, $L$; when convenient, these are abbreviated by $c$ below. Similarly, the incident channel is represented by $c_{0}\equiv\left(  n_{0},I_{0},L_{0}\right)  $. For simplicity, $K_{nI}$ in the incident channel, $K_{n_{0}I_{0}}$, is denoted by $K$ in Eq.~(\ref{3psijm}).

Multiplying the Schr\"{o}dinger equation for $\bar{\Psi}^{\left(  J_TM_{T}\right)
}_{n_0 I_0 L_0}$%
\begin{equation}
\left[  -\frac{\hbar^{2}}{2\mu}\frac{1}{R}\frac{d^{2}}{dR^{2}}R+\frac{\boldsymbol{\hat
{L}}^{2}}{2\mu R^{2}}+A\bar{\tau}+H_{\mathrm{A}}-E\right]  \bar{\Psi
}^{\left(  J_TM_{T}\right)  }_{n_0 I_0 L_0}\left(  \boldsymbol{R},\xi\right)  =0
\end{equation}
by $\left[  i^{P^{\prime}}\Phi_{n^{\prime}I^{\prime}}\left(  \xi\right)
\otimes i^{L^{\prime}}Y_{L^{\prime}}(  \boldsymbol{\hat{R}} )
\right]  _{J_TM_{T}}^{\ast}$ from the left, performing the integration over $\xi$ and $\boldsymbol{\hat{R}}$, one gets%
\begin{equation}
\left[  -\frac{\hbar^{2}}{2\mu}\frac{d^{2}}{dR^{2}}+\frac{\hbar^{2}}{2\mu}
\frac{\left(  L^{\prime}\right)  ^{2}}{R^{2}}-E_{n^{\prime}%
I^{\prime}}\right] \chi_{c^{\prime}c_0}^{\left(  J_T\right)  }\left(
k_{n^{\prime}I^{\prime}},R\right)=-\sum_{c}U^{(J_T)}_{c^{\prime}c}\left(
R\right)  \chi_{cc_0}^{\left(  J_T\right)  }\left(  k_{nI},R\right)  ,
\label{3cceq}%
\end{equation}
where the coupling potential is given by%
\begin{equation}
U^{(J_T)}_{c^{\prime}c}\left(  R\right)  =
\left.\left\langle \left[  i^{P^{\prime}}%
\Phi_{n^{\prime}I^{\prime}}\left(  \xi\right)  \otimes i^{L^{\prime}%
}Y_{l^{\prime}}(  \boldsymbol{\hat{R}} )  \right]  _{J_TM_{T}%
}\right\vert A\bar{\tau}\left\vert \left[  i^{P}\Phi_{nI}\left(  \xi\right)
\otimes i^{L}Y_{L}(  \boldsymbol{\hat{R}} )  \right]  _{J_TM_{T}%
}\right\rangle _{\xi,\boldsymbol{\hat{R}}}.\right.
\label{3ucc}%
\end{equation}
Note that $U^{(J_T)}_{c^{\prime}c}$ is independent of $M_{T}$ as seen 
below (see also Sec. 2.2). The CC equations (\ref{3cceq}) are solved under the boundary condition%
\begin{equation}
\chi_{c^{\prime}c_0}^{\left(  J_T\right)  }\left(  K_{n^{\prime}I^{\prime}%
},R\right)  \rightarrow\frac{i}{2}\left[  KR h_{L^{\prime}}^{\left(  -\right)
}\left(  KR\right)  \delta_{c^{\prime}c_{0}}-\sqrt{\frac{K}{K_{n^{\prime
}I^{\prime}}}}S_{c^{\prime}}^{\left(  J_T\right)  }K_{n^{\prime}I^{\prime}%
}R h_{L^{\prime}}^{\left(  +\right)  }\left(  K_{n^{\prime}I^{\prime}}R\right)
\right]  ,\quad\left(  \text{for }R>R_{\mathrm{N}}\right)  , \label{3open}%
\end{equation}
where $h_{L^{\prime}}^{\left(  -\right)  }$ and $h_{L^{\prime}}^{\left(+\right)  }$ are the incoming and outgoing Hankel functions, respectively. The scattering matrix to the $c^{\prime}$ channel is denoted by $S_{c^{\prime}}^{\left(  J_T\right)  }$, with which scattering observables can be evaluated in the standard manner. It should be noted that we only consider the open channels, for which $E_{nI}>0$. It is known that at low energies, the coupling with the closed channels corresponding to $E_{nI}<0$\ becomes crucial~\cite{austern1987,ogata2016}. We do not apply the MCC framework to that case because the MST itself will be questionable there.

\subsection{Coupling potentials}
\label{secappa3}

Now we describe the form of the coupling potential $U_{c^{\prime}c}$. Let us consider using a G-matrix interaction $g$ as $\tau$. Similarly to $\bar{\tau}$, we use%
\begin{equation}
\bar{g}\equiv\frac{A-1}{A}g=\bar{g}^{\mathrm{dr}}+\bar{g}^{\mathrm{ex}}\hat
{P}^{\mathrm{ex}},
\end{equation}
where $\bar{g}^{\mathrm{dr}}$ and $\bar{g}^{\mathrm{ex}}$\ are the direct and exchange parts, respectively, $\hat{P}^{\mathrm{ex}}$\ is the operator that exchanges the incident nucleon and a nucleon in A. We concentrate on the central part of $U_{c^{\prime}c}$ and do not differentiate proton and neutron in the following formulation. We here introduce the one-body transition density of A defined by%
\begin{equation}
\rho_{n^{\prime}I^{\prime}m_{I}^{\prime}nIm_{I}}^{\mathrm{tr}}\left(
\boldsymbol{r}_{\mathrm{A}}\right)  \equiv\int\Phi_{n^{\prime}I^{\prime}%
m_{I}^{\prime}}^{\ast}\left(  \xi\right)  \sum_{i=1}^{A}\delta\left(
\boldsymbol{r}_{i}-\boldsymbol{r}_{\mathrm{A}}\right)  \Phi_{nIm_{I}}\left(
\xi\right)  d\xi=\sum_{\lambda\mu}
\langle  Im_{I}\lambda\mu|I^{\prime}%
m_{I}^{\prime}\rangle  \rho_{n^{\prime}I^{\prime}nI,\lambda}^{\mathrm{tr}%
}\left(  r_{\mathrm{A}}\right)  Y_{\lambda\mu}^{\ast}\left(  \boldsymbol{\hat
{r}}_{\mathrm{A}}\right)  ,
\label{3trden}
\end{equation}
where $\boldsymbol{r}_{i}$\ is the coordinate of the $i$th nucleon from the c.m. of A. It is understood that the c.m. motion of the nucleus A has been eliminated in $\Phi_{Im_{I}}$\ and the integration over $\xi$ is carried out with the constraint of%
\begin{equation}
\sum_{i=1}^{A}\boldsymbol{r}_{i}=0.
\end{equation}
The transition density is orthonormalized as%
\begin{equation}
\sqrt{4\pi}\int\rho_{n^{\prime}I^{\prime}nI,\lambda}^{\mathrm{tr}}\left(
r_{\mathrm{A}}\right)  r_{\mathrm{A}}^{2}dr_{\mathrm{A}}=\delta_{\lambda
0}\delta_{n^{\prime}n}\delta_{I^{\prime}I}A.
\end{equation}
The one-body nuclear density of A is thus given by%
\begin{equation}
\rho_{nI}\left(  r_{\mathrm{A}}\right)  =\frac{1}{\sqrt{4\pi}}\rho
_{nInI,0}^{\mathrm{tr}}\left(  r_{\mathrm{A}}\right)  .
\end{equation}
It should be noted that the transition density has the following property%
\begin{equation}
\rho_{nIn^{\prime}I^{\prime},\lambda}^{\mathrm{tr}}\left(  r_{\mathrm{A}%
}\right)  =\frac{\hat{I}^{\prime}}{\hat{I}}\left(  -1\right)  ^{I^{\prime}%
-I}\rho_{n^{\prime}I^{\prime}nI,\lambda}^{\mathrm{tr}}\left(  r_{\mathrm{A}%
}\right)  . \label{3trdensym}%
\end{equation}

Let us first consider the direct part of $U_{c^{\prime}c}$, which is written with the transition density by%
\begin{align}
U_{c^{\prime}c,J_T}^{\mathrm{dr}}\left(  R\right)   &  =\int d\boldsymbol{\hat
{R}}\,\sum_{m_{I}^{\prime}m_L^{\prime}}
\langle  I^{\prime}m_{I}^{\prime}%
L^{\prime}m_L^{\prime}|J_TM_{T}\rangle\,
i^{-P^{\prime}}i^{-L^{\prime}}%
Y_{L^{\prime}m_L^{\prime}}^{\ast}(  \boldsymbol{\hat{R}} ) \nonumber\\
&  \times\sum_{m_{I}m_L}
\langle  Im_{I}lm_L|J_TM_{T}\rangle\,
i^{P}i^{L}Y_{Lm_L}(  \boldsymbol{\hat{R}} )
\sum_{\lambda\mu}
\langle  Im_{I}\lambda
\mu|I^{\prime}m_{I}^{\prime}\rangle\,
\mathcal{U}_{n^{\prime}I^{\prime
}nI,\lambda}^{\mathrm{dr}}\left(  \boldsymbol{R}\right)  ,
\end{align}%
\begin{equation}
\mathcal{U}_{n^{\prime}I^{\prime}nI,\lambda}^{\mathrm{dr}}\left(
\boldsymbol{R}\right)  \equiv\int\bar{g}^{\mathrm{dr}}\left(  s,k_{\mathrm{F}%
;n^{\prime}I^{\prime}nI}\left(  \left\vert \boldsymbol{R}+\boldsymbol{s}%
/2\right\vert \right)  \right)  \rho_{n^{\prime}I^{\prime}nI,\lambda
}^{\mathrm{tr}}\left(  r_{\mathrm{A}}\right)  Y_{\lambda\mu}^{\ast}\left(
\boldsymbol{\hat{r}}_{\mathrm{A}}\right)  d\boldsymbol{r}_{\mathrm{A}},
\end{equation}
where%
\begin{equation}
\boldsymbol{s}\equiv\boldsymbol{r}_{\mathrm{A}}-\boldsymbol{R}%
\end{equation}
and $k_{\mathrm{F};n^{\prime}I^{\prime}nI}$ stands for the Fermi momentum defined by%
\begin{equation}
k_{\mathrm{F};n^{\prime}I^{\prime}nI}\left(  r\right)  =\left(  \frac{3\pi
^{2}}{2}\frac{\rho_{n^{\prime}I^{\prime}}\left(  r\right)  +\rho_{nI}\left(
r\right)  }{2}\right)  ^{1/3}.
\end{equation}
After some manipulation, one finds%
\begin{equation}
U_{c^{\prime}c,J_T}^{\mathrm{dr}}\left(  R\right)  =i^{-P^{\prime}-L^{\prime
}+P+L}\sum_{\lambda}X_{I^{\prime}L^{\prime}IL,J_T\lambda}\mathcal{\bar{U}%
}_{n^{\prime}I^{\prime}nI,\lambda}^{\mathrm{dr}}\left(  R\right)  ,
\end{equation}%
with
\begin{equation}
X_{I^{\prime}L^{\prime}IL,J_T\lambda}\equiv\sqrt{2I^{\prime}+1}\sqrt{2L+1}%
\sqrt{2\lambda+1}\frac{\left(  -\right)  ^{I^{\prime}+L-J_T}}{\sqrt{4\pi}%
}W\left(  ILI^{\prime}L^{\prime};J_T\lambda\right)  
\langle  L0\lambda
0|L^{\prime}0\rangle,
\end{equation}%
\begin{equation}
\mathcal{\bar{U}}_{n^{\prime}I^{\prime}nI,\lambda}^{\mathrm{dr}}
\left(
R\right)  \equiv4\pi\sum_{l=0}^{\lambda}\frac{\sqrt{2\left(  \lambda-l\right)
+1}}{\left(  2l+1\right)  \sqrt{2\lambda+1}}\sqrt{_{2\lambda+1}C_{2l}%
}R^{\lambda-l}
\langle  l0,\lambda-l,0|\lambda0\rangle\,
\int\mathcal{F}%
_{n^{\prime}I^{\prime}nI,\lambda l}^{\mathrm{dr}}\left(  R,s\right)
s^{l+2}ds,
\end{equation}%
\begin{equation}
\mathcal{F}_{n^{\prime}I^{\prime}nI,\lambda l}^{\mathrm{dr}}\left(
R,s\right)  \equiv\frac{2l+1}{2}\int_{-1}^{1}\bar{\rho}_{n^{\prime}I^{\prime
}nI,\lambda}^{\mathrm{tr}}\left(  \left\vert \boldsymbol{R}+\boldsymbol{s}%
\right\vert \right)  \bar{g}^{\mathrm{dr}}\left(  s,k_{\mathrm{F};n^{\prime
}I^{\prime}nI}\left(  \left\vert \boldsymbol{R}+\boldsymbol{s}/2\right\vert
\right)  \right)  P_{l}\left(  x\right)  dx,
\end{equation}%
\begin{equation}
\bar{\rho}_{n^{\prime}I^{\prime}nI,\lambda}^{\mathrm{tr}}\left(
r_{\mathrm{A}}\right)  \equiv r_{\mathrm{A}}^{-\lambda}\rho_{n^{\prime
}I^{\prime}nI,\lambda}^{\mathrm{tr}}\left(  r_{\mathrm{A}}\right)  ,
\label{3rhotrbar}%
\end{equation}
where $W\left(  abcd;ef\right)$ is the Racah coefficient, $_{2\lambda
+1}C_{2l}$ is the binomial coefficient, and%
\begin{equation}
x\equiv\frac{\boldsymbol{R}\cdot\boldsymbol{s}}{Rs}.
\end{equation}

To calculate the exchange part of $U_{c^{\prime}c}$, we follow the localization method proposed by Brieva and Rook (BR)~\cite{brieva1977} for NA elastic scattering. The resulting form is%
\begin{equation}
U_{c^{\prime}c,J_T}^{\mathrm{ex}}\left(  R\right)  =i^{-P^{\prime}-L^{\prime
}+P+L}\sum_{\lambda}X_{I^{\prime}L^{\prime}IL,J_T\lambda}\mathcal{\bar{U}%
}_{n^{\prime}I^{\prime}nI,\lambda}^{\mathrm{ex}}\left(  R\right)  ,
\end{equation}%
\begin{equation}
\mathcal{\bar{U}}_{n^{\prime}I^{\prime}nI,\lambda}^{\mathrm{ex}}\left(
R\right)  =4\pi\sum_{l=0}^{\lambda}\frac{\sqrt{2\left(  \lambda-l\right)  +1}%
}{\left(  2l+1\right)  \sqrt{2\lambda+1}}\sqrt{_{2\lambda+1}C_{2l}}%
R^{\lambda-l}
\langle  l0,\lambda-l,0|\lambda0\rangle\,
\int j_{0}\left(
K\left(  R\right)  s\right)  \mathcal{F}_{n^{\prime}I^{\prime}nI,\lambda
l}^{\mathrm{ex}}\left(  R,s\right)  s^{l+2}ds,
\end{equation}%
\begin{align}
\mathcal{F}_{n^{\prime}I^{\prime}nI,\lambda l}^{\mathrm{ex}}\left(
R,s\right)   &  =\frac{2l+1}{2}\int_{-1}^{1}\bar{\rho}_{n^{\prime}I^{\prime
}nI,\lambda}^{\mathrm{tr}}\left(  \left\vert \boldsymbol{R}+\boldsymbol{s}%
/2\right\vert \right)  \bar{g}^{\mathrm{ex}}\left(  s;k_{\mathrm{F};n^{\prime
}I^{\prime}nI}\left(  \left\vert \boldsymbol{R}+\boldsymbol{s}/2\right\vert
\right)  \right) \nonumber\\
&  \times\frac{3j_{1}\left(  k_{\mathrm{F};n^{\prime}I^{\prime}nI}\left(
\left\vert \boldsymbol{R}+\boldsymbol{s}/2\right\vert \right)  s\right)
}{k_{\mathrm{F};n^{\prime}I^{\prime}nI}\left(  \left\vert \boldsymbol{R}%
+\boldsymbol{s}/2\right\vert \right)  s}P_{l}\left(  x\right)  dx,
\end{align}
where $j_{0}$ ($j_{1}$) is the spherical Bessel function of the first kind for order 0 (1). The NA local wave number $K\left(  R\right)  $\ is self-consistently determined to satisfy%
\begin{equation}
\frac{\hbar^{2}}{2\mu}K^{2}\left(  R\right)  +\mathcal{\bar{U}}%
_{n_{0}I_{0}n_{0}I_{0},0}^{\mathrm{dr}}\left(  R\right)  +\mathcal{\bar{U}%
}_{n_{0}I_{0}n_{0}I_{0},0}^{\mathrm{ex}}\left(  R\right)  =E_{n_{0}I_{0}}.
\end{equation}
It should be noted that in Ref.~\cite{amos2000}, the nonlocality in the exchange term was explicitly treated, without the BR localization. 

Thus, the coupling potential of central type is given by%
\begin{equation}
U_{c^{\prime}c,J_T}\left(  R\right)  =i^{-P^{\prime}-L^{\prime}+P+L}%
\sum_{\lambda}X_{I^{\prime}L^{\prime}IL,J_T\lambda}\mathcal{\bar{U}}_{n^{\prime
}I^{\prime}nI,\lambda}\left(  R\right)  ,
\end{equation}%
\begin{equation}
\mathcal{\bar{U}}_{n^{\prime}I^{\prime}nI,\lambda}\left(  R\right)
\equiv\mathcal{\bar{U}}_{n^{\prime}I^{\prime}nI,\lambda}^{\mathrm{dr}}\left(
R\right)  +\mathcal{\bar{U}}_{n^{\prime}I^{\prime}nI,\lambda}^{\mathrm{ex}%
}\left(  R\right). \label{ubarna}%
\end{equation}
Although it is rather trivial, only the channels having the same parity are coupled. When this condition is satisfied, $U_{c^{\prime}c}$ is symmetric under interchange of $c^{\prime}$ and $c$. In numerical calculations of reaction processes, $\bar{\rho}_{n^{\prime}I^{\prime}nI,\lambda}^{\mathrm{tr}}$ of Eq.~(\ref{3rhotrbar}) is the primary input to be prepared with nuclear structure calculations. It will be worth pointing out that $U_{c^{\prime}c}$\ is an energy-dependent and complex potential. It is sometimes argued that the physical origin of the imaginary part of the transition potential, $U_{c^{\prime}c}$ for $c^{\prime}\neq c$, is not clear and it is naively disregarded in some cases. As shown above, the MST clearly gives the origin of the imaginary part of $U_{c^{\prime}c}$. There is no justification to assume $U_{c^{\prime}c}$ to be 
real when $c^{\prime}\neq c$, at least from the viewpoint of the MST.

The spin-orbit part of the coupling potential can be formulated in a similar manner. The explicit form of the diagonal spin-orbit potential for the elastic channel is given in Refs.~\cite{furumoto2008,toyokawa2013}. There are a few applications of the MCC framework including both central and spin-orbit interactions to inelastic scattering~\cite{kelly1992,furumoto2021}.

\subsection{Extended NAF model for alpha-nucleus scattering}
\label{secappa4}

In the extended NAF model, the $\alpha$A coupling potential $U_{c^{\prime} c,J}^{\alpha\mathrm{A}}$ is calculated by folding the NA coupling potential with the one-body density $\rho^{\left( \alpha\right)}$ of $\alpha$; explicit couplings through excited (breakup) channels of $\alpha$ are disregarded because of its large binding energy. It is given with $\mathcal{\bar{U}}_{n^{\prime}I^{\prime}nI,\lambda}$ of Eq.~(\ref{ubarna}) by%
\begin{equation}
U_{c^{\prime}c,J_T}^{\alpha\mathrm{A}}\left(  R\right)  =i^{-P^{\prime
}-L^{\prime}+P+L}\sum_{\lambda}X_{I^{\prime}L^{\prime}IL,J_T\lambda}%
\sum_{\Lambda=0}^{\lambda}U_{n^{\prime}I^{\prime}nI,\lambda\Lambda}%
^{\alpha\mathrm{A}}\left(  R\right)  \frac{4\pi}{\hat{\Lambda}^{2}}%
\sqrt{_{2\lambda+1}C_{2\Lambda}}\left(  -\right)  ^{\Lambda}
\langle
\Lambda0\lambda0|\lambda-\Lambda,0\rangle  ,
\end{equation}%
\begin{equation}
U_{n^{\prime}I^{\prime}nI,\lambda\Lambda}^{\alpha\mathrm{A}}\left(  R\right)
\equiv\int\rho^{\left(  \alpha\right)  }\left(  s_{\alpha}\right)
\mathfrak{\bar{U}}_{n^{\prime}I^{\prime}nI,\lambda\Lambda}\left(  s_{\alpha
},R\right)  s_{\alpha}^{\Lambda}R^{\lambda-\Lambda}s_{\alpha}^{2}ds_{\alpha},
\end{equation}%
\begin{equation}
\mathfrak{\bar{U}}_{n^{\prime}I^{\prime}nI,\lambda\Lambda}\left(  s_{\alpha
},R\right)  =\frac{\hat{\Lambda}^{2}}{2}\int_{-1}^{1}\frac{\mathcal{\bar{U}%
}_{n^{\prime}I^{\prime}nI,\lambda}\left(  \left\vert \boldsymbol{s}_{\alpha
}+\boldsymbol{R}\right\vert \right)  }{\left\vert \boldsymbol{s}_{\alpha
}+\boldsymbol{R}\right\vert ^{\lambda}}P_{\Lambda}\left(  y\right)  dy,
\end{equation}%
\begin{equation}
y\equiv\frac{\boldsymbol{R}\cdot\boldsymbol{s}_{\alpha}}{Rs_{\alpha}},
\end{equation}
where $\boldsymbol{s}_{\alpha}$ is the coordinate of each nucleon in $\alpha$ with respect to the c.m of $\alpha$. Because of its simplicity, this extended NAF model has been applied to various $\alpha$ inelastic scattering processes~\cite{kanada-en'yo2019,kanada-en'yo2019b,kanada-en'yo2020,kanada-en'yo2020b,kanada-en'yo2020c,kanada-en'yo2020d,kanada-en'yo2021,kanada-en'yo2021b}. The CC equations (\ref{3cceq}) can be used as they are because $\alpha$ is a spinless particle.

\subsection{Nucleus-nucleus scattering}
\label{secapp5}

There are several applications of the MCC framework to nucleus-nucleus (AA) scattering~\cite{takashina2010,furumoto2013,minomo2016,furumoto2018}; here we consider target nuclei heavier than $\alpha$. For a complete formulation of the MCC for AA scattering, we need to change the form of the channel wave function to include the spins of both particles. To avoid complexity, we do not show the explicit form of the wave function and coupling potential in this article. Readers are referred to the literature, for instance, Refs.~\cite{furumoto2013,minomo2016}.
\end{appendix}

\bibliography{review-hom}
\bibliographystyle{elsarticle-num}

\end{document}